\documentclass{aa}
\usepackage{graphicx}
\usepackage{txfonts}
\usepackage{color}
\usepackage[dvipsnames]{xcolor}
\usepackage{float}
\usepackage[colorlinks=true,linkcolor=blue,filecolor=blue,citecolor=blue,urlcolor=blue]{hyperref}

\usepackage{multirow}

\usepackage{siunitx}
\usepackage{caption}
\DeclareCaptionFormat{cont}{#1#2#3\par}
 
\newcommand{\citel}[1]{\citeauthor{#1}\,\citeyear{#1}}
\newcommand{\Msun}[1]{\,M$_{\odot}$}
\newcommand{\kmpersec}[1]{\,km.s$^{-1}$}

\begin{document} 
   \titlerunning{Finding the birthplace of HMXBs with {\it Gaia} EDR3}
   \authorrunning{F.\,Fortin et al.}

   \title{Finding the birthplace of HMXBs in the Galaxy using {\bf {\emph Gaia}} EDR3:\\kinematical age determination through orbit integration}

   \author{Francis Fortin \inst{1}
          \and
          Federico Garc\'ia \inst{2}
          \and
          Sylvain Chaty \inst{1}
          }

   \institute{Universit\'e Paris Cit\'e, CNRS, Astroparticule et Cosmologie, F-75013 Paris, France
     \and
     Instituto Argentino de Radioastronom\'ia (CCT La Plata, CONICET; CICPBA; UNLP), C.C.5, (1894) Villa Elisa, Buenos Aires, Argentina
   }

   \date{received 18/05/2022 ; accepted 04/07/2022}

 
  \abstract
   {High-Mass X-ray Binaries (HMXBs) are produced after the first supernova event in a massive binary. These objects are intrinsically young, and can suffer from a significant natal kick. As such, the progenitors of HMXBs are likely to have formed away from the current location of the X-ray emitting systems.}
   {We aim to find the birthplace of the known HMXBs of our Milky Way. Specifically, we want to answer the question whether the formation of HMXBs can be associated to open stellar clusters and/or Galactic spiral structures, and infer from that the time elapsed since the first supernova event.}
    {We use astrometric data from the {\it Gaia} EDR3 to initialize the position and velocity of each known HMXBs from the Galaxy, and integrate their motion back in time. In parallel, we perform the same calculations on a sample of 1381 open clusters detected by {\it Gaia} as well as for four Galactic spiral arms which shape and motion have also been recently modelled using {\it Gaia} data. We report on all the encounter candidates between HMXBs and clusters or spiral arms in the past 100\,Myr.}
   {In our sample of 26 HMXBs, we infer that 7 were born in clusters, 8 were born near a Galactic spiral arm, and conclude that 7 others could have formed isolated. The birthplaces of the remaining 4 HMXBs are still inconclusive due to a combination of great distance, poor astrometric data and lack of known open cluster in the vicinity. We provide the kinematical age since supernova of 15 HMXBs.
   }
   {The astrometry from {\it Gaia} and the orbit integration we employ are effective at finding the birthplaces of HMXBs in the Milky Way. By considering the biases in our data and method, we find it is likely that the progenitors of HMXBs preferentially formed alongside other massive stars in open clusters.}

   \keywords{   X-ray:binaries --
                population --
                neutron star:kicks --
                black hole: kicks --
                optical: astrometry --
                stars: formation --
                stars: evolution
            }

   \maketitle
%

\section{Introduction}

Massive stars in binary systems can evolve through various phases, one of which is the High-Mass X-ray Binary (HMXB) phase. It comes right after the initially most massive star collapses into a compact object -- either a neutron star (NS) or a black hole (BH) -- and starts feeding on its companion star. HMXBs can be the progenitors of double compact object binaries, which radiate gravitational waves during their inspiral and final merger. This scenario can be very altered -- if not completely prevented -- depending on what happens during key evolutionary phases such as the mass transfers episodes, the kicks imparted by supernova events (\citel{lai_pulsar_2001} and \citeauthor{podsiadlowski_effects_2004}\citeyear{podsiadlowski_neutron-star_2005}) or the common envelope phase \citep{ivanova_common_2013}.

Retracing the life of X-ray binaries to find their birthplace is a mean to constrain their evolutionary paths as it provides insights on their formation mechanisms, as well as on their evolution timescales, which can be very different from isolated stellar evolution mostly due to mass transfer. In the case of HMXBs, one would expect them to be formed in the same place as the general population of massive OB stars, i.e. young clusters of stars or Galactic spiral arms, since at least the primary star should be massive \citep{tauris_formation_2006}. However, massive star formation is not necessarily tied to associations or spiral arms, as there are examples of them being born isolated from such Galactic structures (see e.g. \citel{harada_formation_2019}).

Several studies have already pointed out a correlation of HMXBs with OB associations and/or spiral arms, either in close galaxies (e.g. \citel{kaaret_displacement_2004}, \citel{rangelov_connection_2011} and \citel{bodaghee_evidence_2021}) or in the Milky Way (\citel{bodaghee_clustering_2012} and \citel{coleiro_distribution_2013}) on sets of tens of sources. Several studies also target specific HMXBs in order to infer their kinematical age (\citel{ankay_origin_2001}, \citel{ribo_astrobjls_2002}, \citel{mirabel_formation_2003}, \citel{mirabel_microquasar_2004}, \citel{gvaramadze_4u_2011} and \citel{hambaryan_origin_2022}).

Age estimations such as these are precious in the scope of binary evolution: kinematical data can provide ages regardless of the physical evolution of the stars due to binary interaction (mass transfer, wind mass loss, rotation, etc). Hence, they offer good constraints to retrace the history of observed binaries and help to depict the full picture of evolution scenarios for interacting binaries. Having some insight about the nature of the birthplace of HMXBs could also provide interesting constraints for population synthesis models, such as a more accurate spatial distribution of X-ray emitting binaries within modelled galaxies.

The tracing of stellar motion back in time requires integrating their orbit based on the set of currently observed position and velocity; thus the astrometrical data must be of exquisite accuracy. The early third data release of {\it Gaia} ({\it Gaia} EDR3 hereafter, \citel{gaia_collaboration_gaia_2016} and \citel{gaia_collaboration_gaia_2021}) provides just that, as it is the most complete and accurate optical astrometric survey of Galactic sources to date. This is a good opportunity to infer kinematical ages for known HMXBs of the Milky Way.

In this paper, we propose to integrate the motion of known HMXBs in the Milky Way using {\it Gaia} EDR3 to test the hypotheses of them being born in clusters or Galactic spiral arms, and discuss the implications in terms of binary evolution. The information that we use on the position and motion of stellar clusters, as well as the shape of Galactic spiral arms is also resulting from {\it Gaia} EDR3. We work under the hypothesis that both stars in the HMXBs were formed at the same place and time, which is the expected dominant formation channel of binary systems \citep{clarke_theories_1995}, although we do not investigate exactly how the formation of the massive stars occurs (disk fragmentation, pre-main sequence accretion, etc).

We start by presenting the HMXB sample we work with (Section\,\ref{sect:HMXB_sample}). Then, we describe the adopted models for the shape and evolution of the Galactic spiral arms, as well as the method we employ to detect any candidate encounter between HMXBs and those arms and how we determine the corresponding age and peculiar velocity associated using purely astrometrical considerations (Section\,\ref{sect:arm}). We then describe how we use a similar method only applied to a sample of open clusters detected by {\it Gaia} EDR3 in order to identify any interactions with HMXB progenitors (Section\,\ref{sect:oc}). We discuss the accuracy of our method and our general results (Section\,\ref{sect:discuss}), and also provide focused discussions on individual birthplace candidates of binaries by considering extra astrophysical parameters (Section\,\ref{sect:cases}) before concluding (Section\,\ref{sect:conclusion}).

\section{High-Mass X-ray Binary sample}\label{sect:HMXB_sample}

We based ourselves on the observed HMXBs in the Milky Way that we compiled in a previous work \citep{fortin_constraints_2022}. To summarize, the HMXBs were retrieved from \cite{liu_catalogue_2006}, a catalogue dedicated to HMXBs, and completed by more recent discoveries from the fourth catalogue of INTEGRAL sources \citep{bird_ibis_2016}. As a significant number of those sources were still at the stage of candidate HMXBs, we manually checked every catalogued HMXBs to ensure that they have no ambiguity in their identification (high energy to optical--infrared counterparts and early spectral type of the companion). The systems with a confirmed HMXB status were queried in {\it Gaia} EDR3 and, when possible, associated to a single {\it Gaia} counterpart.

Out of the 129 binaries we retrieved in the Milky Way, 94 have a secured HMXB identification. Of those, 80 have an unambiguous counterpart in {\it Gaia} EDR3. Since we are dealing with binary systems, the radial velocity (RV) measured by {\it Gaia} is unlikely to match the true systemic velocity. Hence we searched in the literature which of the HMXBs have a proper radial velocity follow-up using spectroscopy; only 25 have a measure of RV we can use. We made an exception of XTE J0421+560 (CI Cam) which has been determined to be seen pole-on \citep{hynes_spectroscopic_2002}; as such, we reckon we can use the RV measurement from \cite{aret_spectroscopic_2016} for this HMXB, which pushes our sample size to 26 HMXBs. We present the astrometric data we retrieved for those 26 HMXBs in Table\,\ref{tab:astrometry}.

\begin{table*}
\caption{Astrometric information of the 26 HMXBs selected for this study.}\label{tab:astrometry}
\small
\begin{tabular}{lrrrlrrcc}
\hline\hline
HMXB & {\it Gaia} EDR3 ID & RA (ICRS) & Dec (ICRS) & Distance & pmra & pmdec & RV & ref. \\
     &                    & [deg]     & [deg]      & [kpc]    & [mas.yr$^{-1}$] & [mas.yr$^{-1}$] & [\kmpersec\,] & \\
\hline
IGR J00370+6122 & 427234969757165952 & 9.2901 & 61.3601 & 3.40$^{+0.19}_{-0.17}$ & -1.8$\pm$0.01 & -0.53$\pm$0.01 & -80.0$\pm$3.0 & [1] \\
1A 0114+650 & 524924310153249920 & 19.5112 & 65.2916 & 4.48$^{+0.22}_{-0.18}$ & -1.24$\pm$0.01 & 0.76$\pm$0.01 & -31.0$\pm$5.0 & [2] \\
LS I+61 303 & 465645515129855872 & 40.1319 & 61.2293 & 2.50$^{+0.07}_{-0.07}$ & -0.42$\pm$0.01 & -0.26$\pm$0.01 & -41.4$\pm$0.6 & [3] \\
X Per & 168450545792009600 & 58.8462 & 31.0458 & 0.60$^{+0.02}_{-0.01}$ & -1.28$\pm$0.05 & -1.87$\pm$0.03 & 1.0$\pm$0.9 & [4] \\
XTE J0421+560 & 276644757710014976 & 64.9256 & 55.9994 & 4.09$^{+0.28}_{-0.21}$ & -0.47$\pm$0.02 & -0.51$\pm$0.01 & -51.0$\pm$2.0 & [5] \\
1A 0535+262 & 3441207615229815040 & 84.7274 & 26.3158 & 1.79$^{+0.08}_{-0.07}$ & -0.59$\pm$0.03 & -2.88$\pm$0.02 & -30.0$\pm$4.0 & [6] \\
IGR J06074+2205 & 3423526544838563328 & 91.8609 & 22.0966 & 5.99$^{+0.56}_{-0.6}$ & 0.57$\pm$0.02 & -0.61$\pm$0.01 & 18.9$\pm$4.1 & [7] \\
HD 259440 & 3131822364779745536 & 98.2469 & 5.8003 & 1.77$^{+0.09}_{-0.09}$ & -0.03$\pm$0.02 & -0.43$\pm$0.02 & 36.9$\pm$0.8 & [8] \\
IGR J08408-4503 & 5522306019626566528 & 130.1991 & -45.0584 & 2.22$^{+0.13}_{-0.07}$ & -7.46$\pm$0.02 & 6.10$\pm$0.02 & 15.3$\pm$0.5 & [9] \\
Vela X-1 & 5620657678322625920 & 135.5286 & -40.5547 & 1.96$^{+0.06}_{-0.05}$ & -4.82$\pm$0.01 & 9.28$\pm$0.02 & -3.2$\pm$0.9 & [10] \\
2FGL J1019.0-5856 & 5255509901121774976 & 154.7316 & -58.9461 & 4.32$^{+0.21}_{-0.2}$ & -6.45$\pm$0.01 & 2.26$\pm$0.01 & 33.0$\pm$3.0 & [11] \\
Cen X-3 & 5337498593446516480 & 170.3129 & -60.6238 & 6.78$^{+0.63}_{-0.57}$ & -3.12$\pm$0.02 & 2.33$\pm$0.01 & 32.0$\pm$13.0 & [12] \\
1E 1145.1-6141 & 5334851450481641088 & 176.8689 & -61.9537 & 8.10$^{+0.6}_{-0.57}$ & -6.23$\pm$0.01 & 2.36$\pm$0.01 & -13.0$\pm$3.0 & [13] \\
GX 301-2 & 6054569565614460800 & 186.6565 & -62.7704 & 3.60$^{+0.2}_{-0.2}$ & -5.23$\pm$0.02 & -2.07$\pm$0.02 & 4.1$\pm$2.4 & [14] \\
PSR B1259-63 & 5862299960127967488 & 195.6985 & -63.8357 & 2.17$^{+0.06}_{-0.06}$ & -7.09$\pm$0.01 & -0.34$\pm$0.01 & 0.0$\pm$1.0 & [15] \\
4U 1538-522 & 5886085557746480000 & 235.5973 & -52.3860 & 5.61$^{+0.49}_{-0.43}$ & -6.71$\pm$0.01 & -4.11$\pm$0.01 & -158.0$\pm$11.0 & [16] \\
4U 1700-377 & 5976382915813535232 & 255.9866 & -37.8441 & 1.50$^{+0.05}_{-0.06}$ & 2.41$\pm$0.03 & 5.02$\pm$0.02 & -60.0$\pm$10.0 & [17] \\
IGR J17544-2619 & 4063908810076415872 & 268.6053 & -26.3313 & 2.43$^{+0.16}_{-0.13}$ & -0.51$\pm$0.03 & -0.67$\pm$0.02 & -46.8$\pm$4.0 & [18] \\
SAX J1802.7-2017 & 4070968778561141760 & 270.6747 & -20.2882 & 8.53$^{+4.12}_{-2.44}$ & -3.41$\pm$0.12 & -5.49$\pm$0.09 & 51.7$\pm$2.4 & [19] \\
LS5039 & 4104196427943626624 & 276.5628 & -14.8484 & 1.90$^{+0.06}_{-0.05}$ & 7.43$\pm$0.01 & -8.15$\pm$0.01 & 17.3$\pm$0.5 & [20] \\
IGR J18410-0535 & 4256500538116700160 & 280.2518 & -5.5963 & 13.88$^{+3.7}_{-2.71}$ & -2.07$\pm$0.05 & -5.38$\pm$0.04 & 74.4$\pm$2.1 & [21] \\
IGR J18450-0435 & 4258160560148155648 & 281.2566 & -4.5658 & 5.44$^{+0.65}_{-0.64}$ & -1.37$\pm$0.02 & -5.59$\pm$0.02 & 61.0$\pm$1.4 & [21] \\
SS 433 & 4293406612283985024 & 287.9565 & 4.9827 & 7.29$^{+1.19}_{-0.86}$ & -3.03$\pm$0.02 & -4.78$\pm$0.02 & 69.0$\pm$4.7 & [22] \\
Cyg X-1 & 2059383668236814720 & 299.5903 & 35.2016 & 2.15$^{+0.06}_{-0.05}$ & -3.81$\pm$0.01 & -6.31$\pm$0.02 & -7.0$\pm$5.0 & [23] \\
4U 2206+543 & 2005653524280214400 & 331.9843 & 54.5184 & 3.10$^{+0.13}_{-0.14}$ & -4.17$\pm$0.02 & -3.32$\pm$0.01 & -54.5$\pm$1.0 & [24] \\
MWC 656 & 1982359580155628160 & 340.7387 & 44.7217 & 1.98$^{+0.08}_{-0.08}$ & -3.48$\pm$0.02 & -3.16$\pm$0.02 & -14.1$\pm$2.1 & [25] \\
\hline

\end{tabular}
\tablefoot{
All the astrometry but the distances and radial velocities come from the {\it Gaia} EDR3 archives \citep{gaia_collaboration_gaia_2021}. The distances are inferred from {\it Gaia} parallaxes by \cite{bailer-jones_estimating_2021}.

{\bf References. }
[1] \cite{grunhut_orbit_2014}
[2] \cite{koenigsberger_optical_2003}
[3] \cite{aragona_orbits_2009}
[4] \cite{grundstrom_joint_2007}
[5] \cite{aret_spectroscopic_2016}
[6] \cite{hutchings_spectroscopic_1984}
[7] \cite{chojnowski_high-resolution_2017}
[8] \cite{moritani_orbital_2018}
[9] \cite{gamen_eccentric_2015}
[10] \cite{stickland_orbit_1997}
[11] \cite{strader_optical_2015}
[12] \cite{van_der_meer_determination_2007}
[13] \cite{hutchings_orbital_1987}
[14] \cite{kaper_vltuves_2006}
[15] \cite{johnston_radio_1994}
[16] \cite{abubekerov_masses_2004}
[17] \cite{gies_binary_1986}
[18] \cite{nikolaeva_investigation_2013}
[19] \cite{mason_masses_2011}
[20] \cite{casares_new_2011}
[21] \cite{gonzalez-galan_fundamental_2015}
[22] \cite{picchi_optical_2020}
[23] \cite{gies_wind_2003}
[24] \cite{stoyanov_orbital_2014}
[25] \cite{casares_be-type_2014}

}
\end{table*}

\section{Galactic spiral arm encounter candidates}\label{sect:arm}
Here we describe the method we use to determine encounter candidates between Galactic spiral arms and the HMXBs along their orbit in the Milky Way, and how we estimate the age of these encounter candidates.

\subsection{Spiral arm model}\label{subsect:arm:model}
A study by \cite{castro-ginard_milky_2021} built a sample of 2017 open clusters that were either already known to be present in {\it Gaia} DR2 \citep{cantat-gaudin_gaia_2018}, or newly detected in the same data release (\citel{cantat-gaudin_gaia_2019}, \citeyear{cantat-gaudin_painting_2020}, \citel{sim_207_2019} and \citel{liu_catalog_2019}). This sample was updated using the new astrometric parameters available in {\it Gaia} EDR3. Using that sample, they re-determine the current shape of the Perseus, Local, Sagittarius and Scutum Galactic spiral arms. \cite{castro-ginard_milky_2021} also provided an estimation of the angular speed of each spiral patterns $\Omega _{p}$.

We used the provided shape and velocity parameters to integrate back in time the position and movement of each of the four spiral arms. In parallel, we integrated the orbital motion of each HMXB around the Milky Way using their currently observed astrometric parameters from {\it Gaia} EDR3. The integration is performed with the python package \textsc{Galpy} along with the galactic potential solution \textsc{MWPotential2014} from \cite{bovy_galpy_2015} with no additional wrapper. This 2D potential takes into account the contribution of the Galactic bulge, disk and a surrounding dark matter halo; neither the central black hole nor the central bar are part of the model, but we did not investigate any source that comes anywhere close to the Galactic center. Another caveat could be the presence of a Galactic warp, which precession was recently quantified in \cite{poggio_evidence_2020} and may cause a perturbation in the potential that we did not take into account here. The set of parameters we used for the normalisation of the Galactic rotation curve are R$_{\odot}$=8.178\,kpc \citep{gravity_collaboration_geometric_2019} and Z$_{\odot}$=20.8\,pc \citep{bennett_vertical_2019} for the position of the Sun and U$_{\odot}$=11.1, V$_{\odot}$=12.24, W$_{\odot}$=7.25\,\kmpersec\, and V$_{LSR}$=236.26\,\kmpersec\, (\citel{schonrich_local_2010} and \citel{reid_proper_2020}).

Since all astrometric parameters come with uncertainties, a single integration of the orbit of the HMXBs using the mean parameters is unlikely to be representative of the true orbit, especially the more we integrate back in time. As such, we chose to follow a bootstrapping method and integrated each orbit several times using a set of drawn astrometric parameters according to Gaussian distributions, centered on their average and with a standard deviation set on the parameter uncertainty. This results, for each epoch back in time, in a distribution of possible positions within the Milky Way.

\subsection{Age of encounter}\label{subsect:arm:age}

We integrated motions up to 100\,Myr back in time. We estimated that looking further back would not be of use since massive stars composing HMXBs should not live that long: according to the mass-age relation given in \cite{figueiredo_influence_1991}, a 10\Msun\, star should not be able to stay longer than $\sim$80\,Myr on the main sequence. We split the time span into 500 time steps of 0.2\,Myr, and performed 1000 iterations of orbit integration for each binary. For each time step of each iteration, we found the closest point of each spiral arm to the binary by minimising the distance that can be obtained using the equations describing the arms. We note respectively (R$_B$, $\theta_B$) and (R$_G$, $\theta_G$) the galactocentric polar coordinates of a given binary and of all the points constituting a spiral arm. Using the equation linking the radius and the azimuth of the spiral arms given in \cite{reid_trigonometric_2014} that also involves the pitch angle $\psi$ of the galactic arm, the distance between a binary and a spiral arm can be expressed as a function of $\theta_G$ alone:

\begin{equation}
\begin{split}
D^2(t, \theta_G) =&R_B^2(t) + R_{G,ref}^2 e^{-2(\theta_G-\theta_{G,ref}(t))\,tan\,\psi}\\
&-2R_B(t)\,R_{G,ref}\,e^{-(\theta_G-\theta_{G,ref}(t))\,tan\,\psi} cos(\theta_B(t)-\theta_G)
\end{split}
\end{equation}

with $\theta_{G,ref}(t)$ = $\Omega_p$t + $\theta_{G,ref}$ taking into account the evolution of the reference azimuth of the Galactic spiral arms with their rotation around the Milky Way. Finding the value of $\theta_G$ that minimizes D(t,$\theta_G$) is done by the \textsc{minimize} routine of the \textsc{Scipy.optimize} package.

We produced the evolution of the minimum distance distributions with time (time-distance histograms hereafter). We chose to reduce them to relevant percentiles in the corresponding figures found in Appendix\,\ref{appendix}, which are the median and the percentiles corresponding to the 1, 2 and 3\,$\sigma$ of the distributions ([0.16, 0.02275, 0.00135] for the lower bounds and [0.84, 0.97725,0.99865] for the upper bounds).

For each binary with identified encounter candidates, we present the distributions of their distance to Galactic spiral arm projected onto the Galactic Plane (X and Y galactocentric coordinates), since the spiral arms are assumed to be strictly located at Z=0. Encounter candidates were selected by using the 2D time-distance histograms that are produced by the bootstrap procedure. They were convolved with the histogram of the typical width of a galactic spiral arm, that depends on the galactocentric radius of the point considered, which is time-dependent. We used the value of 42\,pc\,kpc$^{-1}$ provided in \cite{reid_trigonometric_2014} for the typical width of all spiral arms. After marginalizing the convolution over the distance, we obtained a probability distribution of binary-spiral arm encounter with time.

An encounter candidate between a spiral arm and a binary is thus identified when the probability distribution shows a single, well-defined peak. When that happens, we defined the encounter age as the centroid of the peak, obtained by fitting a Gaussian to the distribution. The uncertainty we provide on this measurement is obtained using the aforementioned percentiles. The upper and lower bounds of the confidence intervals correspond to the epochs at which the lower 3\,$\sigma$ curve is equal the the distance of the lower 2\,$\sigma$ curve taken at the determined age.

This bootstrap method is a mean to quickly recover encounter candidates between HMXBs and Galactic structures. In the case we would be dealing with an HMXB that truly encountered an arm, the closest physical distance we retrieve would be extremely sensitive to the initial astrometrical parameters, as well as to the number of iterations and the time-step of integration. Hence, this method does not provide an estimation of the distance of encounter; it provides an age at which an encounter is possible given a very specific set of astrometrical parameters drawn from the measurements and their uncertainty. Once candidate encounters are identified, we can discuss their potential as birthplace candidates using our knowledge of individual HMXBs, which is what we do in Section\,\ref{sect:cases}.

\subsection{Peculiar velocity}\label{subsect:arm:vpec}
Once a candidate encounter has been identified, it is possible to get an estimation of the peculiar velocity of a binary in the case it was indeed born at that time and place, then kicked-out after undergoing a supernova event. During the orbit integration, we also kept track of the evolution of the velocity vector of each binary. Hence we could isolate the data points that fell within the age determination, and we computed the peculiar velocity by subtracting the Galactic co-rotation velocity that is given by \cite{bovy_galpy_2015}. Depending on the size of the encounter age confidence interval, we obtained distributions made of ten to seventy-five thousand peculiar velocities, which we also characterized by their median, 16$^{th}$ and 84$^{th}$ percentiles to provide a single estimate of the peculiar velocity at the time of interaction with the spiral arms.

In our sample of 26 HMXBs, we find that 18 have at least one encounter candidate with a Galactic Spiral arm. These encounters are compiled in Table\,\ref{tab:arm}, and the detailed time-distance histograms of each encounter is available in Appendix \ref{appendix}. In these figures the distance is computed in 2D on the Galactic plane, as the Galactic height of the HMXBs stay contained within $\pm$500\,pc above or below it, which is compatible with the Galactic disk.

\section{Open cluster encounter candidates}\label{sect:oc}
In this section we describe the methods employed to look for encounter candidates between known open clusters of stars and our set of HMXBs. We first build the astrometric data set of the open clusters, then integrate their motion back in time and compare their orbit to each of our binaries. We isolate, among the hundreds of clusters, a selection of candidates that may be the birthplace of the binaries, provide an estimation of their age since supernova as well as of the resulting peculiar velocity.

\subsection{Open cluster data set}\label{subsect:oc:data}
The set of open clusters that is used in \cite{castro-ginard_milky_2021} to trace the Galactic spiral arms can also be used in itself. Since the authors do not provide the astrometric data set of each open cluster, we had to rebuild it following their procedure. First, we started by retrieving all the members of the 2017 open clusters listed by \cite{cantat-gaudin_painting_2020} from a table available on the Centre de Donn\'ees astronomiques de Strasbourg\footnote{\url{http://cdsarc.u-strasbg.fr/viz-bin/cat/J/A+A/640/A1}}. We updated each of the members astrometric parameters from {\it Gaia} DR2 to {\it Gaia} EDR3 by querying their {\it Gaia} identifiers or their position in the case the former changed between release. We retained only the members for which the \textit{parallax\_over\_error} parameters from {\it Gaia} EDR3 is greater than 1, to discard sources with poor parallax determinations. This resulted in a homogeneous data set comprised of the sky position ($\alpha$, $\delta$) and the proper motion ($\mu_{\alpha}$, $\mu_{\delta}$) of the members of open clusters. We retrieved the distances and radial velocities separately. Estimations of the distance to {\it Gaia} EDR3 sources are taken from \cite{bailer-jones_estimating_2021}. The estimations of the radial velocities of the open clusters are directly provided by \cite{tarricq_3d_2021}; this however limited the open cluster sample to 1381, since not all clusters found in \cite{cantat-gaudin_painting_2020} have a radial velocity estimation. We computed average astrometric parameters by taking the median and 1\,$\sigma$ percentiles of the distributions across the members of each open cluster.

\subsection{Binary-cluster encounter candidates}\label{subsect:oc:candidates}
We followed the same method that is used in Section\,\ref{sect:arm}, however this time we also had to integrate the motion of the open clusters. To get a full picture of the possible orbits according to the uncertainties on the astrometric parameters, we performed the same bootstrapping for open clusters on a grid of 500 time steps of 0.2\,Myr and 1000 iterations.

We produced, for each HMXBs, a time-distance histogram using each of the clusters potential orbits. To identify any encounter candidate, we first compared the lower 3\,$\sigma$ bound of the distance evolution to the typical radius of the open clusters, which we computed using their distance and the standard deviation of the sky positions of their members. However, some clusters can have very low physical sizes (down to 1 to 2\,pc), and it is unlikely that we are able to retrieve such close-by approaches with our method. Instead, we chose to investigate any encounter candidate that crossed below the threshold of 3 times the maximum cluster radius found in our sample, which is 20\,pc. While this will raise the chances of false positive, we prefer to bring the probability of a false negative as close to zero as possible.

As we already mentioned in Section\,\ref{subsect:arm:age}, the minimum distance we retrieve using the 3\,$\sigma$ lower bound is not an accurate measurement of the true distance of approach in the case of an actual encounter. Because our method explores every combination of 6 different astrometrical parameters along with their statistical and systematic uncertainties, it is expected that only a small fraction of the bootstrap iterations results in close approaches, hence the use of the 3\,$\sigma$ lower bound to identify candidate encounters and thin-out clusters that are not worth investigating further. Then, their validity as birthplace candidates is discussed in Section\,\ref{sect:cases}.

For those clusters, we defined the age of encounter in the same manner as we did for encounters with Galactic spiral arms. The time-distance histograms were convolved with a Gaussian of standard deviation equal to the typical radius of the cluster, then marginalized over distance and fitted with a Gaussian to retrieve the most likely encounter age. Uncertainties on that age are obtained using the 3\,$\sigma$ and 2\,$\sigma$ lower bounds of the time-distance histograms as well.

Finally, we further filtered these encounter age results by comparing them to the cluster ages determined in \cite{cantat-gaudin_painting_2020} by isochrone fitting. Since currently observed HMXBs were generated by massive stars, if a cluster is indeed associable to a binary then it should not be that much older; hence we discarded clusters that are older than 100\,Myr. Moreover, if a cluster is younger than the earliest encounter estimate with a binary, then we assumed the encounter is spurious and discarded it. All the unambiguous HMXB to open cluster encounter candidates that we found using this method are summarized in Table\,\ref{tab:cluster}, and the associated time-distance histograms can be found in Appendix \ref{appendix}.

\section{General discussion}\label{sect:discuss}

\subsection{Accuracy of the encounter detection}\label{subsect:accurasim}

To quantify the ability of our method to find encounter candidates between HMXBs and clusters that are truly associated, we performed a set of simulations involving dummy clusters and binaries.

We first initialized a cluster at a random position in the Galaxy, and set its velocity as the Galactic co-rotation speed plus a random peculiar velocity. We chose to draw the direction of the peculiar velocity isotropically in the sphere, and its magnitude is drawn from a Gaussian distribution with a 15\kmpersec\,width according to the typical velocity dispersion in clusters reported by \cite{gieles_velocity_2010}. The cluster effective radius was randomly picked following the distribution in radii in our sample of open clusters.

We then generated an HMXB in the vicinity of that cluster at a random position picked from a Gaussian of width corresponding to the radius of the cluster, clipped at 3$\sigma$. We also imprinted to it a peculiar velocity drawn the same way as we do for the cluster, although the average of the Gaussian drawing was set at the cluster's velocity instead of the Galactic co-rotation speed. The width, still set at 15\kmpersec\,,is probably overestimated as \cite{gieles_velocity_2010} also argue that it might be broadened by the orbital motion of binaries present in the clusters.

We integrated the motions of the binary and the cluster forward in time, up until a randomly drawn time that corresponds to the lifetime of the primary star in the binary. At this point, it goes through a supernova event and kicks the system away. To model the effect of the supernova kick, we added to the HMXB's velocity a random value drawn from a uniform distribution between 0--100\kmpersec\,taken isotropically on the sphere.

We resumed the integration of motion forward in time, up until an age that is again randomly drawn and corresponds to the age since supernova we aim to recover. The Zero-Age Main-Sequence (ZAMS hereafter) lifetime of the primary and the age since supernova are drawn so that their addition, representing the full lifetime of the binary, does not go beyond 100\,Myr.

\begin{figure}
    \includegraphics[width=\columnwidth]{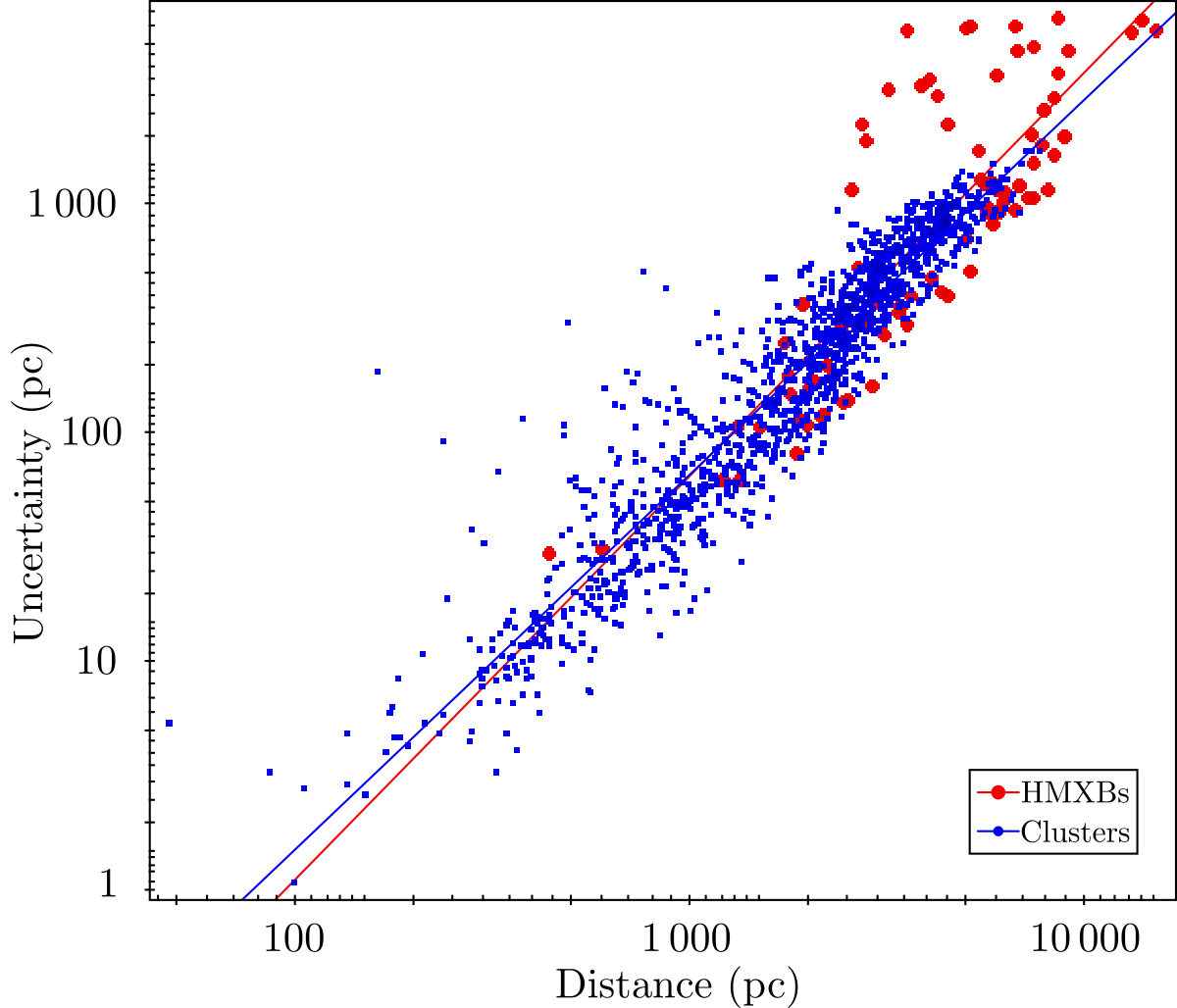}
    \caption{Logarithmic regression between the distance inferred from {\it Gaia} parallax and the associated uncertainty. The HMXB sample is constituted of 80 sources.}
    \label{fig:logdist}
\end{figure}

At this point, we took a snapshot of the 6-D data of both binary and cluster, and generated a set of simulated parameters emulating the results of {\it Gaia} observations. We first drew the uncertainties on those parameters according to the distribution we get from our sample of binaries and clusters. For distances, there is a strong correlation between the average and the uncertainty, that can be roughly fitted in the log-log plane by a line parametrized as in Eq.\ref{eq:logdist}. We fitted (m=1.76; c=-3.47) for the set of HMXBs, and (m=1.64; c=-3.09) for the set of clusters (see Figure\,\ref{fig:logdist}).

\begin{equation}\label{eq:logdist}
    log(D) = log(\sigma_D)\times m + c
\end{equation}

\begin{figure}
    \centering
    \includegraphics[width=\columnwidth]{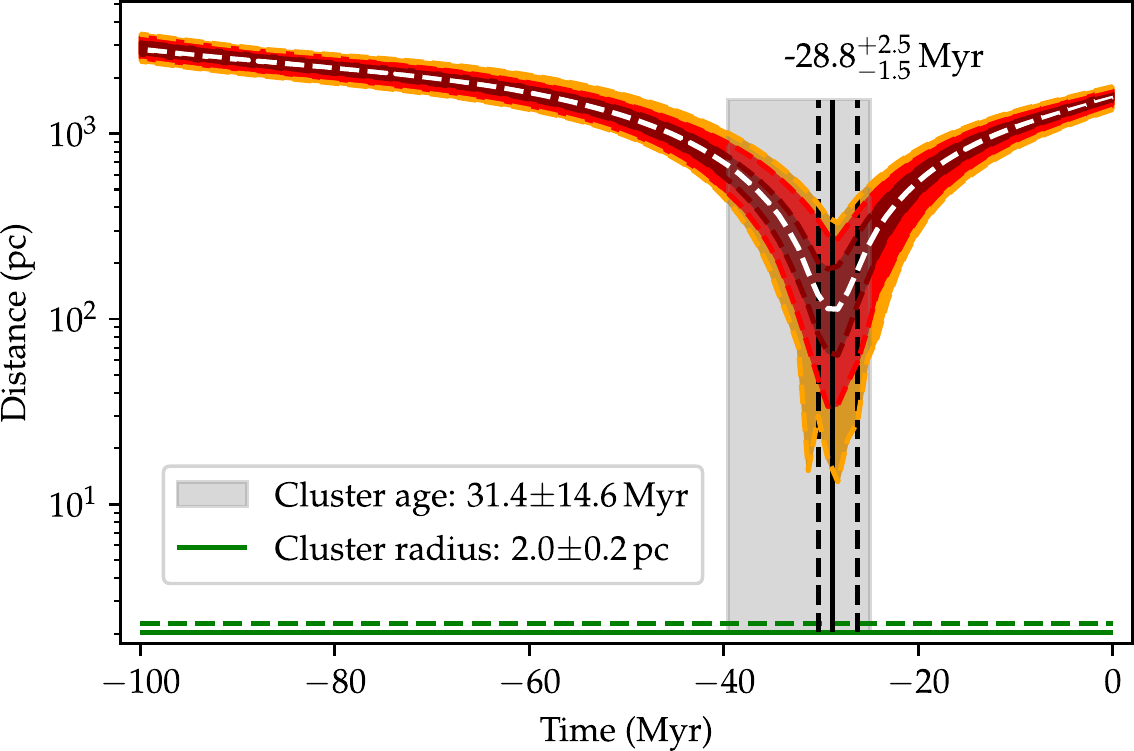}
    \caption{Simulated time-distance histogram between a dummy binary located at 950\,pc and its birth cluster with good astrometrical constraints. All of the median (dotted white) and the 1 to 3\,$\sigma$ intervals (dark red to yellow) admit a single minimum that coincides with the age since supernova, which was set at -29\,Myr.}
    \label{fig:sim_good}
\end{figure}

\begin{figure}
    \centering
    \includegraphics[width=\columnwidth]{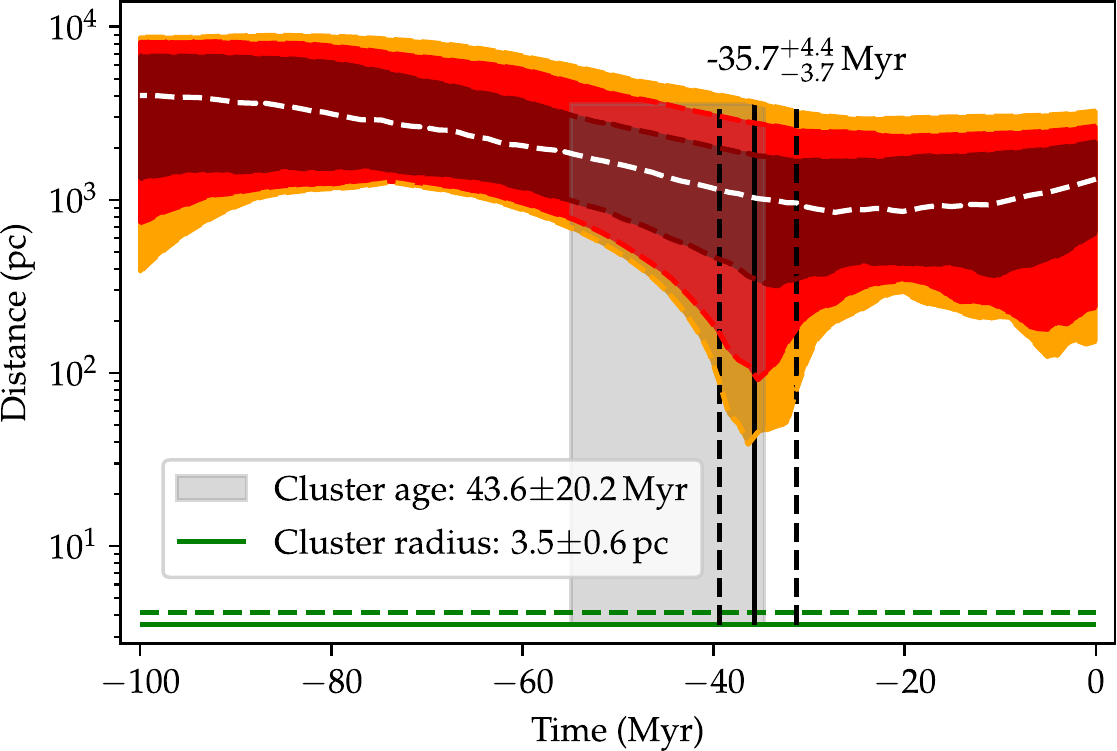}
    \caption{Simulated time-distance histogram between a dummy binary located at 1.4\,kpc and its birth cluster with average astrometrical constraints. The local minimum around -5\,Myr is ignored by our method, and even though the median of the distance (dotted white) has no clearly identifiable minimum, the 3\,$\sigma$ threshold of the distance distribution (yellow) still provides a good indicator for the age since supernova set at -37\,Myr.}
    \label{fig:sim_ave}
\end{figure}

We then drew the average of each parameter in a Gaussian which width is proportional to 5\% of the uncertainty, to allow for a systematic error to be present in the simulated data. We then applied our method of encounter detection described in Section\,\ref{subsect:oc:candidates} to this dummy data set. We repeated the whole process 1\,000 times to obtain an estimation of the false detection rate, and quantified the reasons when the detections fail. In 89.4\% of the outcomes, the true birth date of the binary lies within 1$\sigma$ of our estimation, shown as vertical plain and dotted black lines in Figure\,\ref{fig:sim_good} and \ref{fig:sim_ave}.

\begin{figure}
    \centering
    \includegraphics[width=\columnwidth]{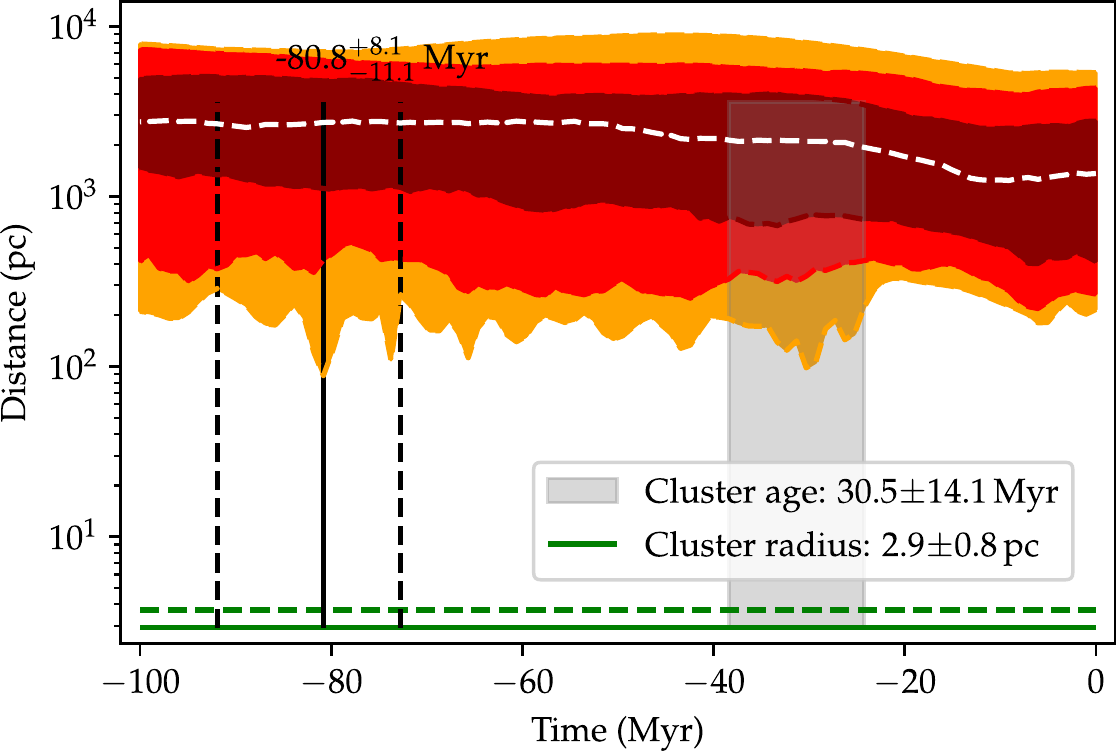}
    \caption{Simulated time-distance histogram between a dummy binary and its birth cluster with bad astrometrical constraints. The greatest source of uncertainty comes from the distance to the binary (6$\pm$1.5\,kpc). The evolution of the distance between them does not change much with time; such a flat probability distribution does not allow us to retrieve the age since supernova, which was set at -7\,Myr.}
    \label{fig:sim_bad}
\end{figure}

In the remaining 10.6\% so-called failed attempts, there is a notable decrease in the general astrometric parameter quality. This produces flattened time-distance histograms as shown in Figure\,\ref{fig:sim_bad}, throwing our age estimations way off. In both the simulated binaries and clusters, failed attempts have an average SNR on the distance measurement that is half of the one in successful attempts. This is also the case for the distribution in uncertainties for proper motion and radial velocity, although it is more pronounced in the set of simulated clusters than binaries. All in all, failed detections begin to happen for far away binaries ($>$5\,kpc) and/or binaries with overall poor astrometric quality. In our case, since far away binaries do not have any nearby clusters to be confronted to, the limiting factor to finding cluster encounters will not be the quality of the astrometric data, but the number of clusters observed by {\it Gaia} and detected in the latest data release.

\subsection{Encounter rate vs cluster spatial density}
\begin{figure}[hbt!]
    \centering
    \includegraphics[width=\columnwidth]{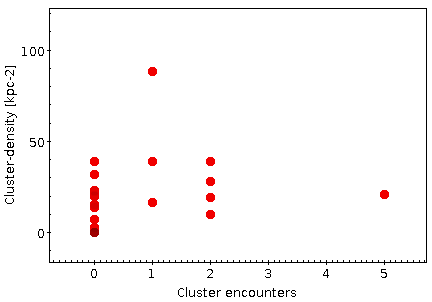}
    \caption{Number of detected encounter candidates with open clusters vs the local cluster density associated to each HMXB in a 1\,kpc radius.}
    \label{fig:enc_vs_density}
\end{figure}

Our method produces false positives, since we have cases in which binaries have multiple encounter candidates with different clusters. We computed the average cluster density in a radius of 1\,kpc around each binary of our data set and confronted it to the number of encounters we detect for each binary. As shown in Figure\,\ref{fig:enc_vs_density}, there is no clear correlation between the cluster density and the encounter rate. In fact, we have both extreme cases: Cyg X-1 is the binary with the highest local cluster density despite having a single cluster encounter candidate, and LS\,I+61\,303 has the highest encounter number (4) despite being located in a relatively uncrowded region.

\subsection{Determining initial mass of both stars in HMXBs}
In the case a cluster with a measured age can be unambiguously identified as the birthplace of an HMXB, we have access to two timescales: the time spent in the ZAMS by the primary star, and the age since its supernova event (assuming it indeed went through one). As a first approach, we can use these timescales to roughly estimate the ZAMS mass of both stars through simple mass-luminosity relation of type $L \propto M^{\alpha}$ which can be reformulated into a ZAMS mass-age relation (eq. \ref{eq:age}).

\begin{equation}\label{eq:age}
\frac{M}{M\sun} = \left[ 10^{-4}\left(\frac{t_{ZAMS}}{Myr} \right) \right]^{\frac{1}{1-\alpha}}
\end{equation}

The parameters of the mass-luminosity relationship for massive stars (M$\geq$10\,M\sun) are yet to be unambiguously set, as several variants of the equations are proposed in the literature. We settled for a very simplistic approach and worked with a fixed value of $\alpha=3.125$, which is the average value derived by \cite{figueiredo_influence_1991} for 5\,M\sun\,$\leq$\,M\,$\leq$\,60\,M\sun.

On top of the uncertainties brought by the mass-luminosity relationship, there are caveats to the masses determined by this method. Firstly, the primary star should always be of high mass; without the knowledge of its metallicity, it is hard to know exactly how massive it was at birth since much can be lost even during main sequence through stellar wind. Secondly, the initial mass transfer from the primary to the secondary may also impact the time to supernova, although in a more limited way in case B Roche Lobe overflow. One of the biggest uncertainty is the amount of mass that was effectively transferred from the primary to the secondary (noted Xfer hereafter), which can be inferred by comparing its estimated ZAMS mass and its current mass; without advanced hydrodynamical simulations, we can only provide an upper limit to the secondary ZAMS mass if we assume it formed at the same time as the stellar cluster, as well as the corresponding lower limit on the mass transferred to bring it to its current mass. All of those parameters are compiled alongside the cluster encounter candidates in Table\,\ref{tab:cluster}.

\subsection{Comparison with previous studies}

The most notable works linking HMXBs with star forming structures in the Milky Way are from \cite{bodaghee_clustering_2012} and \cite{coleiro_distribution_2013} (B12 and CC13 hereafter). Both propose to study the clustering of HMXBs and OB associations, and mostly view the HMXBs and clusters as a population. The former adopt a statistical clustering method, and the latter propose age determinations and migration distances by finding spatial offsets between Galactic spiral arms and their HMXB sample.

Since we use a sample of HMXBs with a secured identification, and which have both a {\it Gaia} counterpart and a proper radial velocity measurement, this study has few HMXBs in common with B12 and CC13. The limiting factors of those studies are the spatial constraints on the HMXBs. The distance to the binaries in CC13 were estimated by fitting their optical--infrared spectral energy distribution with blackbodies, assuming that the bulk of the emission comes from the companion star. Hence, the distance estimate is heavily influenced by the accuracy of the spectral type inferred on the companion stars of HMXBs. We find that the relative difference between distances from CC13 and the ones we have using \cite{bailer-jones_estimating_2021} is 35\% in average, or 1\,400\,pc in absolute units. B12 anticipated that {\it Gaia} would significantly improve our knowledge in the spatial distribution of HMXBs. Hence, we are confident that we bring results on age and peculiar velocity that are much more secure on individual HMXBs thanks to {\it Gaia} data.

The average time since supernova in HMXBs from the Milky Way is 15$\pm$10\,Myr from our results, which is compatible with the 4\,Myr inferred by B12 (which is rather low) and almost compatible with the value we can derive from Table\,4 of CC13 at 49$\pm$23\,Myr. We note that this value from CC13 is more in line with the maximum age since supernova we retrieve from our sample (38$\pm$4\,Myr for HD\,259440, or 35$^{+19}_{-15}$\,Myr for 1E\,1145.1-6141 in the case it is associated to the Local arm). Along with the shortest estimation of time since supernova of $\sim$2\,Myr (4U\,1700-377 and Cen\,X-3), this provides an interesting time span between the supernova event and the onset of the X-ray emitting phase. Assuming that the X-ray phase is very short-lived compared to these timescales, it shows that there can be systems that almost instantly begin accreting after the supernova, while some others might have to wait a couple dozen million years for that to happen probably due to their Roche lobe geometry.

The peculiar velocities that result from our birthplace candidates are in overall good agreement with the values that we previously determined in \cite{fortin_constraints_2022}, although a few HMXBs (X\,Per, HD\,259440, 4U\,1700-377 and 4U\,2206+543) do have a higher peculiar velocity than expected. Except from 4U\,2206+543, we have an open cluster as birthplace candidate for those HMXBs, hence the peculiar velocities we derive are likely to be more accurate than the previous estimations since they were taken using the co-rotating Galactic velocity as reference. However, even in the present case where we use the velocity of the birth cluster as a reference, the true peculiar velocity of the HMXBs will still be blurred by a random dispersion imprinted at birth. Taking that into consideration, the actual difference with previous measurements becomes marginal. To quantify this, we have tested all of those HMXBs using the method employed in \cite{fortin_constraints_2022} to infer the neutron star natal kick, only using the peculiar velocity distributions obtained in the present study. We did not find any significant difference in the neutron star natal kicks obtained with the new values of peculiar velocity. Hence, bringing more observational constraints on natal kicks is unlikely to come from better constrained peculiar velocity measurements. Instead, observational natal kicks may benefit from a larger sample size to help understanding the different explosion mechanisms e.g. asymmetry, electron capture or stripped events.


\begin{table}
\caption{Encounter age with Galactic spiral arms and corresponding peculiar velocity.}\label{tab:arm}

\begin{tabular}{llcc}
\hline\hline
HMXB & Arm & Encounter age   & $v_{pec}$  \\
     &     & [Myr]           & [km\,s$^{-1}$] \\
\hline
IGR J00370+6122& Perseus & -56$^{+19}_{-13}$ & 43$^{+3}_{-3}$ \\[1ex]

1A 0114+650& Scutum & -99$^{+5}_{-5}$ & 47.9$^{+0.4}_{-0.4}$ \\[1ex]

X Per& Local & -3.4$^{+2.0}_{-1.8}$ & 10$^{+1}_{-1}$ \\[1ex]

XTE J0421+560& Scutum & -96$^{+7}_{-4}$ & 29.4$^{+0.6}_{-0.6}$ \\[1ex]

1A 0535+262& Perseus & -18$^{+6}_{-10}$ & 49.7$^{+0.4}_{-0.7}$ \\[1ex]

IGR J06074+2205& Scutum & -81$^{+11}_{-11}$ & 22$^{+1}_{-1}$ \\[1ex]

IGR J08408-4503& Local & -40$^{+3}_{-2}$ & 36$^{+2}_{-2}$ \\[1ex]

Vela X-1& Perseus & -77$^{+6}_{-7}$ & 74$^{+1}_{-1}$ \\
& Local & -28$^{+1}_{-1}$ & 47$^{+1}_{-1}$ \\[1ex]

2FGL J1019.0-5856& Sagittarius & -10$^{+1}_{-3}$ & 42$^{+1}_{-1}$ \\[1ex]

Cen X-3& Sagittarius & -1.9$^{+1.9}_{-5.3}$ & 78$^{+11}_{-4}$ \\[1ex]

1E 1145.1-6141& Local & -35$^{+15}_{-19}$ & 81$^{+7}_{-11}$ \\
& Sagittarius & -15$^{+14}_{-19}$ & 77$^{+14}_{-9}$ \\
& Perseus & -80$^{+26}_{-16}$ & 71$^{+2}_{-4}$ \\[1ex]

GX 301-2& Perseus & -83$^{+11}_{-8}$ & 43$^{+2}_{-3}$ \\
& Scutum & -76$^{+12}_{-10}$ & 40$^{+4}_{-3}$ \\[1ex]

4U 1538-522& Perseus & -4.4$^{+2.8}_{-2.2}$ & 81$^{+8}_{-8}$ \\
& Local & -15$^{+4}_{-7}$ & 89$^{+12}_{-6}$ \\[1ex]

4U 1700-377& Perseus & -92$^{+12}_{-7}$ & 77$^{+2}_{-2}$ \\
& Local & -41$^{+2}_{-1}$ & 89.4$^{+0.1}_{-0.2}$ \\
& Sagittarius & -31$^{+3}_{-3}$ & 82$^{+2}_{-2}$ \\[1ex]

IGR J17544-2619& Perseus & -100$^{+8}_{-8}$ & 47.3$^{+0.4}_{-0.5}$ \\
& Scutum & -54$^{+9}_{-7}$ & 49.8$^{+0.3}_{-1.3}$ \\[1ex]

IGR J18450-0435& Perseus & -23$^{+9}_{-10}$ & 59$^{+2}_{-2}$ \\
& Local & -57$^{+28}_{-12}$ & 59$^{+3}_{-3}$ \\
& Sagittarius & -89$^{+13}_{-9}$ & 61$^{+7}_{-6}$ \\
& Scutum & -17$^{+11}_{-8}$ & 62$^{+2}_{-3}$ \\[1ex]

SS\,433 & Sagittarius & -20 -- 0 & 32$_{-9}^{+8}$ \\[1ex]

4U 2206+543& Perseus & -40$^{+19}_{-35}$ & 38.3$^{+0.7}_{-0.6}$ \\[1ex]

\hline

\end{tabular}

\end{table}

\begin{table*}
\centering
\caption{Encounter age with open clusters and corresponding peculiar velocity.}\label{tab:cluster}
\begin{tabular}{llccccc}

\hline\hline
{HMXB} & {Cluster} & {Encounter age}   & {$v_{pec}$} & {M1$_{\mathrm{ZAMS}}$} & {M2$_{\mathrm{ZAMS}}$} & {Xfer} \\
     &         & {[Myr]} & {[km\,s$^{-1}$]} & {[\Msun\,]} & {[\Msun\,]} & {[\Msun\,]} \\
\hline
1A 0114+650&UBC 600&-4.4$^{+4.4}_{-3.4}$&37$^{+11}_{-10}$ &13$^{+1}_{-1}$&12$^{+2}_{-1}$&4$\pm$3 \\
&NGC\,457&-17$^{+4}_{-4}$&50$^{+14}_{-11}$ &40$^{+120}_{-12}$&18$^{+3}_{-2}$&... \\

LS I+61 303&IC 1805&-5$^{+5}_{-15}$&9$^{+13}_{-6}$&55$^{+\infty}_{-30}$&31$^{+11}_{-5}$&...\\
&UBC 191&-16$^{+6}_{-6}$&12$^{+5}_{-3}$ &39$^{+120}_{-13}$&19$^{+2}_{-2}$&... \\
&NGC 884&-15$^{+4}_{-8}$&14$^{+3}_{-3}$ &51$^{+110}_{-22}$&20$^{+3}_{-2}$&... \\
&NGC 957&-15$^{+7}_{-8}$&18$^{+4}_{-4}$ &15$^{+2}_{-1}$&12$^{+5}_{-1}$&1$\pm$3 \\
&ASCC 9&-10$^{+5}_{-19}$&33$^{+30}_{-16}$ &18$^{+10}_{-4}$&15$^{+2}_{-1}$&... \\

X\,Per&ASCC 13&-24$^{+8}_{-9}$&33$^{+4}_{-4}$ &28$^{+120}_{-8}$&15$^{+2}_{-1}$&1$\pm$4 \\

HD\,259440&Stock 10&-38$^{+4}_{-4}$&31$^{+1}_{-1}$ &12.7$^{+0.5}_{-0.5}$&10$^{+1}_{-1}$&6$\pm$3 \\
&COIN-Gaia 28&-47$^{+32}_{-25}$&11$^{+4}_{-4}$ &28$^{+130}_{-14}$&15$^{+2}_{-1}$&4$\pm$3 \\

4U\,1700-377&UBC 323&-1.9$^{+0.5}_{-0.5}$&110$^{+26}_{-25}$ &31$^{+1}_{-1}$&27$^{+4}_{-3}$&19$\pm$6 \\
&NGC 6231&-2.1$^{+0.5}_{-0.7}$&62$^{+12}_{-8}$ &24.6$^{+0.6}_{-0.6}$&23$^{+3}_{-2}$&23$\pm$6 \\

IGR J17544-2619&UBC 571&-5.4$^{+1.0}_{-0.8}$&37$^{+3}_{-4}$ &14.4$^{+0.2}_{-0.1}$&14$^{+2}_{-1}$&10$\pm$3 \\
&UBC 540&-28$^{+4}_{-8}$&54$^{+10}_{-10}$ &14$^{+1}_{-1}$&11$^{+2}_{-1}$&12$\pm$3 \\

Cyg X-1&NGC 6871&-4.4$^{+4.4}_{-8.0}$&26$^{+16}_{-12}$ &56$^{+110}_{-28}$&32$^{+4}_{-3}$&... \\

\hline
\end{tabular}
\end{table*}

\section{Results on individual sources}\label{sect:cases}
In this section, we discuss each encounter candidates we detected between our sample of HMXBs and Galactic spiral arms and/or stellar clusters. To assess their potential as birthplace candidates, we assume in the following that they are true encounters and compare the resulting timescales with the knowledge already available on the HMXBs. We select the encounters that make good birthplace candidates and compile our results in Table\,\ref{tab:birthplaces}.

\subsection{IGR J00370+6122}
There is a single encounter with the Perseus arm 56\,Myr ago. Even taking into account the large uncertainties on that measurement, the encounter is not compatible with the ZAMS lifetime of the secondary star in IGR J00370+6122 based on its mass (22\Msun\, taken from \citel{grunhut_orbit_2014}, corresponding to t$_{ZAMS}$=14\,Myr).

Despite the HMXB being located in a region with a cluster density of almost 39\,kpc$^{-2}$, we do not find any cluster encounter. This makes IGR J00370+6122 a sound candidate for being born isolated from stellar formation structures.

\subsection{1A 0114+650}
We detect a single encounter with the Scutum arm around 100\,Myr ago, at the limit of our age range. The encounter is not compatible with the ZAMS lifetime of the secondary based on its mass of 16\Msun\, \citep{hu_evolution_2017}, corresponding to t$_{ZAMS}$=28\,Myr.

Two clusters are encountered at -17\,Myr (NGC 457) and at -4.4\,Myr (UBC 600). Both are compatible age-wise with the current parameters of the secondary star. The time-distance histograms would be in favor of UBC\,600 being the best birthplace candidate even though the absolute closest approaches are the same in both cases. 1A 0114+650 approaches NGC\,457 at about twice its typical radius at best ($\sim$20\,pc), while it reaches below the radius of UBC\,600 (19.2\,pc).

\subsection{LS I+61 303}
This gamma-ray binary does not have any encounter with the Galactic spiral arms in the last 100\,Myr. While the cluster density in a 1\,kpc radius of 21\,kpc$^{-2}$ around this system fits with the average in our sample, we detect five encounters with different clusters at epochs between -5 and -16\,Myr. The quality of the time-distance histograms of those encounters does not allow us to easily pinpoint a single, unambiguous birthplace candidate. In all of the potential orbits of LS I+61 303, the closest approach happens with NGC\,884 at -15.4\,Myr. The least convincing encounters are the ones with IC\,1805 and ASCC\,9, which do not produce well-defined peaks in the time-distance histograms, as we would expect from our simulations in Section\,\ref{subsect:accurasim} in the case they would indeed be associated.

We note that the three best birthplace candidates (NGC\,884, NGC\,957 and UBC\,191) all give age estimates compatible with the maximum lifetime of the 12.5\Msun\, secondary star (based on its spectral type, \citel{martins_new_2005}), which amounts to about 47\,Myr.

\cite{mirabel_microquasar_2004} suggested that LS\,I+61\,303 might originate from the IC\,1805 association, and give a kinematical age that is compatible to what we find on that cluster. Our data suggest that IC\,1805 is not the best birthplace candidate.

\subsection{X\,Per}
A lower bound on the age of the companion at 6\,Myr is given by \cite{fabregat_astrophysical_1992} from the spectral type (O9.5III, \citel{slettebak_spectral_1982}). There is a single encounter with the Local arm at 3.4$\pm$2\,Myr, which is slightly too early to be compatible with this age estimation.

There is however a cluster encounter candidate, ASCC 13, crossing the path of X Per 24.4$\pm$9\,Myr ago, compatible with the maximum age of 29.5\,Myr attainable by the 15.5\Msun\, secondary \citep{lyubimkov_fundamental_1997}. The cluster is measured to be 33$\pm$15\,Myr old. Using the age-mass relation in Eq.\,\ref{eq:age}, the primary would have been roughly 27.6\Msun\, at birth. The maximum ZAMS mass of the secondary is 14.7\Msun, meaning it should have gained at least about 1\Msun\, of material from the primary during initial mass transfer, bringing it to its current mass of 15.5\Msun\,.

Based on those results, ASCC\,13 is the best birthplace candidate for the Be X-ray binary X\,Per.

\subsection{XTE J0421+560 (CI Cam)}
We do not have a reliable mass measurement for this source. Considering that the secondary star is a B0-2\,I[e] supergiant \citep{hynes_spectroscopic_2002}, the encounter with the Scutum arm we detect at -95\,Myr\, is very likely to be spurious. While we do not detect any cluster encounter either, it could very well be due to the local cluster density around XTE J0421+560 being very low (2.54\,kpc$^{-2}$). Hence, we find neither evidence for a birthplace candidate, nor for an isolated formation of this binary.

\subsection{1A 0535+262}
This binary had a single encounter with the Perseus arm 18.3\,Myr\, ago. The secondary is weighed at 20\Msun\, \citep{okazaki_natural_2001}, which translates into a ZAMS lifetime of about 17.2\,Myr. Considering the uncertainties on those values, there are possible scenarios of this binary indeed being formed in the Perseus arm from the encounter we detect 18.3\,Myr ago.

There are no cluster encounter despite this binary being located in a region with a cluster density of 23\,kpc$^{-2}$.

\subsection{IGR J06074+2205}
The single encounter with the Scutum arm at 80.8\,Myr is not in line with the measured mass of the secondary (14.6\Msun\, estimated from its spectral type using \citel{porter_rotational_1996}), which provides a maximum ZAMS lifetime of 33\,Myr. Hence, this encounter is likely to be spurious.

We do not detect any open cluster encounter either, which is not surprising considering the binary is located far away in a region devoid of open clusters detectable by {\it Gaia}. Without any nearby known clusters to even attempt finding a candidate, we cannot put any reasonable constraints on the origin of IGR J06074+2205.

\subsection{HD 259440}
We do not detect any significant encounter between HD 259440 and any Galactic spiral arm. We do however detect two cluster encounters, with Stock\,10 and COIN\,Gaia\,28 at respectively -38.1 and -46.8\,Myr, although the latter is not well constrained. The time-distance histogram with the Stock\,10 cluster is much more convincing, with all of the median and the 1 to 3$\sigma$ curves showing a global minimum near the estimated encounter age. We thus discard the COIN\,Gaia\,28 encounter as a potential candidate, leaving Stock\,10 as the preferred birthplace of HD\,259440.

Considering the currently measured mass of the secondary star in HD\,259440 (15.7$\pm$2.5\Msun\,, \citel{aragona_hd_2010}) and the timescales associated to the encounter with Stock\,10, it is likely that at least 40\% of its current mass was gained during initial mass transfer. This would also imply that the primary star was stripped from 6.2\Msun\, off of its ZAMS mass of 12.7\Msun. While no pulse period was detected in HD\,259440, the stellar masses at play largely favor the presence of a neutron star in this HMXB.

\subsection{IGR J08408-4503}
The secondary in this binary is quite massive, measured at 33\Msun\, \citep{gamen_eccentric_2015}. This does not match the single encounter we find with the Local arm almost 40\,Myr ago, as the secondary should not have been able to survive at its current weight more than $\sim$6\,Myr. Even though it is located in a region populated by almost 20 clusters per kiloparsec squared, we do not find any encounter candidates for IGR\,J08408-4503. We thus reckon that this HMXB may have formed isolated.

\subsection{Vela X-1}
There are two encounters with Galactic arms (Perseus and Local) at respectively 77 and 27.9\,Myr. Given the relatively high mass of the secondary in Vela X-1 (26$\pm$1\Msun\, \citel{falanga_ephemeris_2015}), these encounters are not good birth site candidates as the secondary would be about three times older than its estimated lifetime of 10\,Myr.

We do not find any cluster encounter candidates either for Vela X-1, in a region where the cluster density is slightly above 20\,kpc$^{-2}$. The supergiant HMXB might have been formed isolated, although it is still possible that our open cluster data is lacking.

\subsection{2FGL J1019.0-5856}
There is a single, well-defined encounter with the Sagittarius arm a little over 10\,Myr ago. The 23\Msun\, secondary \citep{waisberg_echelle_2015,strader_optical_2015} has an estimated ZAMS lifetime of 12.8\,Myr, which is perfectly compatible with this spiral arm being the birthplace of 2FGL J1019.0-5856. We do not find any encounter with stellar clusters from {\it Gaia}.

\subsection{Cen X-3}
The single encounter with the Sagittarius arm 1.9\,Myr ago is compatible with the ZAMS lifetime of the 20.2\Msun\, \citep{van_der_meer_mass_1999} secondary star ($\sim$17\,Myr). 
We do not detect any encounters with open clusters, since there are none  present in the {\it Gaia} data near Cen X-3.

The exact evolutionary stage of the companion star in Cen X-3 is not well-known, as the measures of its spectral type range from O9\,III-V \citep{vidal_spectroscopic_1974} to O6-7\,II-III \citep{ash_mass_1999}. If the donor is already set towards the giant phase, then the system could not have formed much earlier than 17\,Myr ago in the case no mass was gained by the secondary before the HMXB phase.

\subsection{1E 1145.1-6141}
The secondary in this binary weighs 14$\pm$4\Msun\, \citep{hutchings_orbital_1987} corresponding to a ZAMS lifetime between 15 and 75\,Myr. The Perseus arm encounter at 80\,Myr might be spurious even though the age is compatible, since the minimum approach distance is quite high compared to the other encounters. The other two encounters we detect, with the Local and Sagittarius arms at 34.9 and 15.4\,Myr respectively, are more compatible with the lifetime of the secondary. The peculiar velocities that result from these three encounters are compatible with one another.

We do not have any known clusters in the region neighbouring 1E 1145.1-6141, hence we cannot check for any birthplace candidates apart from spiral arms. Thus, the hypothesis that the HMXB formed in a cluster cannot be discarded yet.

\subsection{GX\,301-2}
This HMXB has one of the most massive secondary in our sample (43$\pm$10\Msun\,, \citel{kaper_vltuves_2006}), so it is likely that the supernova event occurred no more that a few million years ago. The encounters with the Perseus and the Scutum arms are both unlikely to be associated with GX\,301-2, as they happened more than 75\,Myr ago.

The open cluster density around GX\,301-2 leans towards the low end in our sample ($\sim$ 13.7\,kpc$^{-2}$); we do not detect any significant encounter with any cluster from our list. Since we have been able to detect multiple encounters at lower cluster densities (in the case of the binary 1A\,0114+650 for instance), for now we categorize GX\,301-2 in the isolated formation candidates, with the caveat of a low neighbouring cluster density.

\subsection{PSR\,B1259-63}
We have not detected any significant encounter whatsoever with this binary system. This is despite PSR\,B1259-63 having very well-determined astrometric parameters, as well as being located in a region with a cluster density of almost 32\,kpc$^{-2}$, which is three times higher than the lower density at which we found previous cluster encounters.

For now, we reckon that PSR\,B1259-63 is a good candidate for going through an isolated formation.

\subsection{4U 1538-522}
We detect two encounters with the Perseus and Local arms at 4.4 and 14.3\,Myr, which are both compatible with the ZAMS lifetime of the secondary set at 12--27\,Myr from its current mass estimation (20\Msun\,, \citel{falanga_ephemeris_2015}). It is noteworthy that even though the age since supernova is quite different for both encounters, they both result in a similar magnitude in the peculiar velocity (90-100\kmpersec\,) which is on the high end of the peculiar velocity range of HMXBs that we measured in a previous study \citep{fortin_constraints_2022}.

4U 1538-522 lies in a region devoid of open clusters to test against, as such we cannot fully discard the hypothesis that the binary formed in one.

\subsection{4U 1700-377}

\cite{ankay_origin_2001} suggest that this HMXB originates from the OB association of Sco OB1, of which NGC 6231 is the core. This is based on astrometric data from Hipparcos, and more recent data from {\it Gaia} DR2 corroborates this finding \citep{van_der_meij_confirming_2021}. The companion star in 4U 1700-377 is an early supergiant (O6\,Ia, \citel{sota_galactic_2011}) weighing 46$\pm$5\Msun\, \citep{abubekerov_mass_2004}. This implies that the compact object formed no more than a couple million years ago, otherwise the companion star would have evolved into a compact object as well.

The orbit of the binary crosses the Perseus, Local and Sagittarius arms at epochs greater than 30\,Myr. These encounters are very likely to be spurious, as it is unlikely that the system is so old.

We detect two cluster encounters (UBC\,323 and NGC\,6231) at very similar epochs, both less than 3\,Myr including uncertainties. The clusters are respectively 8.9 and 13.2\,Myr old. The timescales are a much better match with the spectral type of the companion star than with the Galactic spiral arm associations. Although both cluster encounters result in similar timescales for the binary, they produce very different peculiar velocity in the case they indeed are the birthplace of 4U 1700-377: about 60\kmpersec\, for NGC\,6231 and more than 110\kmpersec\, for UBC\,323.

Given the current estimated mass of the companion star in 4U 1700-377 and the estimated age since supernova associated to both NGC\,6231 and UBC\,323, we infer that the companion star has accreted at least 18 to 23\Msun\, of material during an initial mass transfer phase. This poses a problem in the NGC\,6231 association, since the inferred primary mass (24.6\Msun\,) is very close to the mass needed for the companion to be gained (23.4\Msun\,) to bring it to its current mass. While the uncertainties on these values technically allow for a scenario where the primary almost entirely depletes into its companion, the initial mass and mass transfer inferred using the UBC\,323 association are much less restrictive in that matter, especially considering that the mass transfer we give is a lower limit. While we cannot discard the core of NGC\,6231 as a potential birthplace, we reckon that UBC\,323 is a more reasonable candidate for the birthplace of 4U\,1700-377.

\subsection{IGR J17544-2619}
There are two detected encounters with the Perseus ($\sim$100\,Myr) and the Scutum arm (55\,Myr) which do not match our lifetime estimation of IGR J17544-261: the secondary star, which current mass is of 23$\pm$2\Msun\, \citep{bikmaev_spectroscopic_2017}, should not be older than 15\,Myr.

There are however two cluster encounters at 27.7 (UBC\,540) and 5.4\,Myr (UBC\,571). Even considering the confidence intervals, the former is slightly too old to be a good candidate for a birthplace. The latter, much younger, is more fit to be the birthplace of IGR J17544-2619, in which case the primary of 14.4\Msun\, would have depleted at least 9.5\Msun\, into its companion during initial mass transfer, and likely suffered through a stripped supernova event.

\subsection{SAX\,J1802.7-2017}
Because the distance to SAX\,J1802.7-2017 is very high and not well-constrained (8.5$^{+4}_{-2}$\,kpc, \citel{bailer-jones_estimating_2021}), there are no known clusters to compare its orbit to and we cannot put any relevant constraint on potential encounters with Galactic spiral arms.

\subsection{LS\,5039}
We do not detect any encounter with Galactic spiral arms, nor with any open clusters despite LS\,5039 being located within the Galactic plane in a region with a cluster density of 22.3\,kpc$^{-2}$, which is slightly above-average for our sample (see Figure\,\ref{fig:enc_vs_density}).

The 6-D data we retrieved for this system also happen to be on the top of our sample accuracy-wise. Based on our simulations performed in Subsection\,\ref{subsect:accurasim}, if the binary was indeed born in a nearby open cluster from our sample, we should have been able to detect an unambiguous encounter. Moreover, the spectral type and mass of the companion in the binary (ON6.5\,V and 23\Msun\,, \citel{townsend_orbital_2011} and \citel{casares_possible_2005}) tend to discard the case where we would have put too much constraints on the maximum age of the binaries (100\,Myr) and missed an encounter that happened before that. This is further supported by \cite{casares_possible_2005} who conclude that LS\,5039 actually might bear a black hole, and hence the binary was initially formed by a couple of very massive stars.

\cite{ribo_astrobjls_2002} propose association between LS\,5039 and supernova remnants as well as a HI cavity nearby; however the authors worked with a distance to the system of 2.9$\pm$0.3\,kpc, which is way off from what can be inferred from {\it Gaia} EDR3 (1.90$\pm$0.06\,kpc). \cite{moldon_origin_2012} also attempted to localize the birthplace of LS\,5039 using astrometry from radio and optical observations without much success, since they also struggled with the lack of constraints on the distance to the source. Our failure to obtain any birthplace candidate for the gamma-ray binary using {\it Gaia} EDR3 data, which will probably be the most accurate astrometric data for the years to come, is another hint in favor of an isolated formation scenario for LS\,5039.

\subsection{IGR\,J18410-0535}

Similarly to IGR\,J06074+2205 and SAX\,J1802.7-2017, we cannot constrain the origin of IGR\,J18410-0535 as the binary is likely located well further away than 10\,kpc \citep{bailer-jones_estimating_2021}. There are no evidence of encounters with Galactic spiral arms, and the local cluster density from the {\it Gaia} data is null.

\subsection{IGR J18450-0435}
This binary is the only one to have encounters with all four Galactic spiral arms we consider in this study. According to its spectral type given by \cite{coe_discovery_1996} (O9.5\,I), the companion star in IGR\,J18450-0435 should weigh around 29.6\Msun\, using the tables from \cite{martins_new_2005}. This puts heavy constraints on the maximum formation age of the binary between 6--9.5\,Myr. Only the encounters with the Perseus and Scutum arms at 17 and 22\,Myr are valid in this scope. They both produce peculiar velocities close to 60\kmpersec.

There are no encounters with any known cluster since no data is available in the region neighbouring IGR J18450-0435. While there are encounters with Galactic spiral arms, they provide very loosely constrained age estimations. Our data would suggest spiral arms as birthplace candidates, however we cannot rule out a formation inside a cluster for IGR\,J18450-0435.

\subsection{SS\,433}
The candidate black hole HMXB SS\,433 is located quite far away from us ($>$6\,kpc, \citel{bailer-jones_estimating_2021}), hence there are no known clusters from the {\it Gaia} data we can confront it to. Furthermore, we do not find any firm evidence of encounters with Galactic spiral arms. The time-distance histogram with the Sagittarius arm does allow SS\,433 to have encountered it in the past 20\,Myr, but we cannot isolate a precise epoch. This does match the information brought by the measured mass of the secondary (21$\pm$1\Msun\,, \citel{bowler_ss_2018}), which puts a constraint on the maximum age since supernova at 15$\pm$1.5\,Myr. Thus, we propose that SS\,433 formed near the Sagittarius spiral arm, with the caveats of needing more precise astrometry to constrain the exact age, and have clusters to confront it to.

\subsection{Cyg X-1}
\cite{mirabel_formation_2003} argue that the binary was born about 5\,Myr ago, and the formation of the black hole released only a very little amount of mass (about 1\Msun\,). We detect a single encounter, with the open cluster NGC\,6871 at an age of 4.4\,Myr, although this value is poorly constrained. The cluster formed at around 6.3\,Myr. This would be perfectly compatible with Cyg X-1 being born in that cluster, since if the primary star was massive enough to give birth to a black hole it should have done it in a very short amount of time, which would be about 2\,Myr in this case. This corresponds to an average ZAMS primary mass of 56\Msun\, which is on the higher end of the mass range in our sample of binaries; however since the age to the supernova is poorly constrained and very close to the formation age of the associated cluster, the initial mass of the primary can span from 28 to more than 160\Msun\,.

\subsection{4U 2206+543}
A recent paper from \cite{hambaryan_origin_2022} argue that 4U\,2206+543 can be associated to the Cepheus OB1/NGC 7380 association and provide a kinematical age of 2.8$\pm$0.4\,Myr.  We do not have NGC 7380 in our sample of stellar clusters because of the lack of an average radial velocity measurement.

We detect a single encounter with the Perseus arm around 21--74\,Myr ago, which is just in line with the measured mass of the secondary star (18\Msun\, \citel{zorec_evolutionary_2005}) providing a maximum ZAMS lifetime between 15--32\,Myr. The evolution of the distance between the binary and the spiral arm within the Galactic plane (X and Y axes) is not very well constrained. The minimal approach also coincides with a crossing of the Galactic plane (Z axis) 44\,Myr ago; while this could have been an extra argument in favor of this encounter between the Perseus arm and 4U 2206+543, the binary's orbit does not extends past a few dozen parsecs away from the Galactic plane over the last 100\,Myr. Thus, considering the information on the mass of the companion star of 4U 2206+543, we propose that a possible encounter with the Perseus arm happened between -21 and -32\,Myr.

\subsection{MWC\,656}
We do not detect any encounters between this black hole HMXB and Galactic spiral arms, nor with any open cluster to our knowledge. Similarly to GX\,301-2 or LS\,5039, the nearby cluster density is neither low nor high in comparison with our sample (15.6\,kpc$^{-2}$), and we have example of cluster encounters at lower densities. For now, we reckon that we can make the hypothesis of an isolated formation for MWC\,656.

\section{Conclusion}\label{sect:conclusion}

We studied a sample of 26 HMXBs that have a full set of 6-D data provided by {\it Gaia} EDR3 (position and proper motion) combined with various radial velocity measurements from the literature to trace back their motion in time. The detection of encounter candidates with Galactic spiral arms and/or stellar clusters can provide birthplace candidates to those HMXBs as well as the time elapsed since the first supernova event (Table\,\ref{tab:birthplaces}). We discussed the legitimacy of those encounters using our knowledge of the current spectral type and mass of the donor stars.

After a careful review of each encounter candidate, we determined that there are no conflicting birthplace candidates in our sample. When multiple birthplace candidates exist for a single HMXB, they all are of the same nature (cluster or spiral arm) and resulted in similar age estimations. Either the HMXBs have an associated cluster (7), an associated spiral arm (8) or we conclude that they may have formed isolated (7). For the latter case, we cannot provide any age estimation. For the remaining 4 HMXBs (XTE\,J0421+560, IGR\,J06074+2205, SAX\,J1802.7-2017 and IGR\,J18410-0535), the data we have did not allow us to prefer one of the three formation scenarios over the others. There are 14 sgHMXBs and 12 BeHMXBs in our HMXB sample. Among the sgHMXBs (BeHMXBs), 4 (3) are associated to clusters, 4 (4) are associated with Galactic spiral arms and 3 (4) are candidates for being formed isolated.

\begin{table*}[h]
    \caption{Inferred birthplace of HMXBs observed by {\it Gaia} and corresponding age since the first supernova event.}\label{tab:birthplaces}
    \centering
    \begin{tabular}{llllll}
    \hline\hline
        HMXB & Birthplace & \multicolumn{2}{c}{t$_{SN}$}     & \multicolumn{2}{c}{v$_{pec}$}\\
             &            & \multicolumn{2}{c}{[Myr]}        & \multicolumn{2}{c}{[\kmpersec\,]} \\
             &            & This work & CC13 & This work & F22 \\
    \hline
        IGR J00370+6122   & isolated     & ... &  & & $26.5^{+3.5}_{-3.6}$ \\
        1A 0114+650       & cluster      & 4.4$_{-4.4}^{+3.4}$ & & 37$^{+11}_{-10}$ & $37.5^{+5.0}_{-4.9}$ \\
        LS I+61 303       & cluster      & 15 -- 16 & & 12 -- 18 & $5.5^{+0.7}_{-0.5}$\\
        X Per             & cluster      & 24$_{-8}^{+9}$ & & 33$^{+4}_{-4}$ & $13.7^{+0.1}_{-0.1}$\\
        XTE J0421+560     & ...$^*$      & ... &  & & $13.8^{+1.5}_{-1.2}$\\
        1A 0535+262       & arm          & 18$_{-6}^{+10}$ & 80 & 49.7$^{+0.4}_{-0.7}$ & $40.4^{+3.9}_{-3.9}$\\
        IGR J06074+2205   & ...$^*$      & ... \\
        HD 259440         & cluster      & 38$_{-4}^{+4}$ &  & 31$^{+1}_{-1}$ & $7.5^{+0.8}_{-0.8}$\\
        IGR J08408-4503   & isolated     & ... & & & $39.6^{+2.2}_{-2.0}$ \\
        Vela X-1          & isolated     & ... & & &  $57.5^{+1.7}_{-1.6}$\\
        2FGL J1019.0-5856 & arm$^*$      & 10$_{-1}^{+3}$ & & 42$^{+1}_{-1}$ & $33.6^{+1.9}_{-1.8}$\\
        Cen X-3           & arm$^*$      & 1.9$_{-1.9}^{+5.3}$ & & 78$^{+11}_{-4}$ &  \\
        1E 1145.1-6141    & arm$^*$      & 15 -- 35 & & 77 -- 81 & $50.3^{+10.6}_{-7.8}$ \\
        GX 301-2          & isolated$^*$ & ... & & & $56.3^{+3.4}_{-3.2}$\\
        PSR B1259-93      & isolated     & ... & 60 & & $26.0^{+1.2}_{-1.2}$\\
        4U 1538-522       & arm$^*$      & 4.4 -- 15 & 20 & 80 -- 90 & $73.4^{+9.5}_{-8.0}$\\
        4U 1700-377       & cluster      & 1.9$_{-0.5}^{+0.5}$ & 80 & 110$^{+26}_{-25}$ & $67.5^{+6.5}_{-5.6}$\\
        IGR J17544-2619   & cluster      & 5.4$_{-1.0}^{+0.8}$ & & 37$^{+3}_{-4}$ & $44.6^{+3.9}_{-4.0}$ \\
        SAX J1802.7-2017  & ...$^*$      & ... \\
        LS 5039           & isolated     & ... &  & & $89.1^{+2.8}_{-2.6}$\\
        IGR J18410-0535   & ...$^*$      & ... & 60 \\
        IGR J18450-0435   & arm$^*$      & 17 -- 23 & & 59 -- 62 & \\
        SS 433            & arm$^*$      & $<$\,20 & & 32$_{-9}^{+8}$ & \\
        Cyg X-1           & cluster      & 4.4$_{-4.4}^{+8.0}$ & & 26$^{+16}_{-12}$ & \\
        4U 2206+543       & arm          & 21 -- 32 & & 38.3$^{+0.7}_{-0.6}$ & $21.9^{+0.7}_{-0.6}$\\
        MWC 656           & isolated     & ... \\
    \hline
    
    \end{tabular}
    \tablefoot{$^*$: caveat for binaries located in regions with low cluster density ($<$10\,kpc$^{-2}$). CC13: \cite{coleiro_distribution_2013}, F22: \cite{fortin_constraints_2022}}
\end{table*}

The results from orbit integration using {\it Gaia} data produce quite homogeneous statistics over the formation scenarios we considered. There is an obvious caveat in the sample size which is too low to allow us to make definitive claims in terms of favored formation scenarios and about any differences between sgHMXBs and BeHMXBs. We also have a bias with the number of known {\it Gaia} open clusters, which decreases drastically with distance. In the case of a far away HMXB actually born in a cluster, it significantly lowers the chance of our cluster sample to contain the true birthplace. Hence, it is possible that the true ratio between HMXBs born in clusters vs HMXBs born near spiral arms actually leans towards the former, while our results are compatible with a 1:1 ratio. Out of the 8 HMXBs which lie in regions devoid of open clusters, 5 have at least one spiral arm as a birthplace candidate; the birthplace of the remaining 3 are inconclusive. If all of those binaries were actually born in a stellar cluster, we would then observe a 4:1 ratio in favor of formation within clusters versus spiral arms.

Open cluster to HMXB associations allowed us to retrace the history of binaries and to provide estimation of the ZAMS mass of both components as well as of the mass transfer from primary to secondary. In the case of HD\,259440, the only HMXB associated to a cluster without a detected spin-period, the mass history lead us to suggest that the binary indeed has a neutron star accretor. The information about evolution timescales and spatial distribution that results from inferring the history of observed HMXBs in the Milky Way may be of interest in the scope of population synthesis models and the modeling of the X-ray luminosity of galaxies (see e.g. \citel{zuo_population_2014}). We also reckon that a finer modeling of their evolution using more accurate hydrodynamical simulations might be facilitated when looking to establish a link between the current population of HMXBs and the population of gravitational wave mergers, and it would be a great opportunity to combine observation and simulation to test our knowledge on various key mechanisms that dictate the life of binaries, e.g. the mass transfer, the supernova events and the common envelope phase \citep{fragos_complete_2019}.

As concluded in a previous study, we stress the fact that further optical--infrared observations of HMXBs are needed to complete their set of known parameters; in our case, better astrometric data is unlikely to be obtained in the foreseeable future, but deriving more radial velocities could significantly bump our sample size up and allow us to finally study observed HMXBs as a population, and not only as individual peculiar sources.

\begin{acknowledgements}
The authors were supported by the LabEx UnivEarthS: Interface project I10 "Binary rEvolution: from binary evolution towards merging of compact objects".
SC is grateful to the CNES (Centre National d'\'Etudes Spatiales) for the funding of MINE (Multi-wavelength INTEGRAL Network). The authors would like to thank Alexis Coleiro for useful discussions and reading of the manuscript.

This work has made use of data from the European Space Agency (ESA) mission
{\it Gaia} (\url{https://www.cosmos.esa.int/gaia}), processed by the {\it Gaia}
Data Processing and Analysis Consortium (DPAC,
\url{https://www.cosmos.esa.int/web/gaia/dpac/consortium}). Funding for the DPAC
has been provided by national institutions, in particular the institutions
participating in the {\it Gaia} Multilateral Agreement.

{\em Software:} {\sc matplotlib} \citep{hunter_matplotlib_2007}, {\sc NumPy} \citep{van_der_walt_numpy_2011}, {\sc scipy} \citep{jones_scipy_2001} and {\sc Python} from \url{python.org}

\end{acknowledgements}

\bibliographystyle{aa}
\bibliography{biblio}

\appendix
\onecolumn
\section{Time-distance histograms of encounter candidates}\label{appendix}
\begin{figure}[!htb]
    \centering
    \includegraphics[width=0.4\columnwidth]{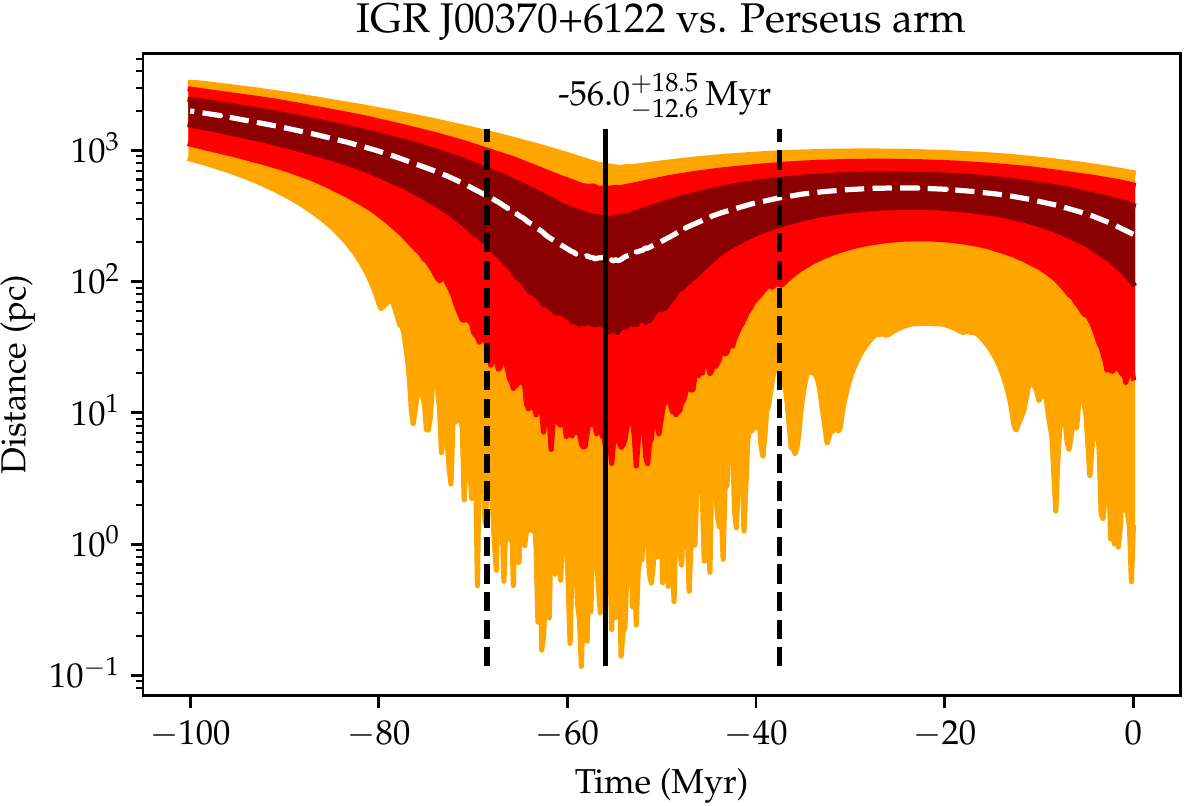}
    \includegraphics[width=0.4\columnwidth]{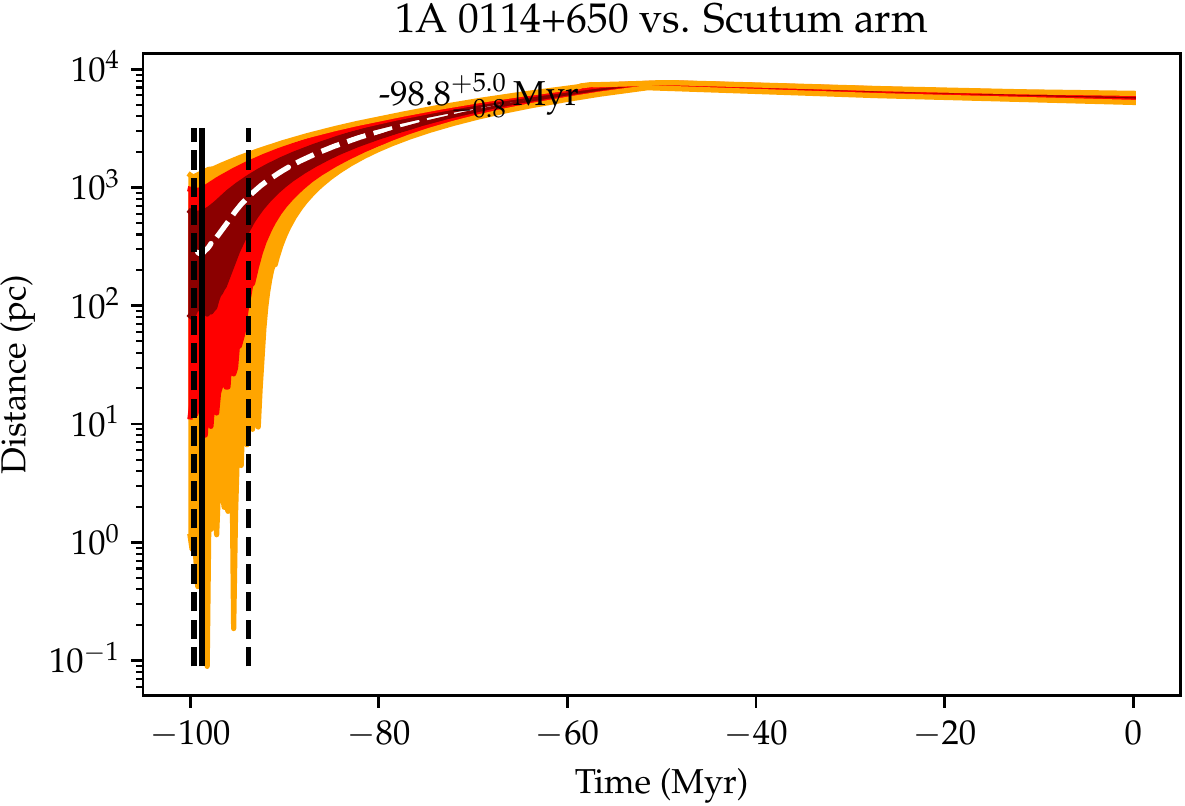}
    \includegraphics[width=0.4\columnwidth]{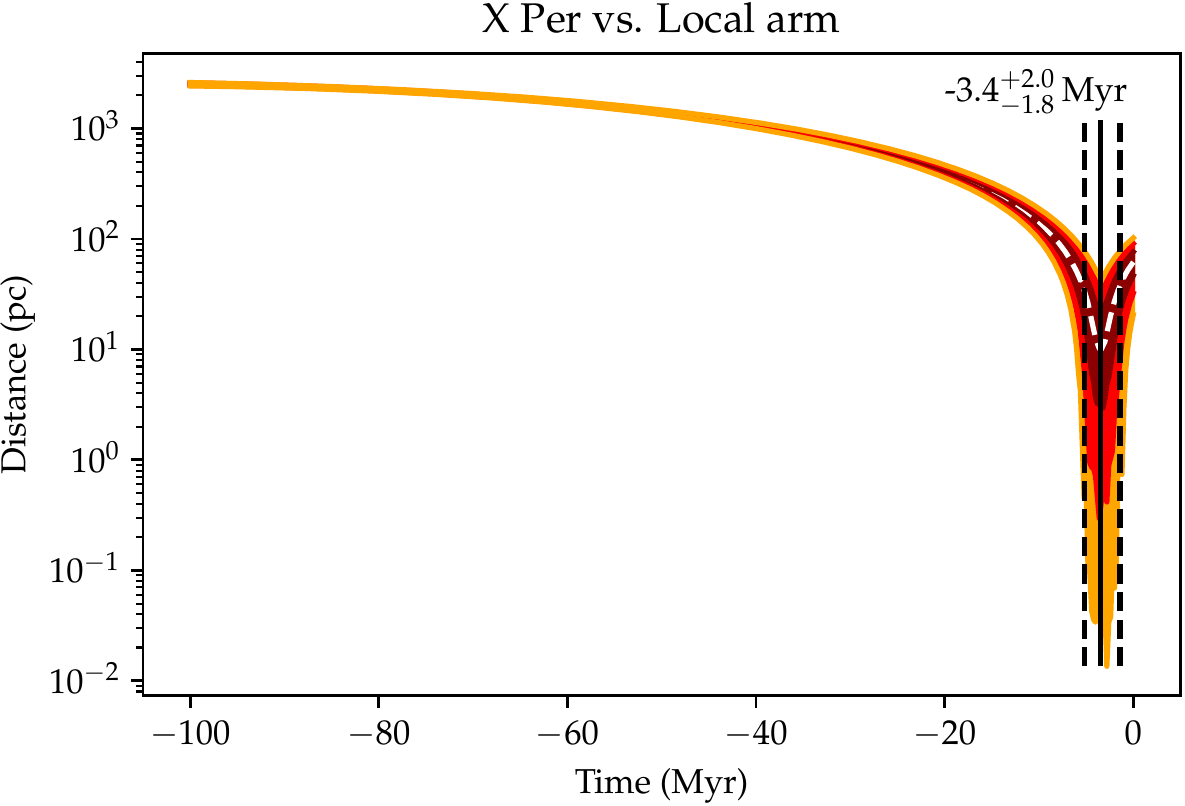}
    \includegraphics[width=0.4\columnwidth]{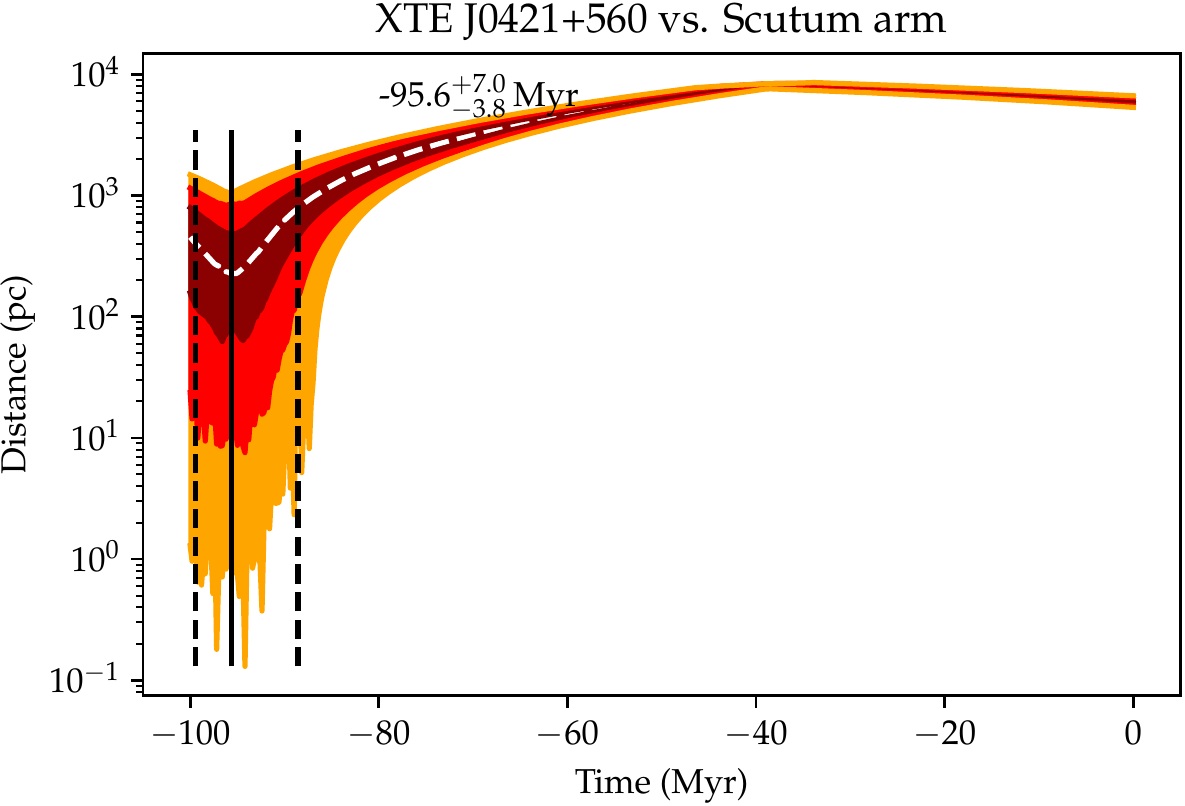}
    \includegraphics[width=0.4\columnwidth]{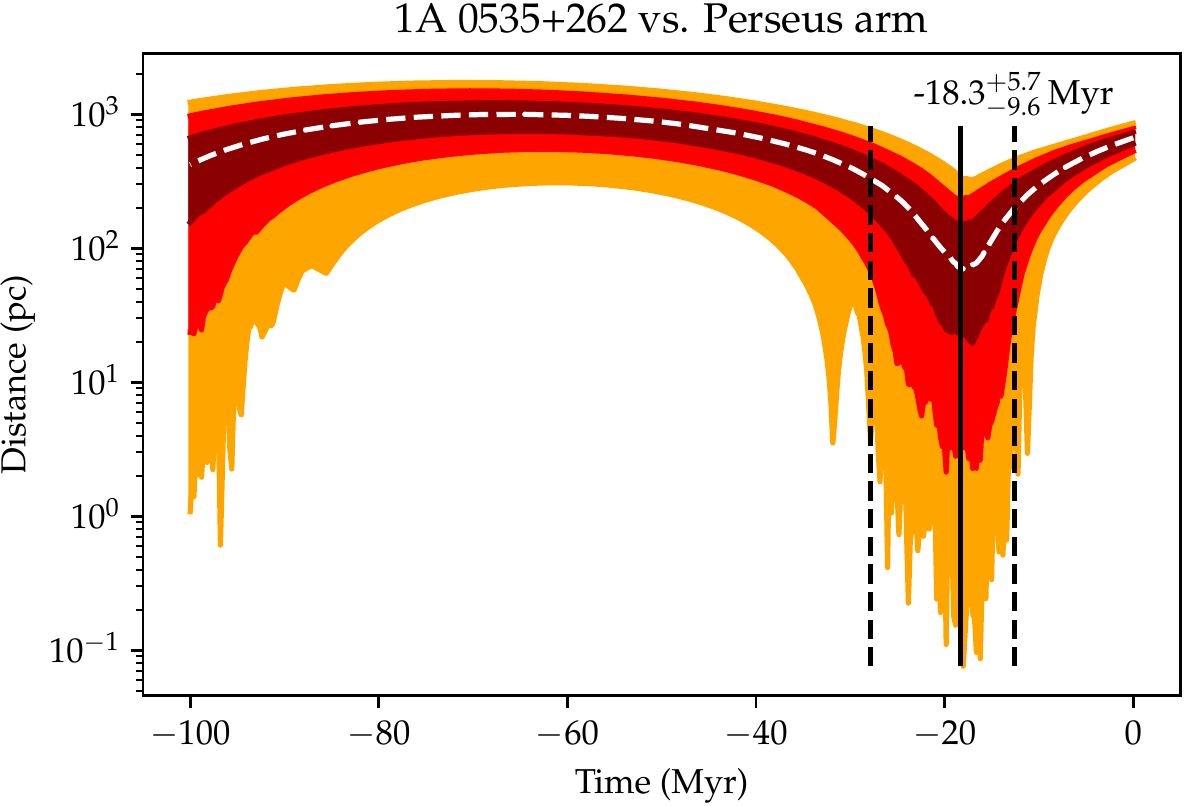}
    \includegraphics[width=0.4\columnwidth]{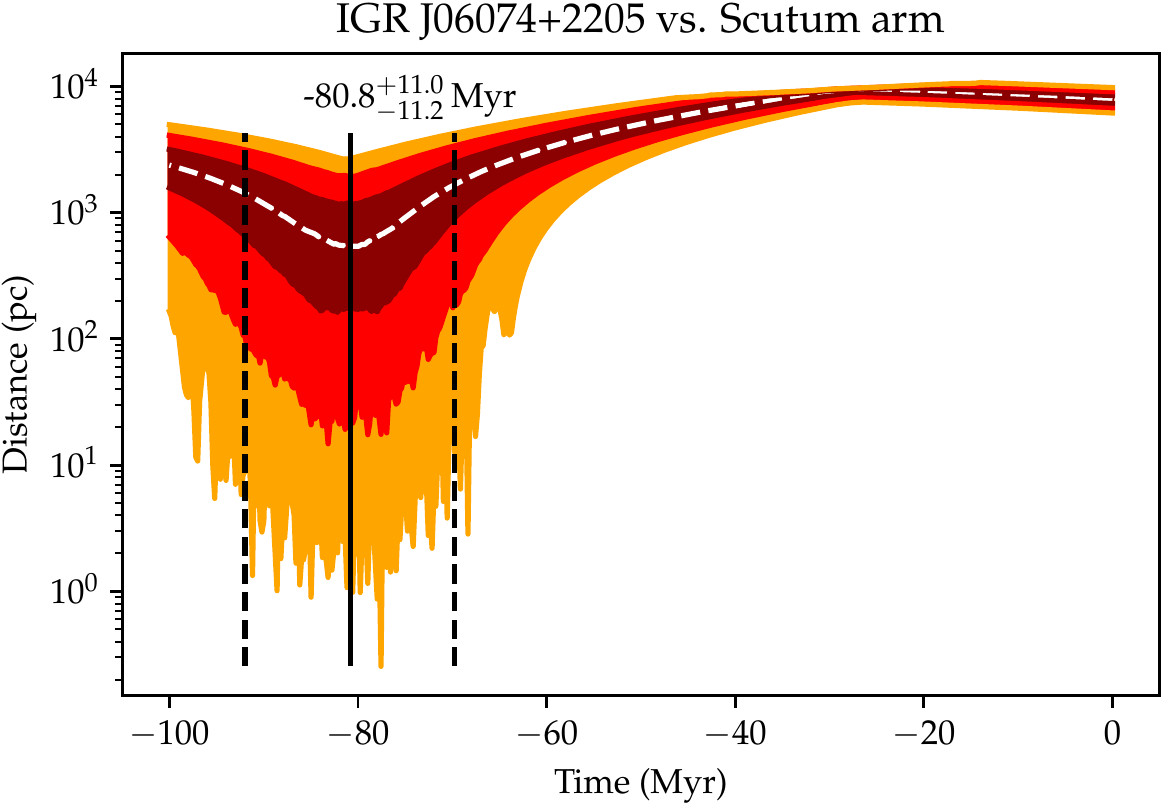}
    \includegraphics[width=0.4\columnwidth]{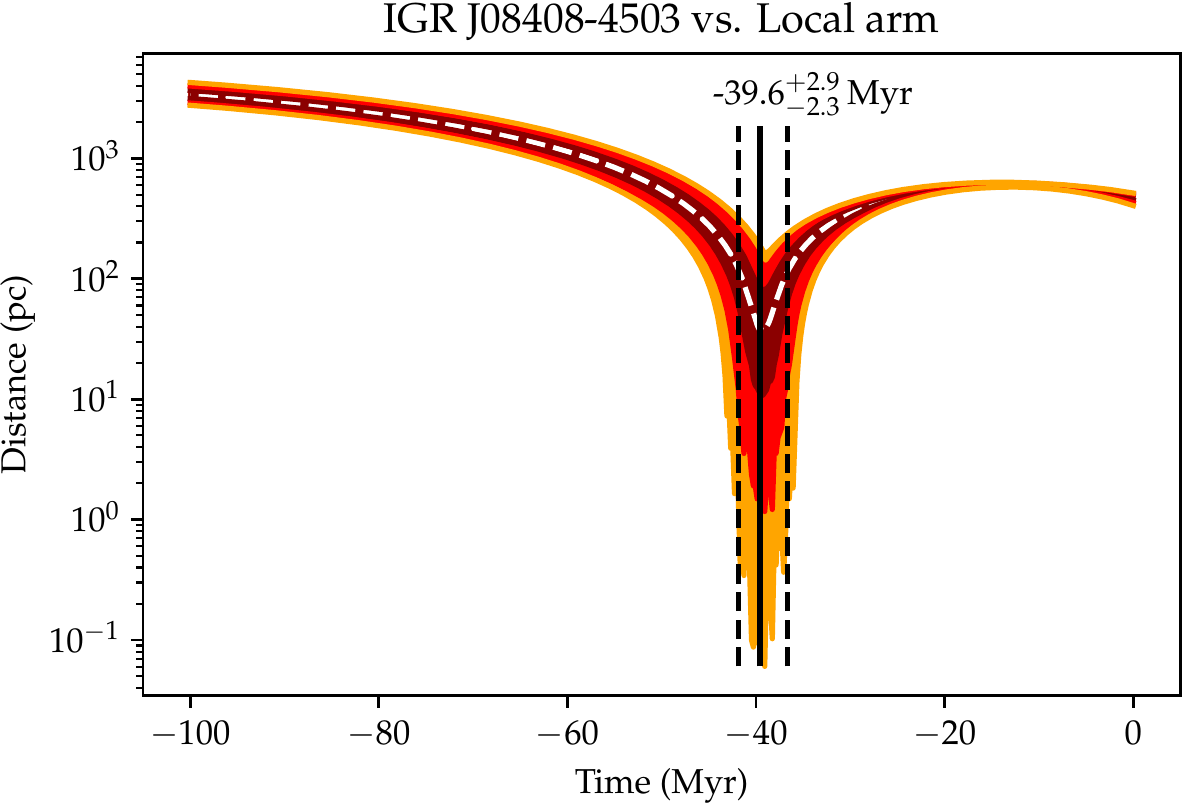}
    \includegraphics[width=0.4\columnwidth]{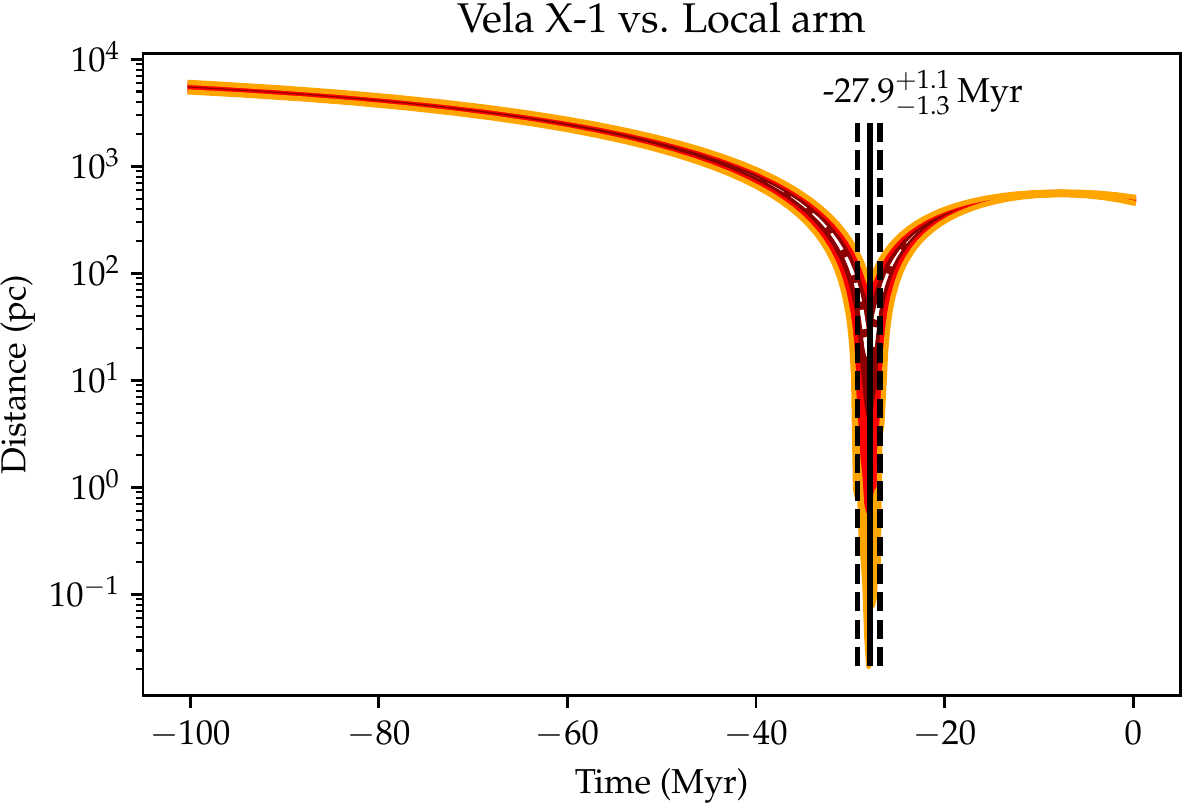}
    
    \caption{Time-distance histograms of encounter candidates between HMXBs and Galactic spiral arms in the past 100\,Myr. Dotted white lines represent the median distance, and the 1 to 3$\sigma$ intervals are represented from dark red, to red, to orange. The distance is the one in the plane of the Galaxy (XY plane). The black vertical plain and dotted lines provide our estimation of the age since supernova in each binary-arm encounter candidate.}
    \label{fig:HMarm_1}
\end{figure}

\begin{figure}[!htb]
    \ContinuedFloat
    \captionsetup{list=off,format=cont}
    \centering
     
    \includegraphics[width=0.4\columnwidth]{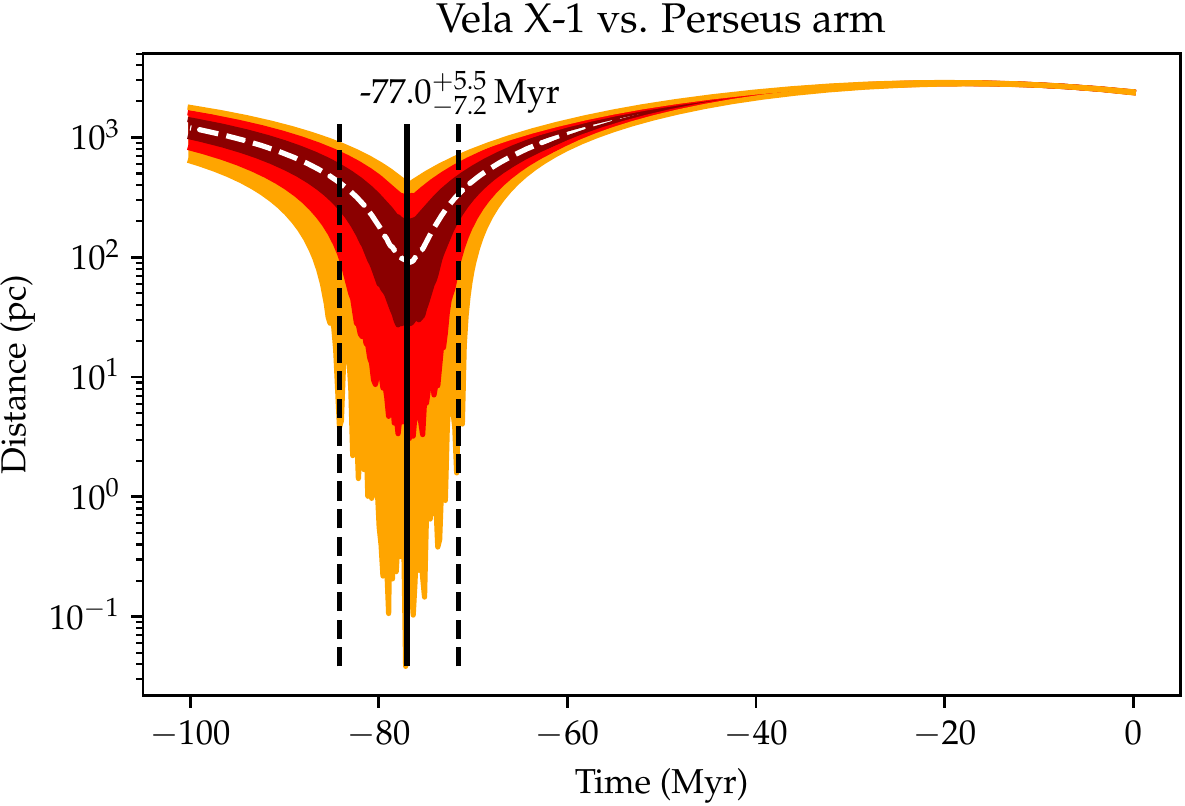}
    \includegraphics[width=0.4\columnwidth]{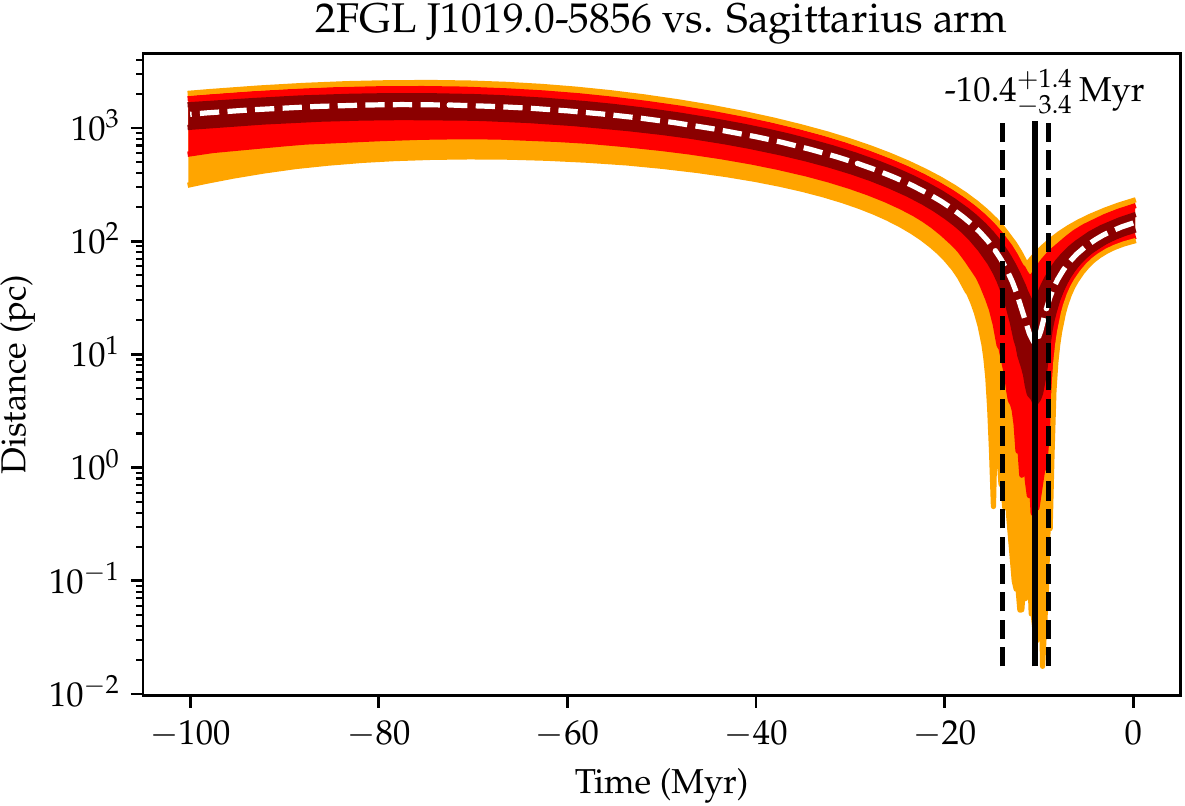}
    \includegraphics[width=0.4\columnwidth]{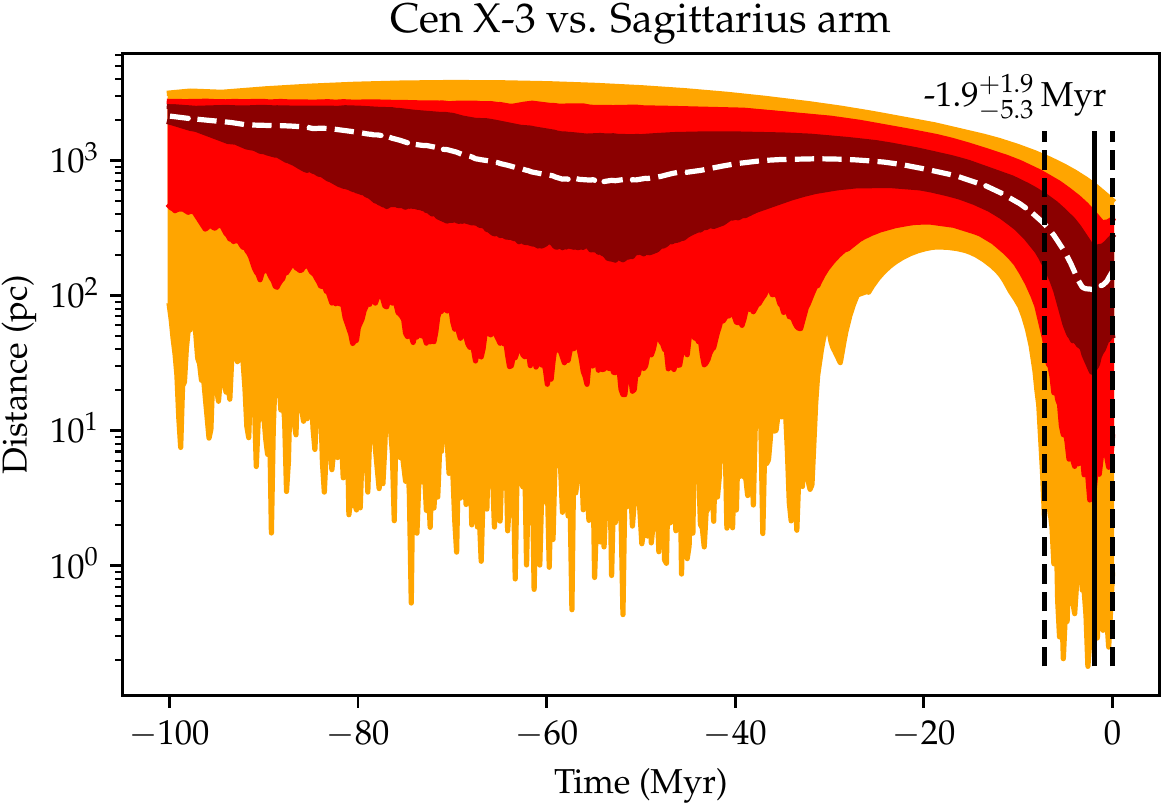}
    \includegraphics[width=0.4\columnwidth]{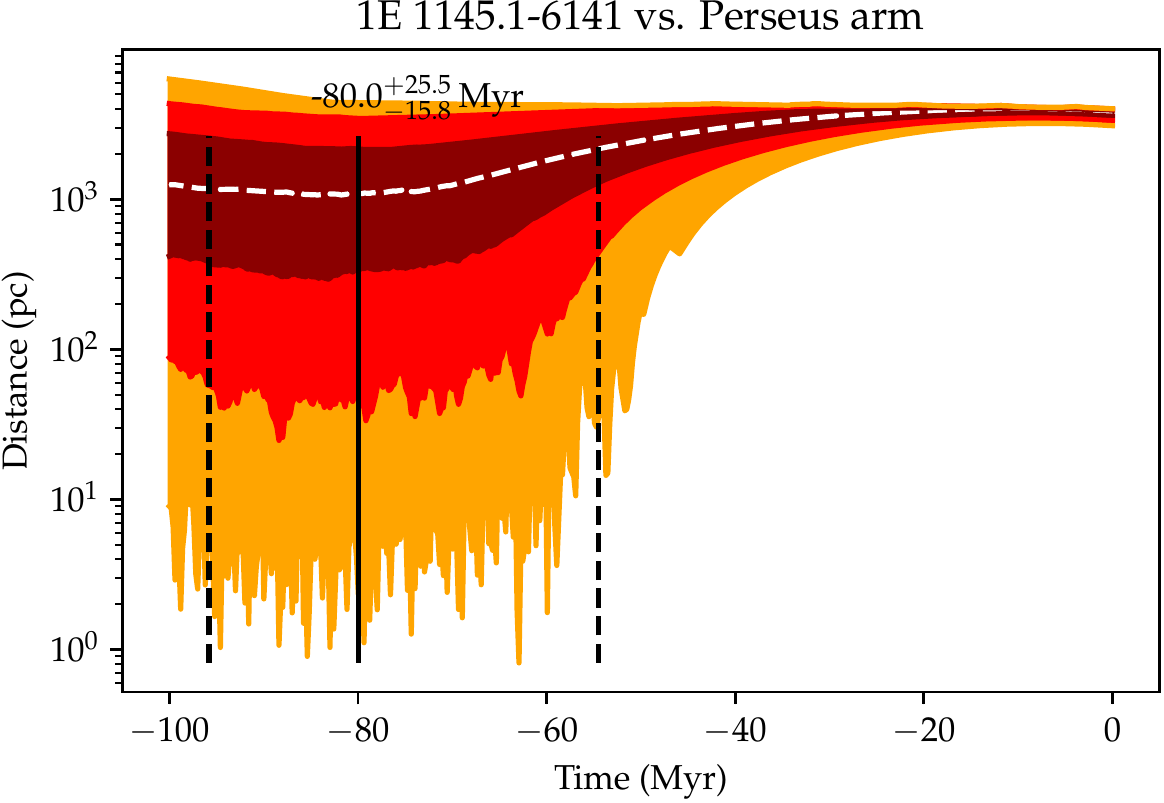}
    \includegraphics[width=0.4\columnwidth]{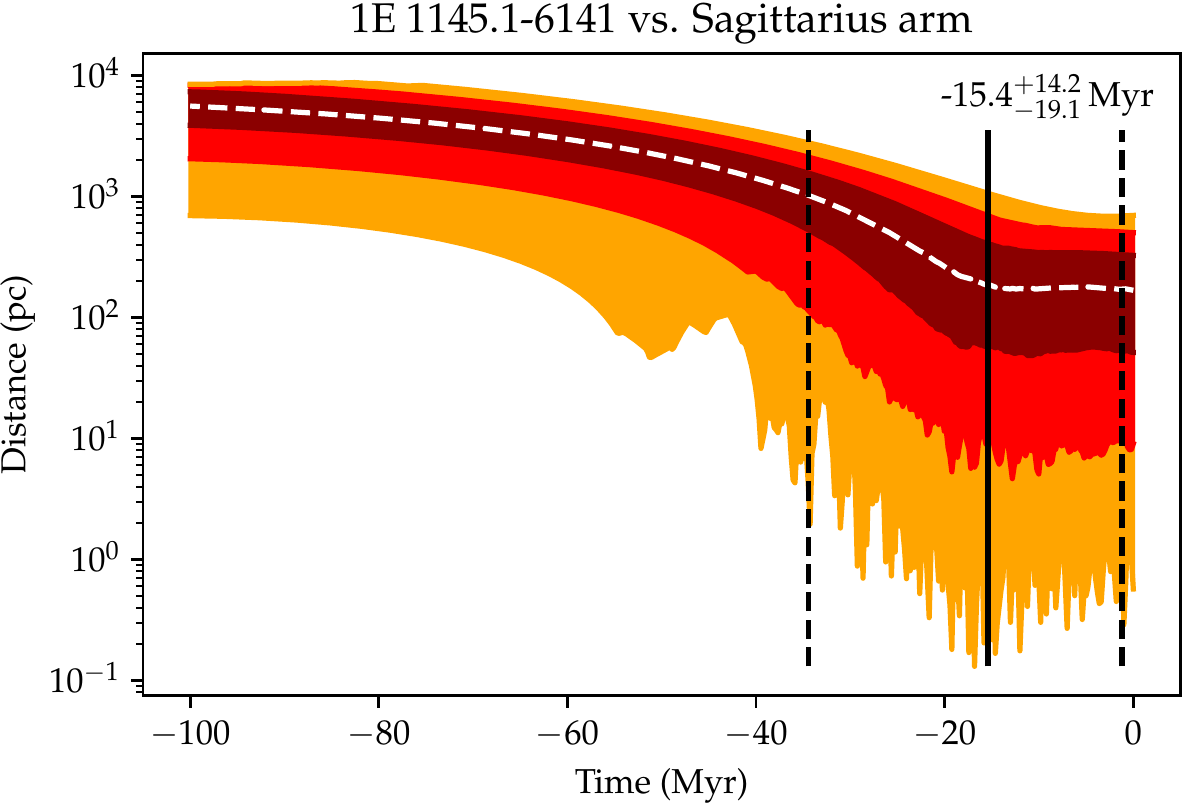}
    \includegraphics[width=0.4\columnwidth]{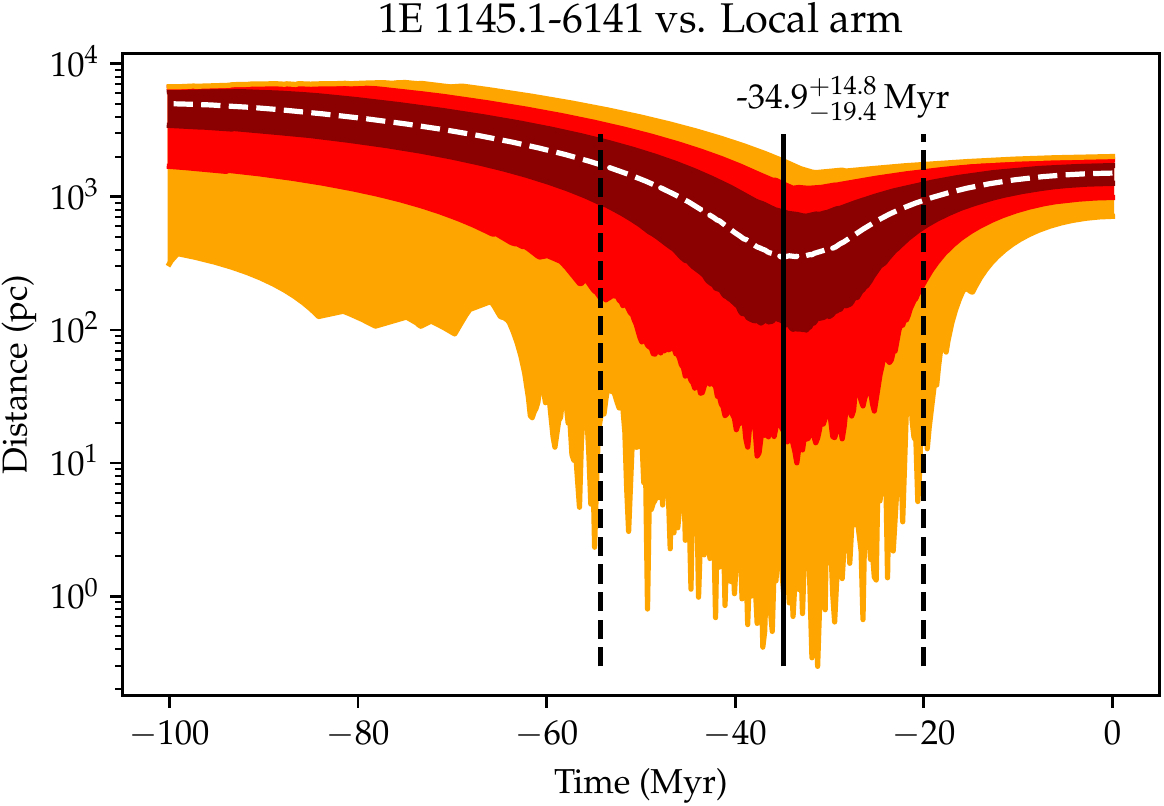}
    \includegraphics[width=0.4\columnwidth]{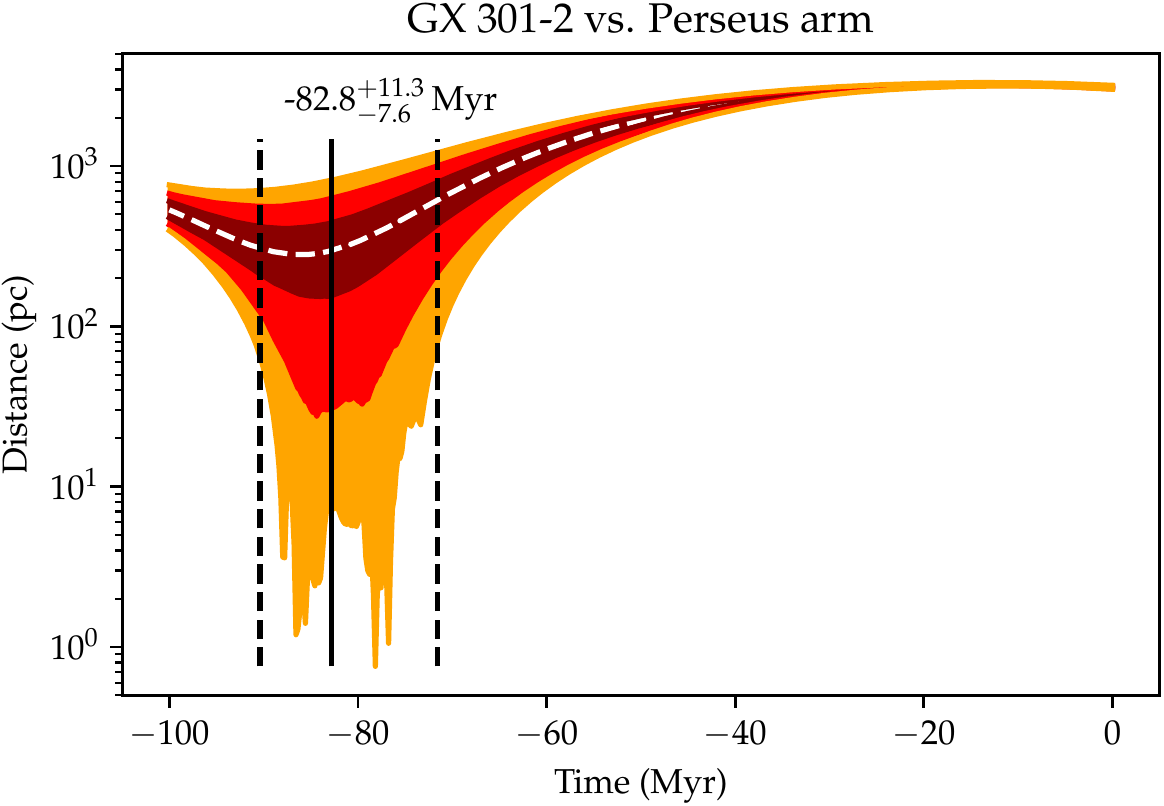}
    \includegraphics[width=0.4\columnwidth]{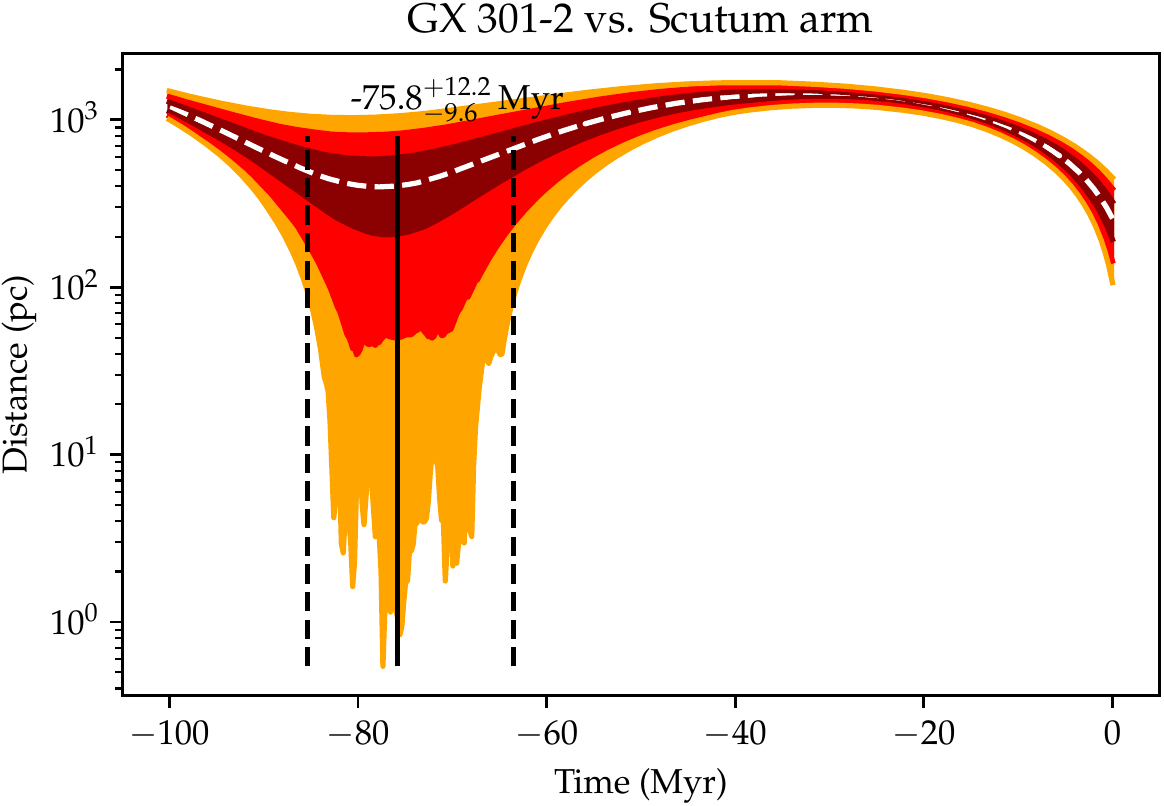}

    \caption{continued.}
    \label{fig:HMarm_2}
\end{figure}

\begin{figure}[!htb]
    \captionsetup{list=off,format=cont}
    \ContinuedFloat
    \centering
    \includegraphics[width=0.4\columnwidth]{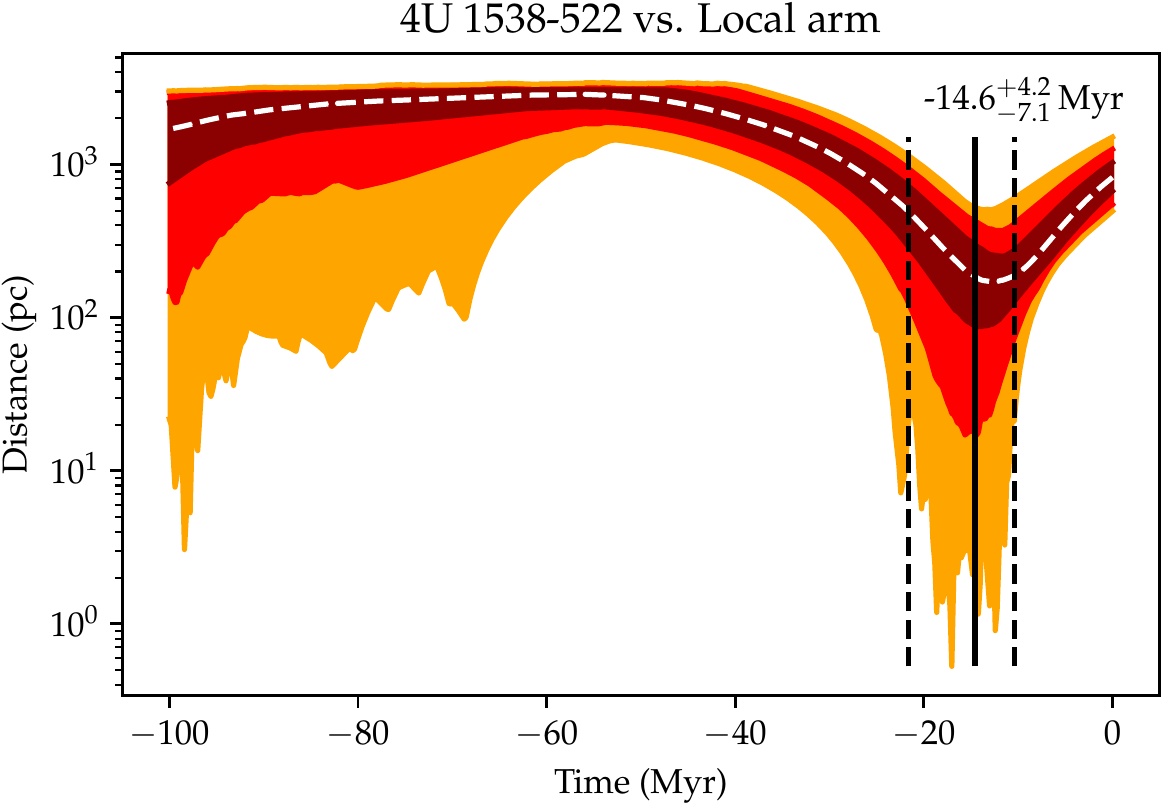}
    \includegraphics[width=0.4\columnwidth]{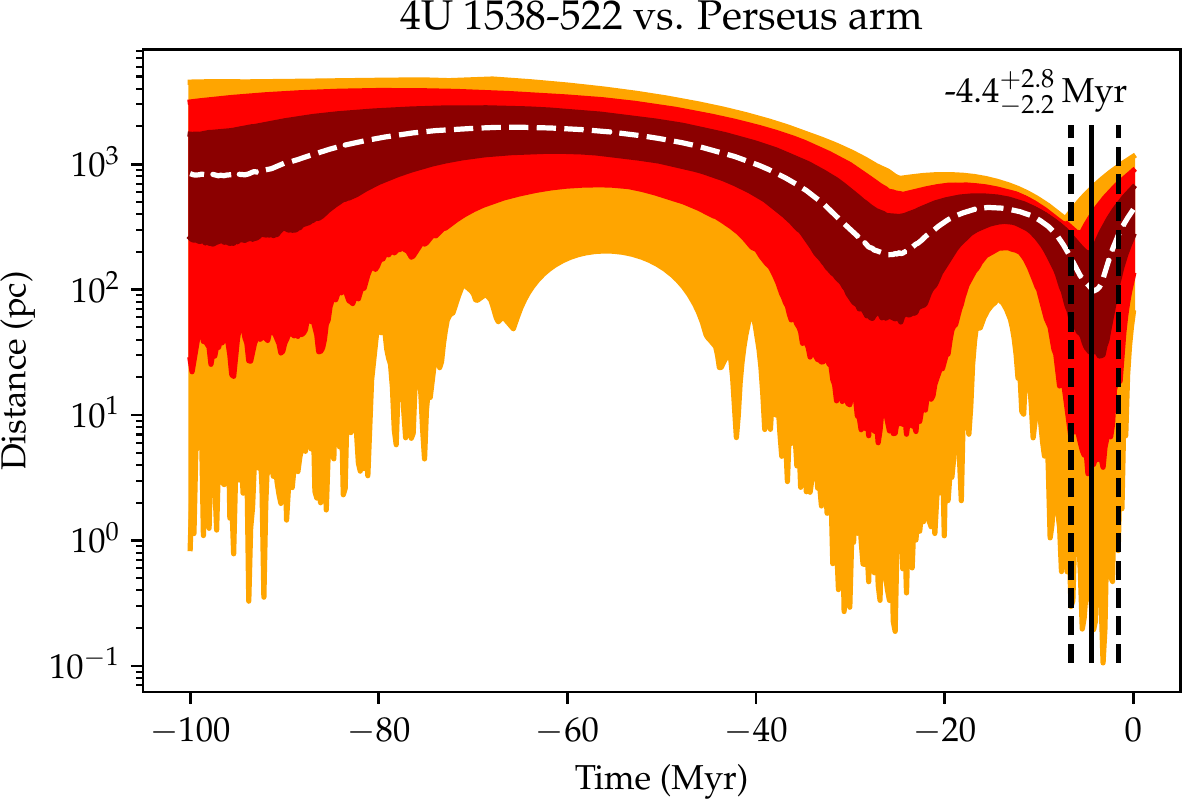}
    \includegraphics[width=0.4\columnwidth]{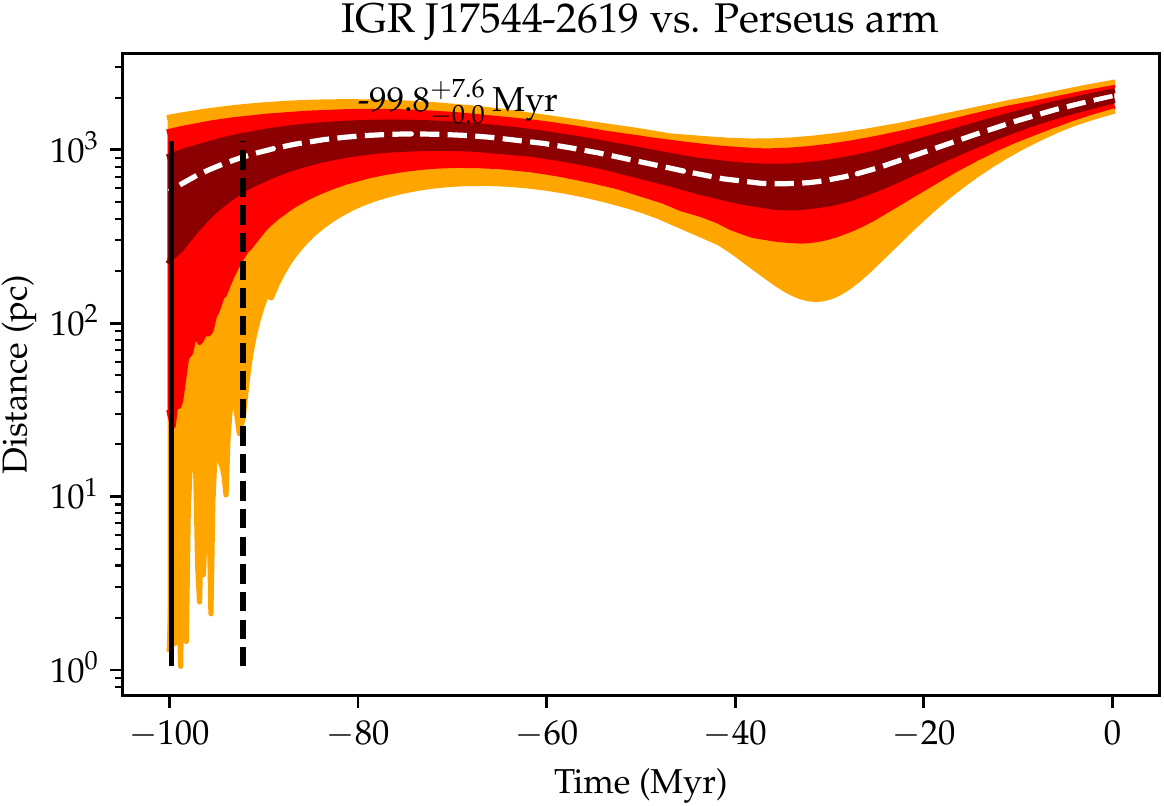}
    \includegraphics[width=0.4\columnwidth]{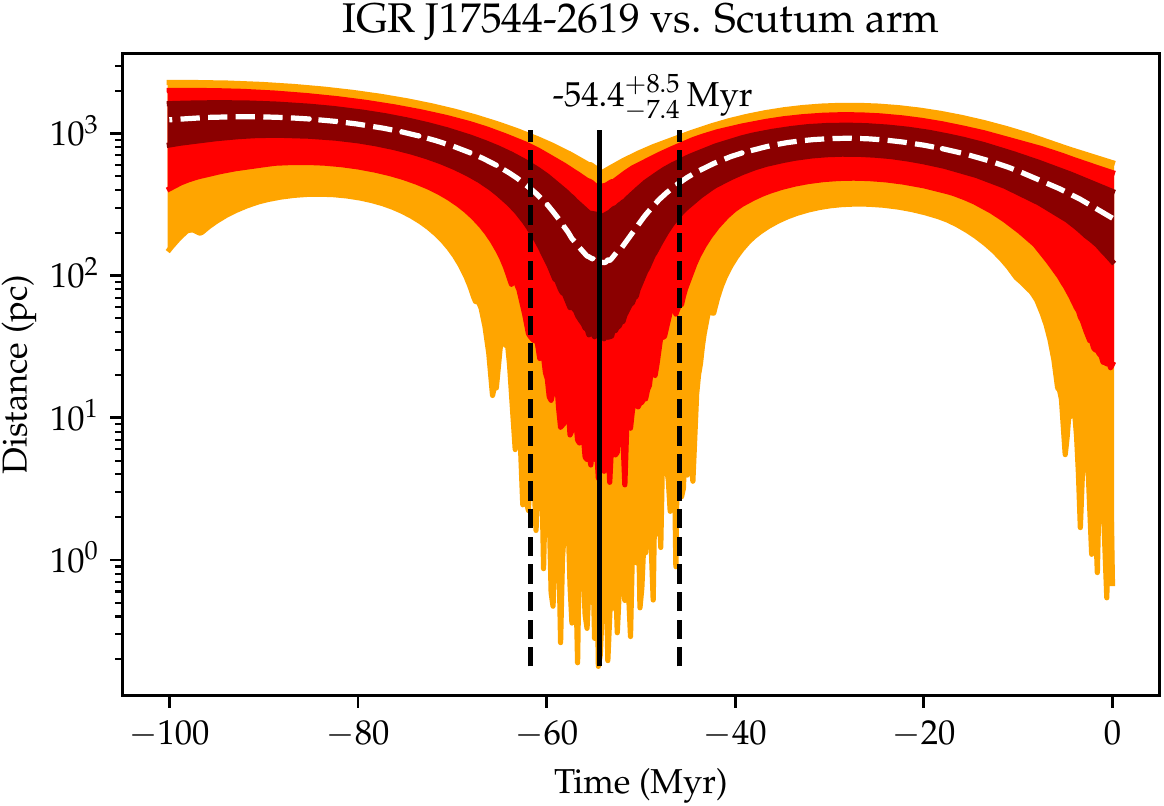}
    \includegraphics[width=0.4\columnwidth]{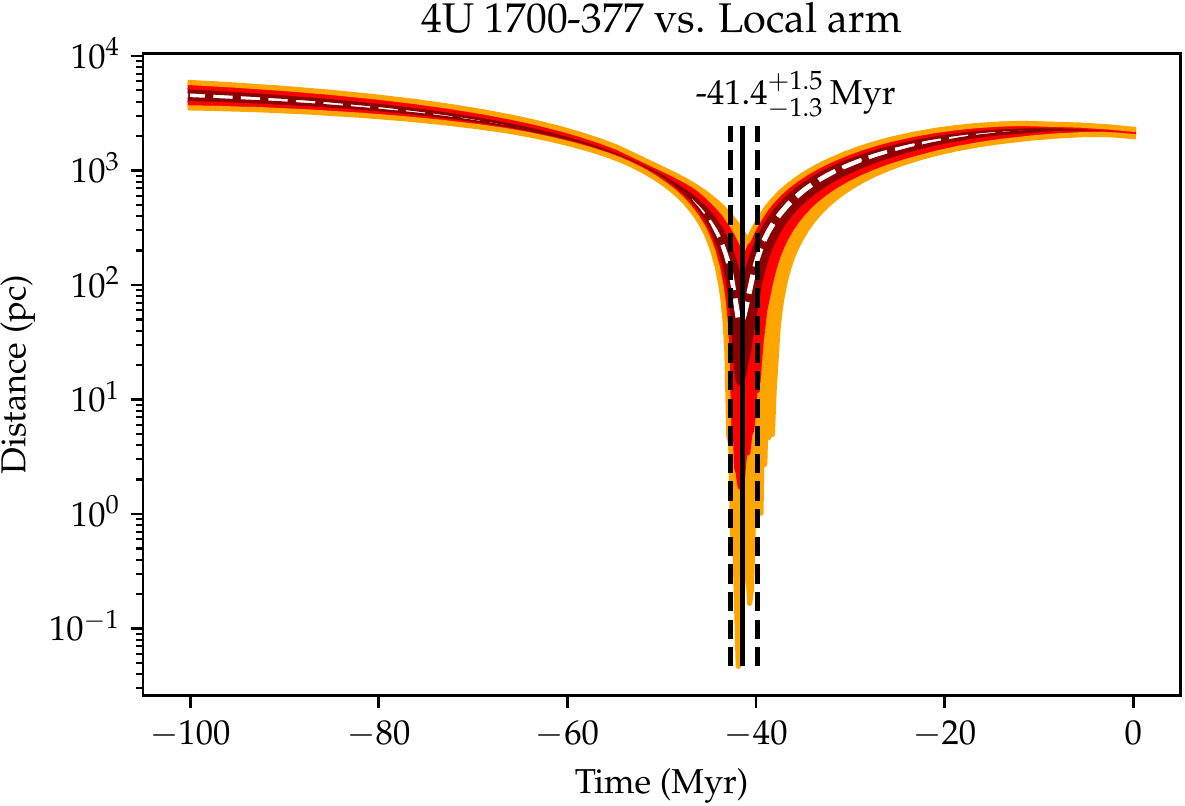}
    \includegraphics[width=0.4\columnwidth]{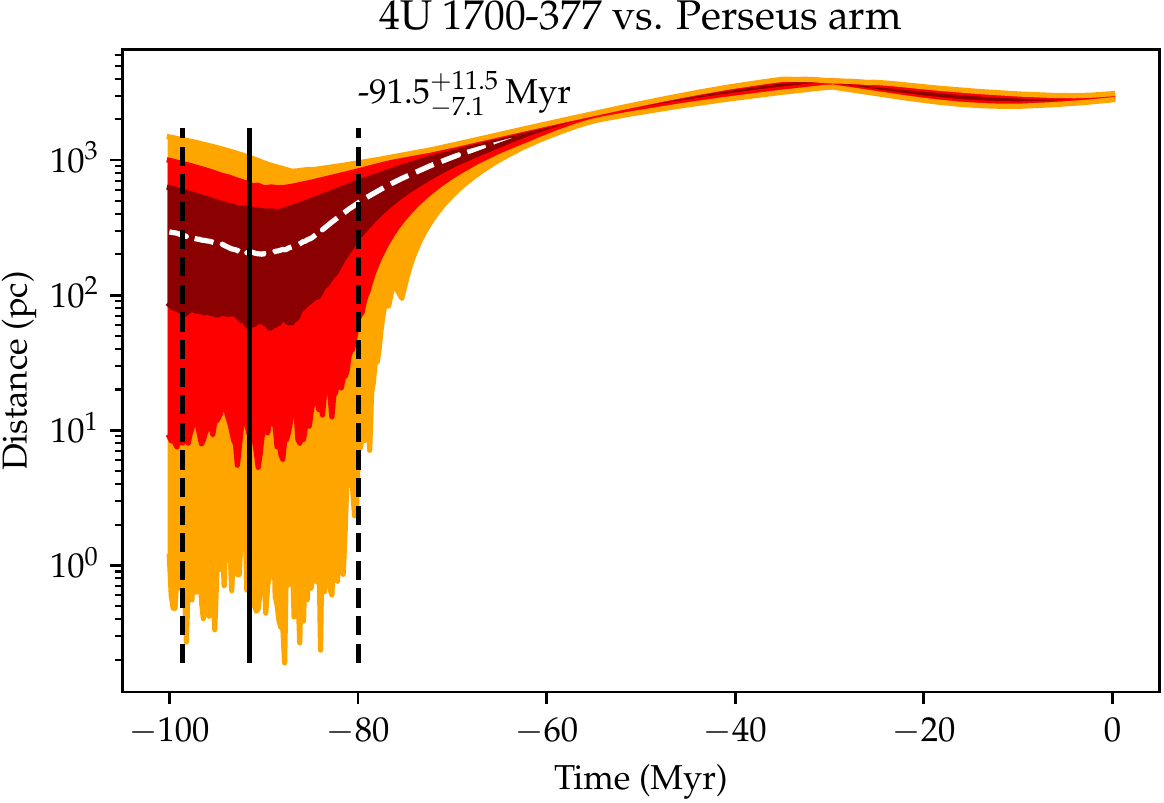}
    \includegraphics[width=0.4\columnwidth]{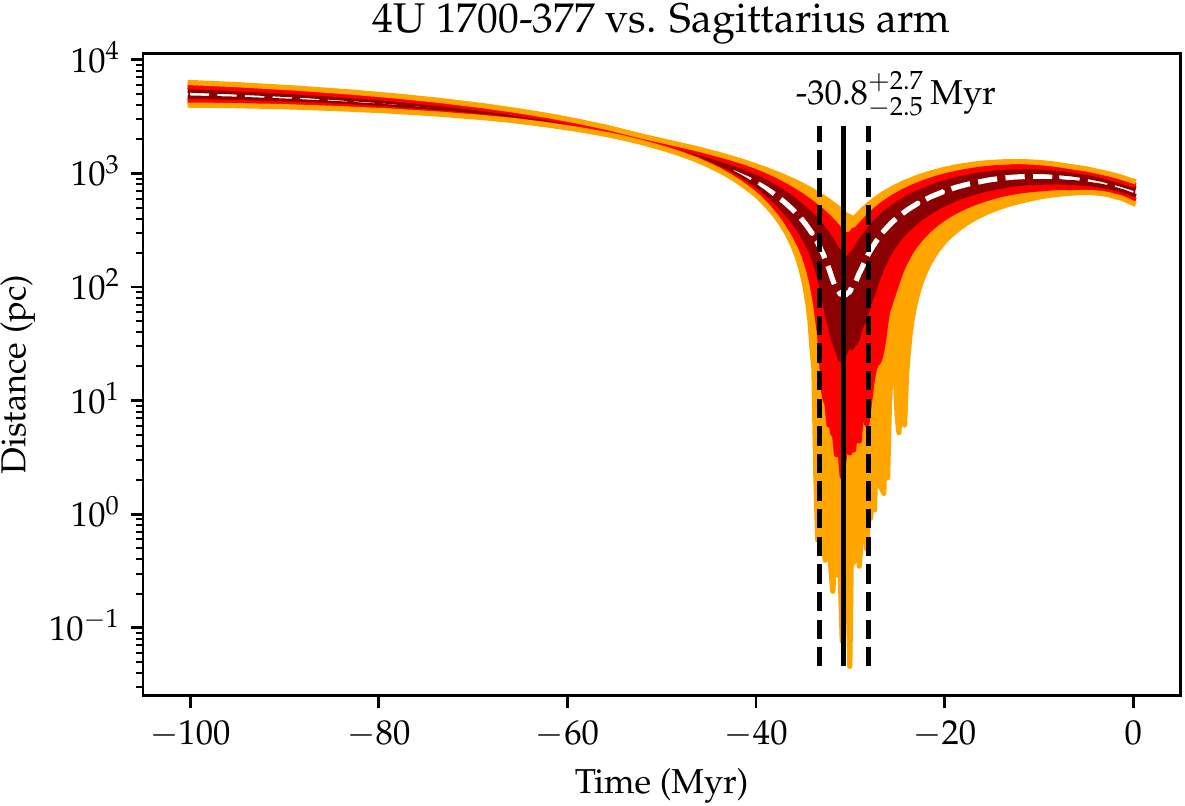}
    \includegraphics[width=0.4\columnwidth]{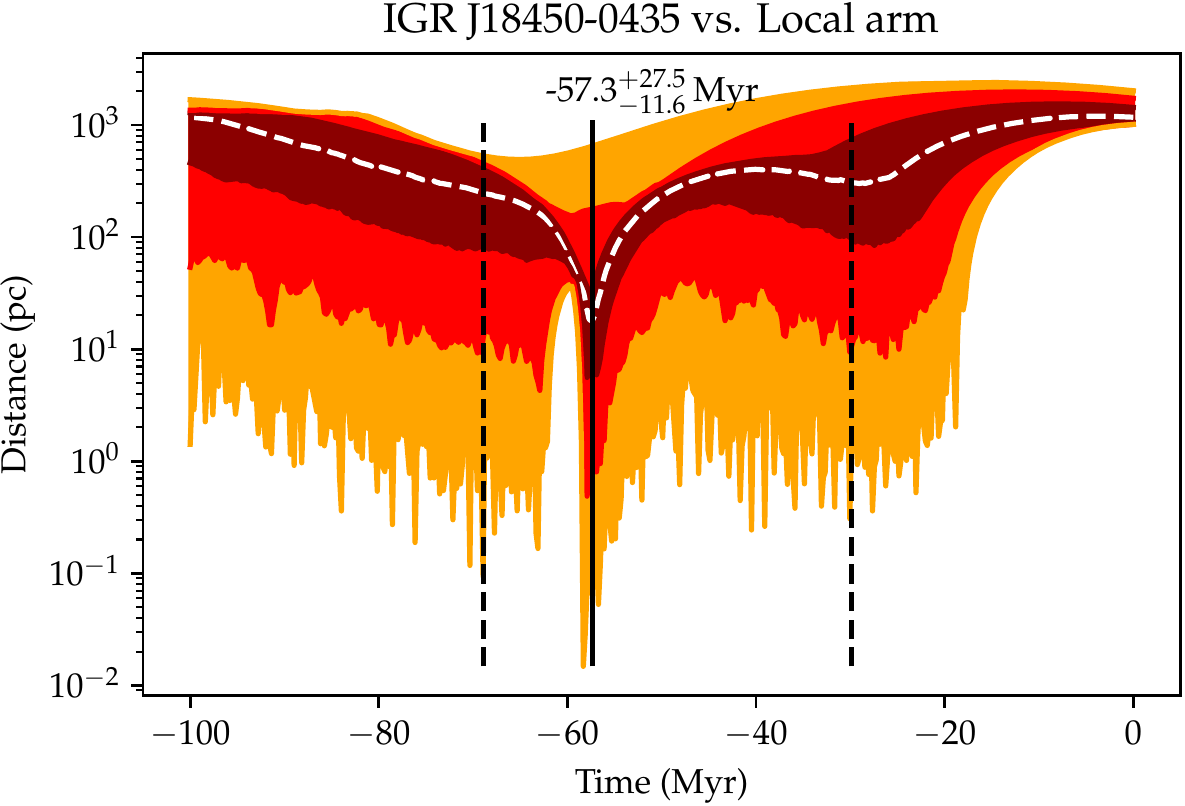}

    \caption{continued.}
    \label{fig:HMarm_3}
\end{figure}

\begin{figure}[!htb]
    \captionsetup{list=off,format=cont}
    \ContinuedFloat
    \centering
    \includegraphics[width=0.4\columnwidth]{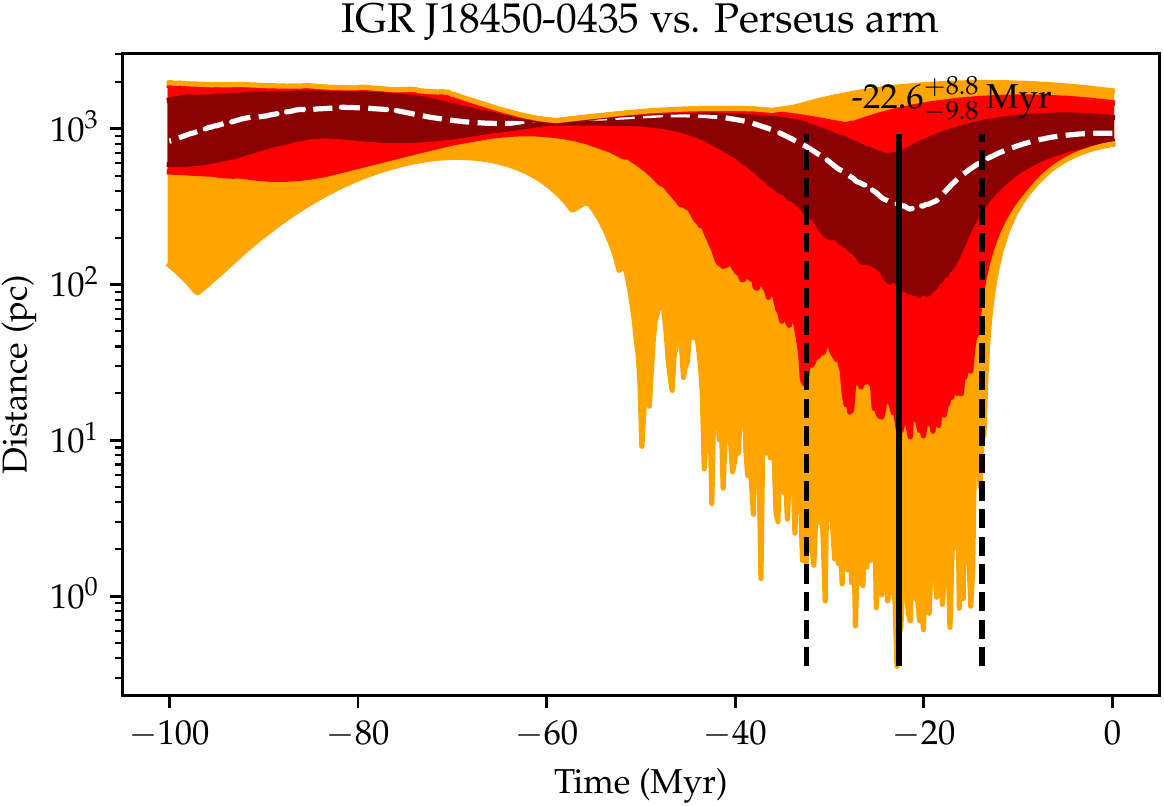}
    \includegraphics[width=0.4\columnwidth]{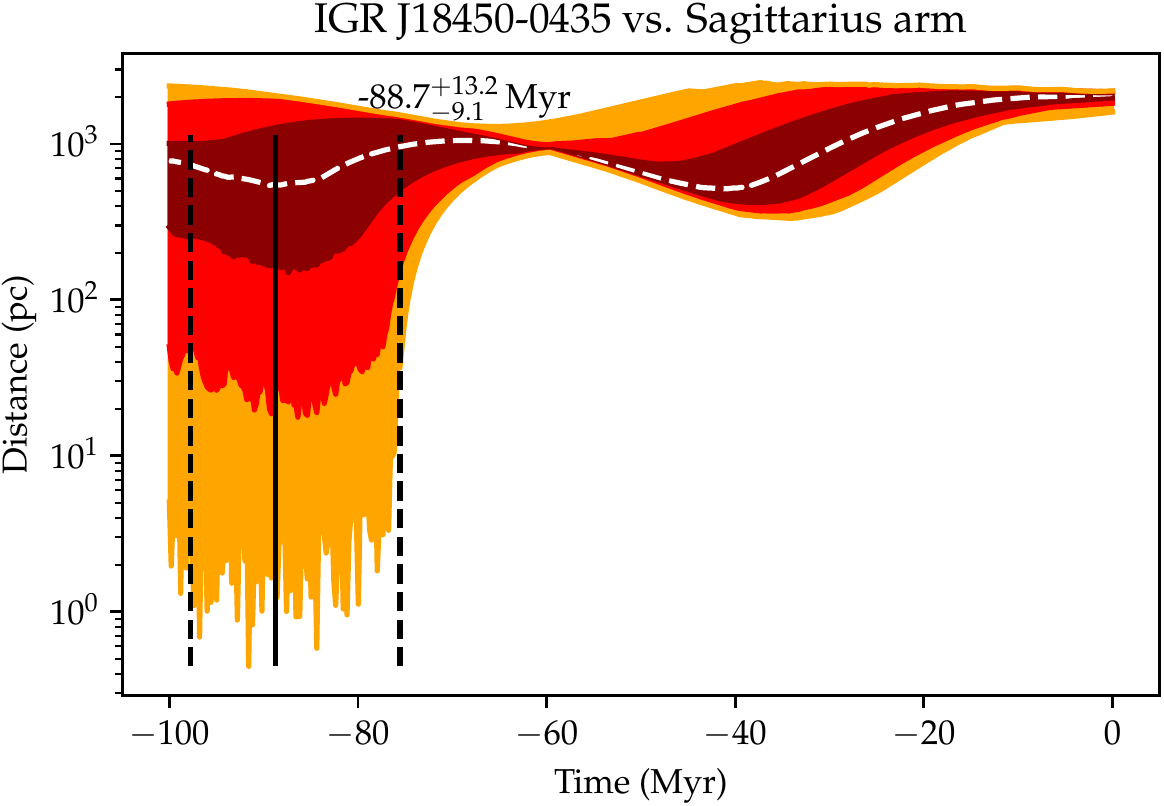}
    \includegraphics[width=0.4\columnwidth]{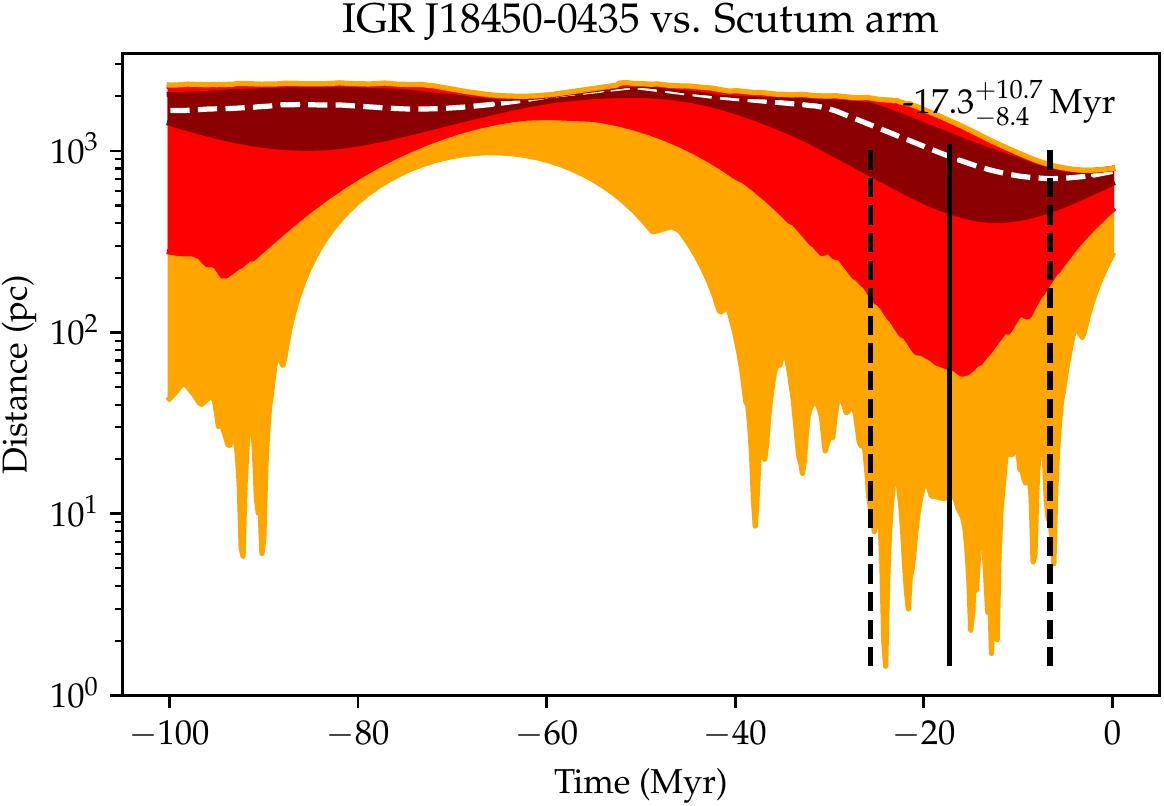}
    \includegraphics[width=0.4\columnwidth]{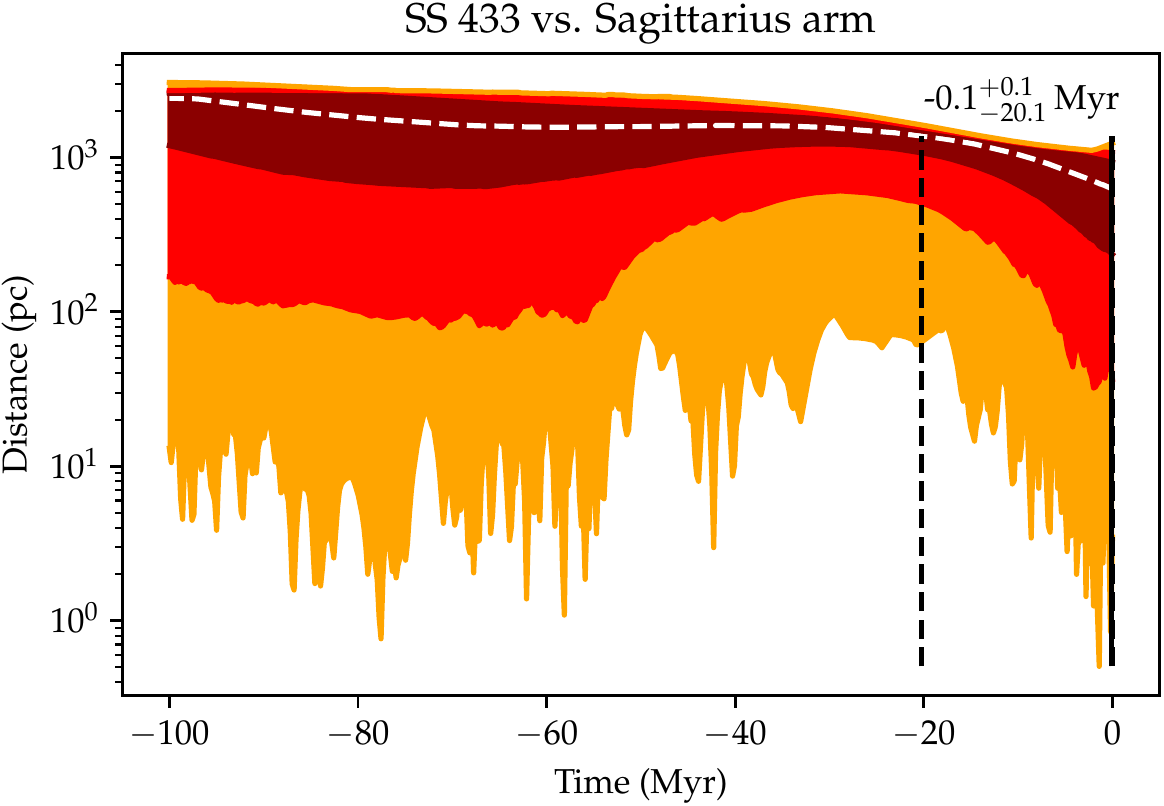}
    \includegraphics[width=0.4\columnwidth]{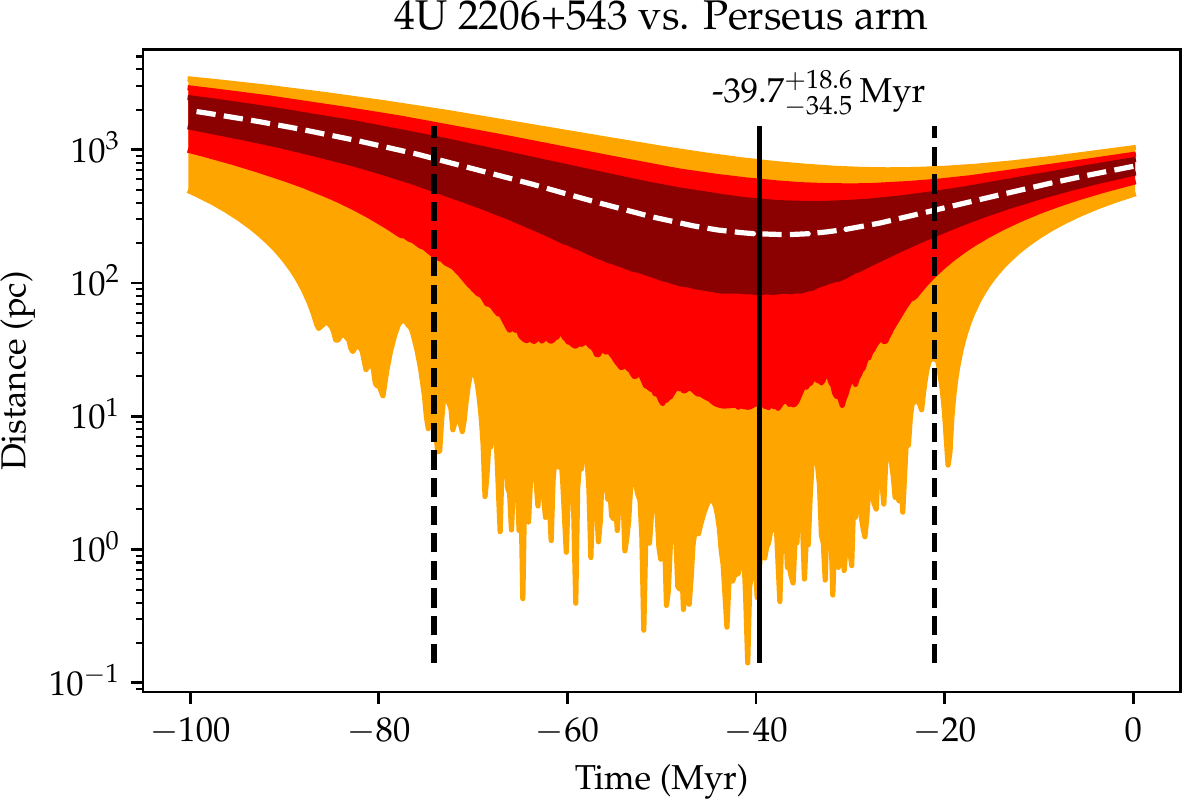}
    \caption{continued.}
    \label{fig:HMarm_4}
\end{figure}

\begin{figure}[!htb]
    \centering
    \includegraphics[width=0.4\columnwidth]{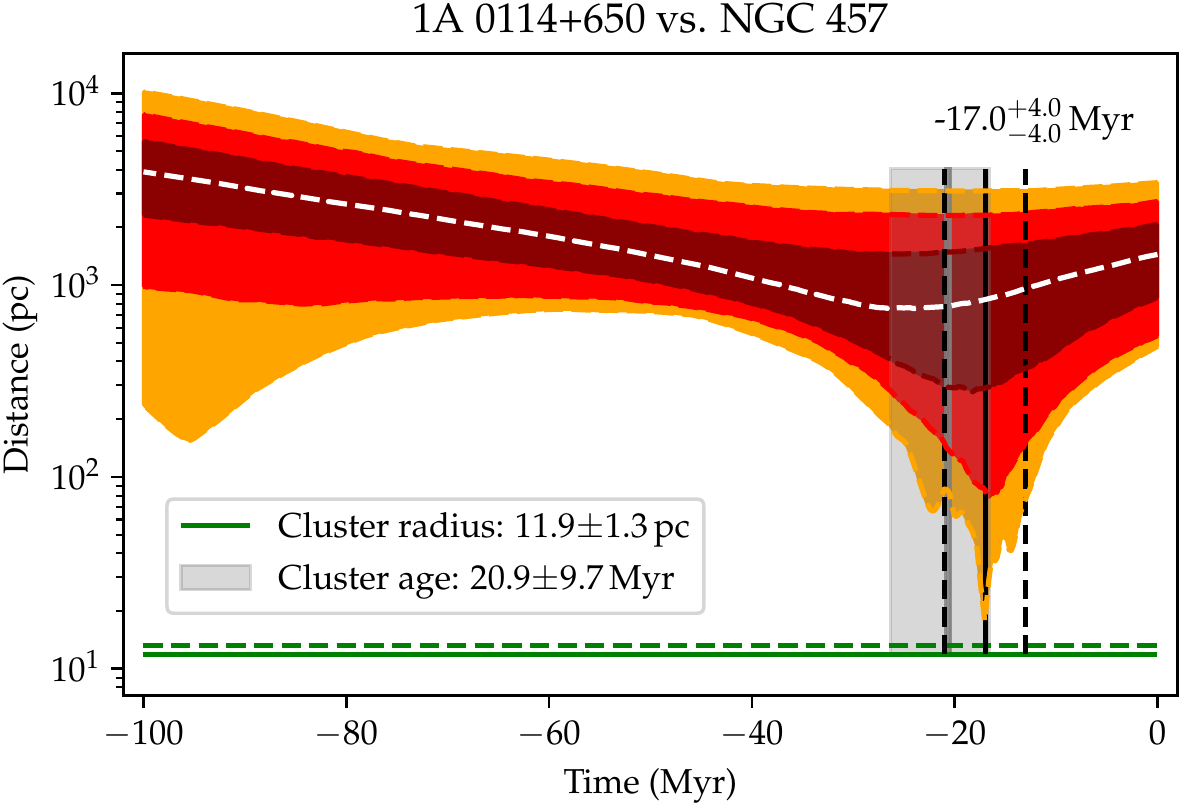}
    \includegraphics[width=0.4\columnwidth]{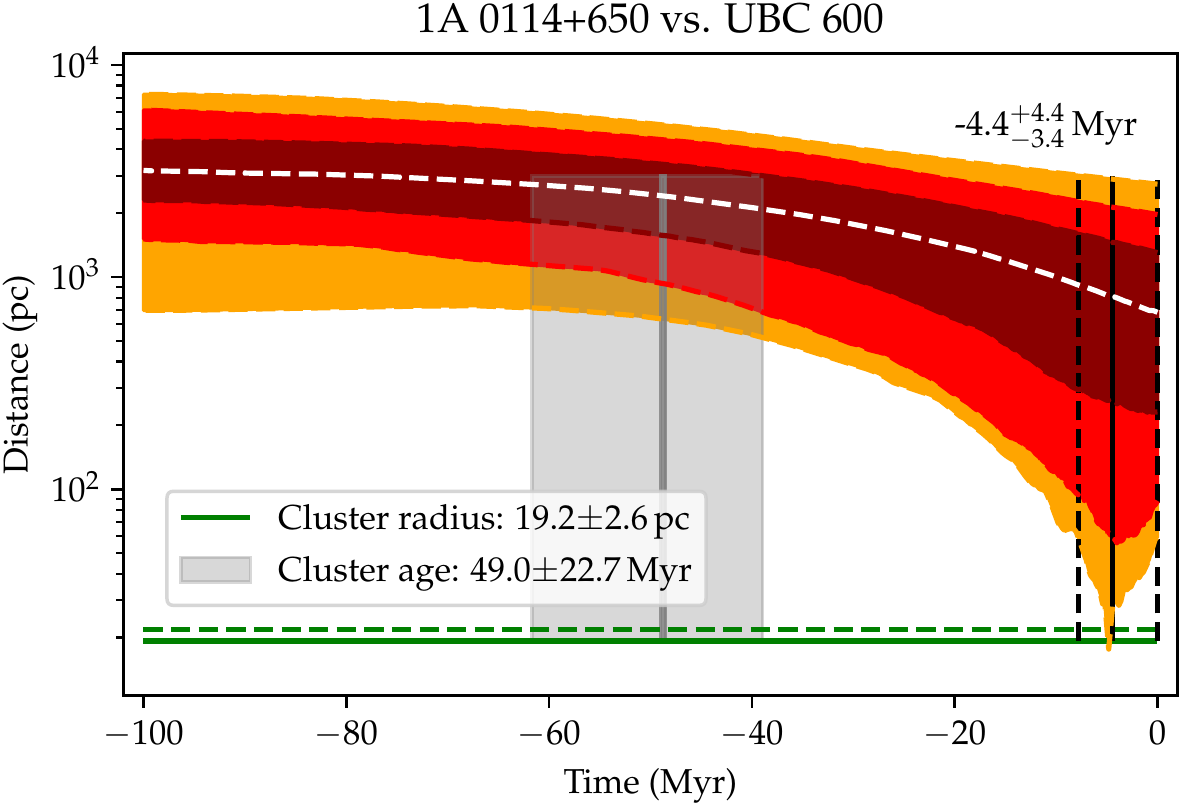}
    
    \includegraphics[width=0.4\columnwidth]{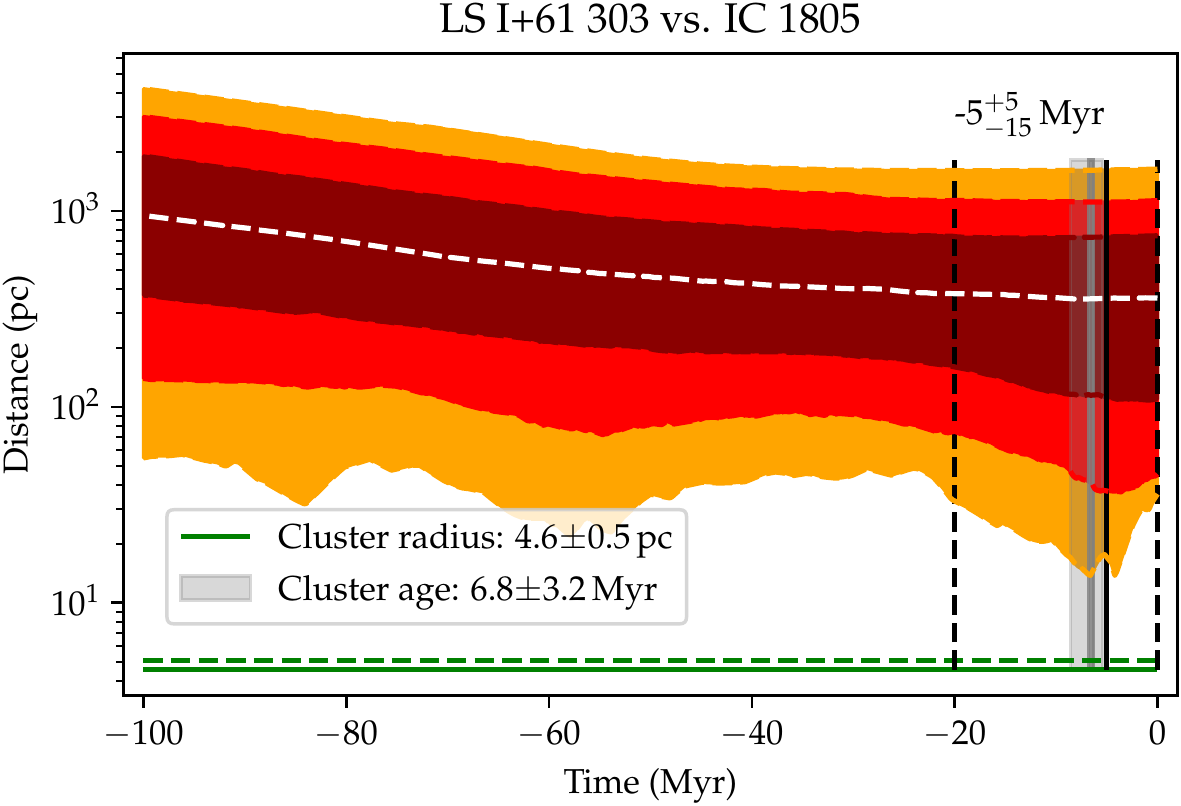}
    \includegraphics[width=0.4\columnwidth]{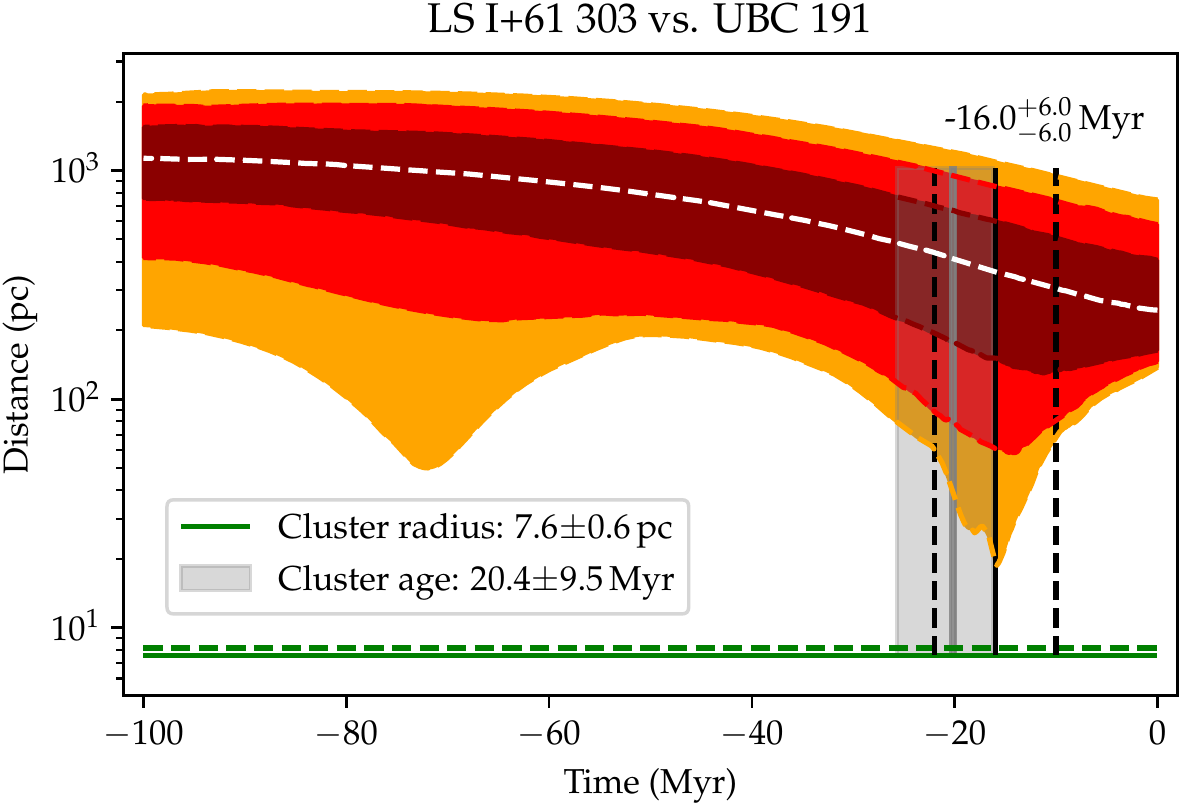}
    \includegraphics[width=0.4\columnwidth]{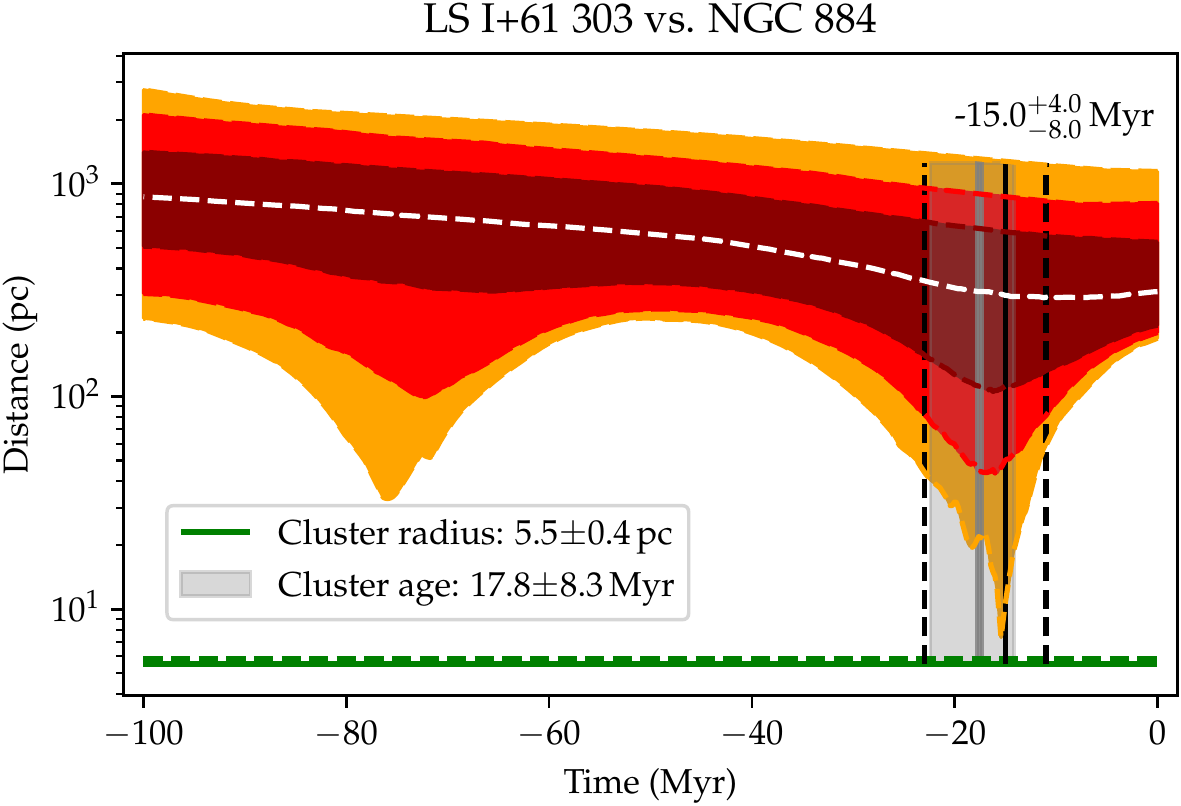}
    \includegraphics[width=0.4\columnwidth]{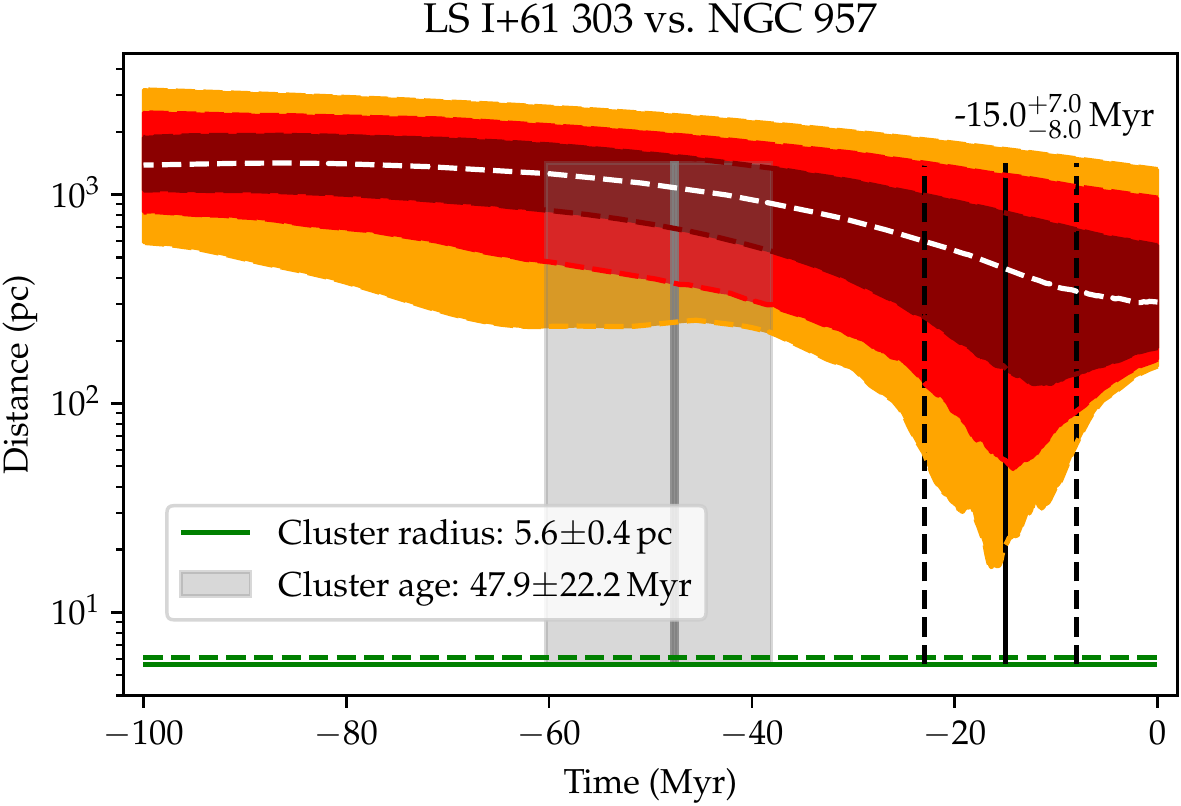}
    \includegraphics[width=0.4\columnwidth]{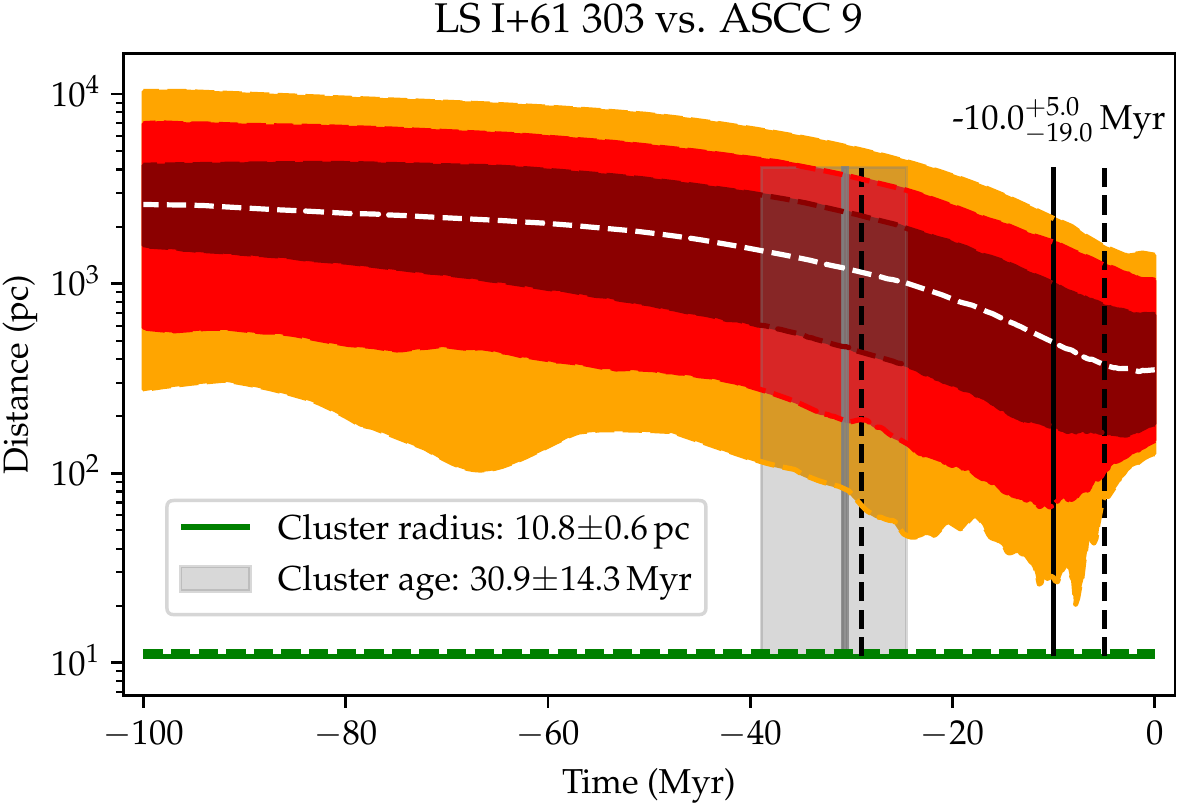}
    \includegraphics[width=0.4\columnwidth]{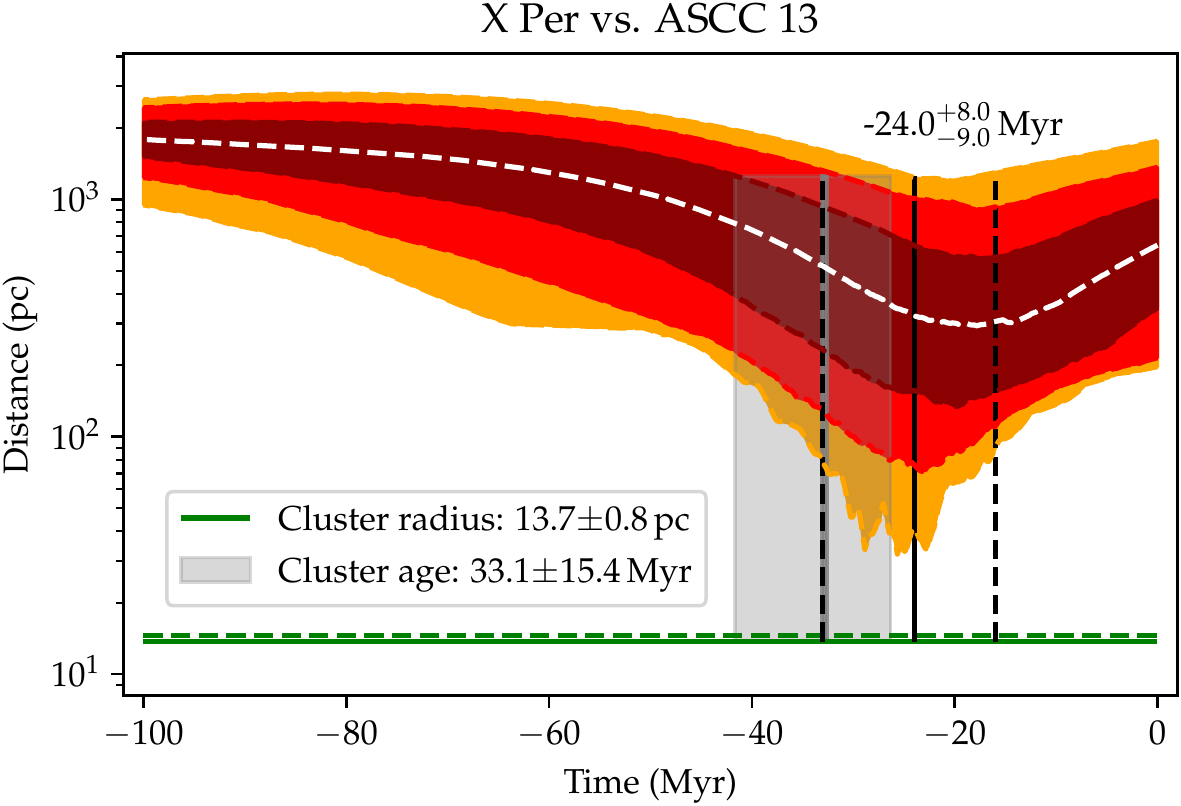}
    
    \caption{Time-distance histograms of encounter candidates between HMXBs and open clusters in the past 100\,Myr. Dotted white lines represent the median distance, and the 1 to 3$\sigma$ intervals are represented from dark red, to red, to orange. In green is indicated the physical radius of the clusters computed from their extension in the plane of the sky and their parallax. Their age range obtained through isochrone fitting is indicated in grey. The black vertical plain and dotted lines provide our estimation of the age since supernova in each binary-cluster encounter candidate.}
    \label{fig:HMclu_1}
\end{figure}

\begin{figure}[!htb]
    \ContinuedFloat
    \captionsetup{list=off,format=cont}
    \centering
    \includegraphics[width=0.4\columnwidth]{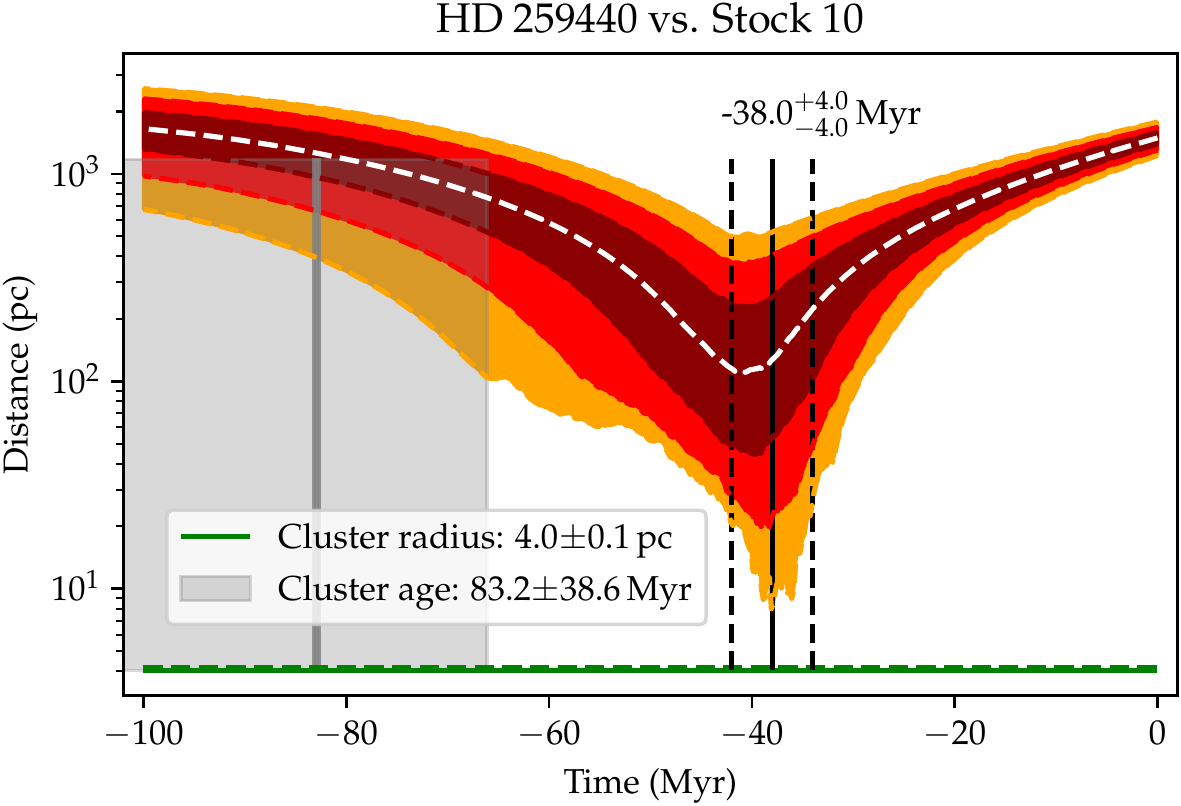}
    \includegraphics[width=0.4\columnwidth]{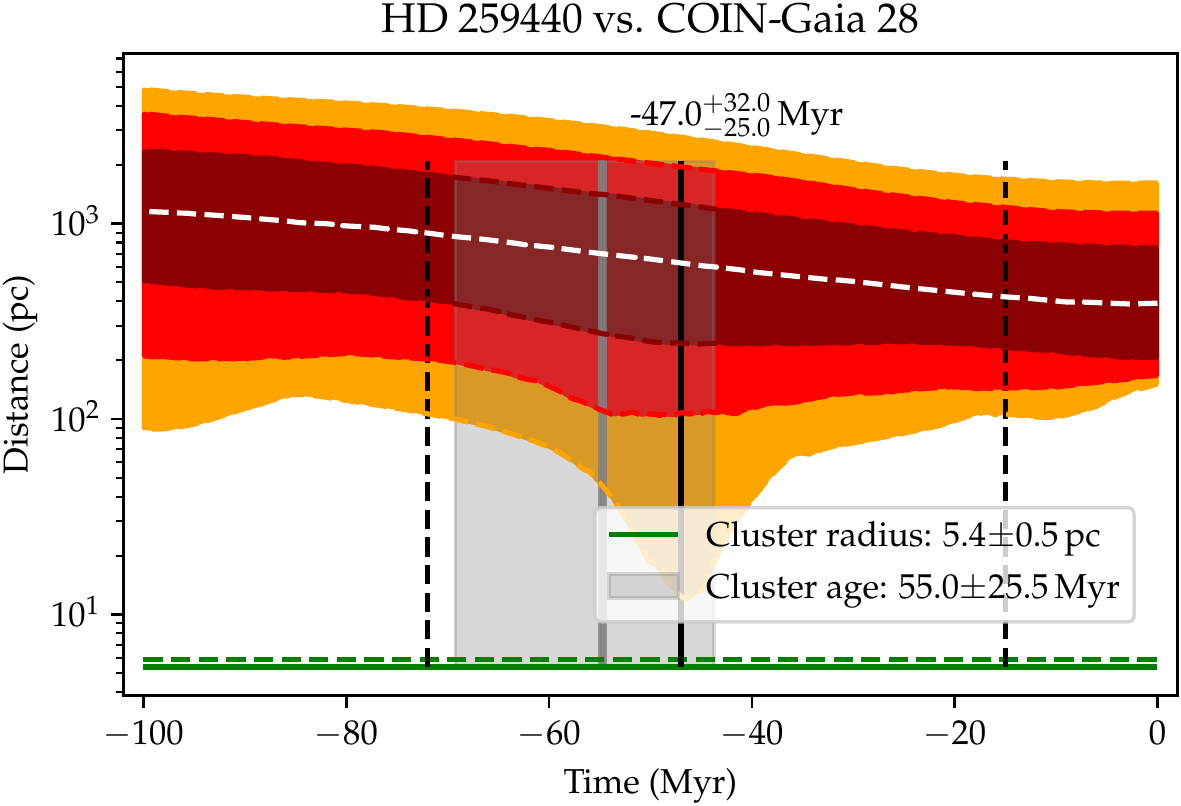}
    
    \includegraphics[width=0.4\columnwidth]{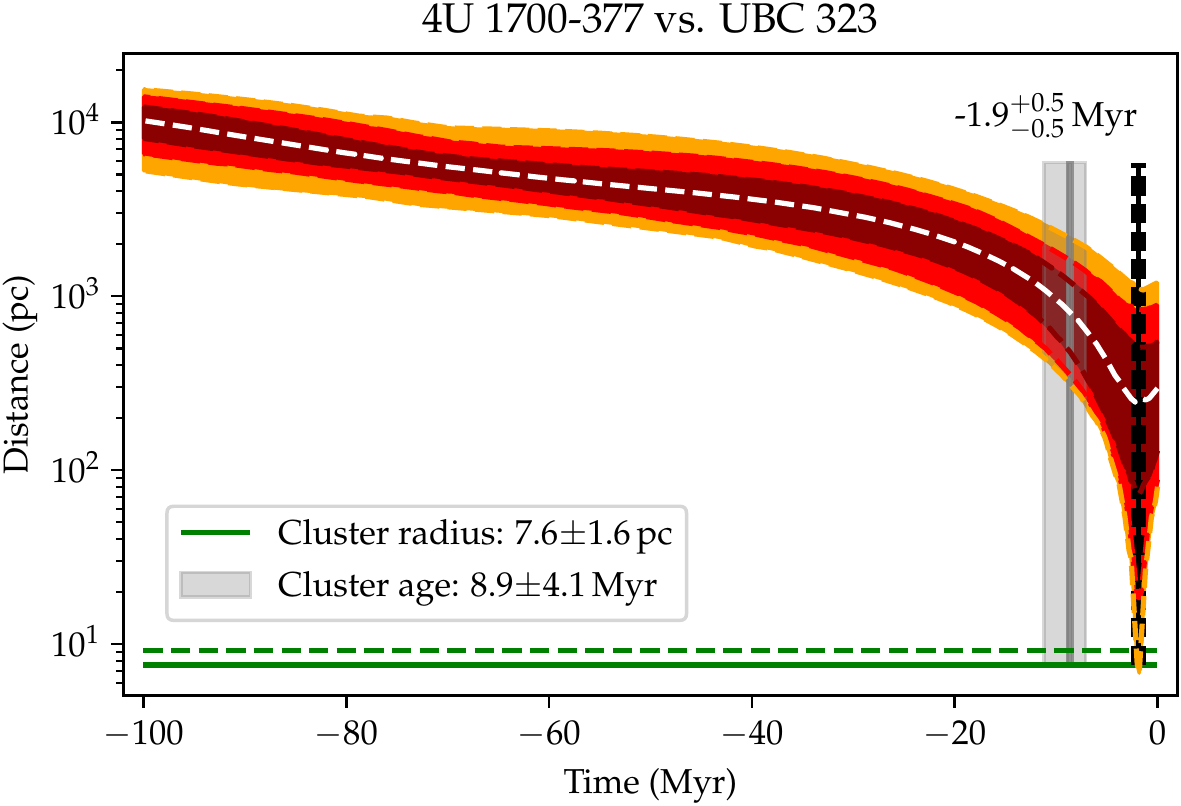}
    \includegraphics[width=0.4\columnwidth]{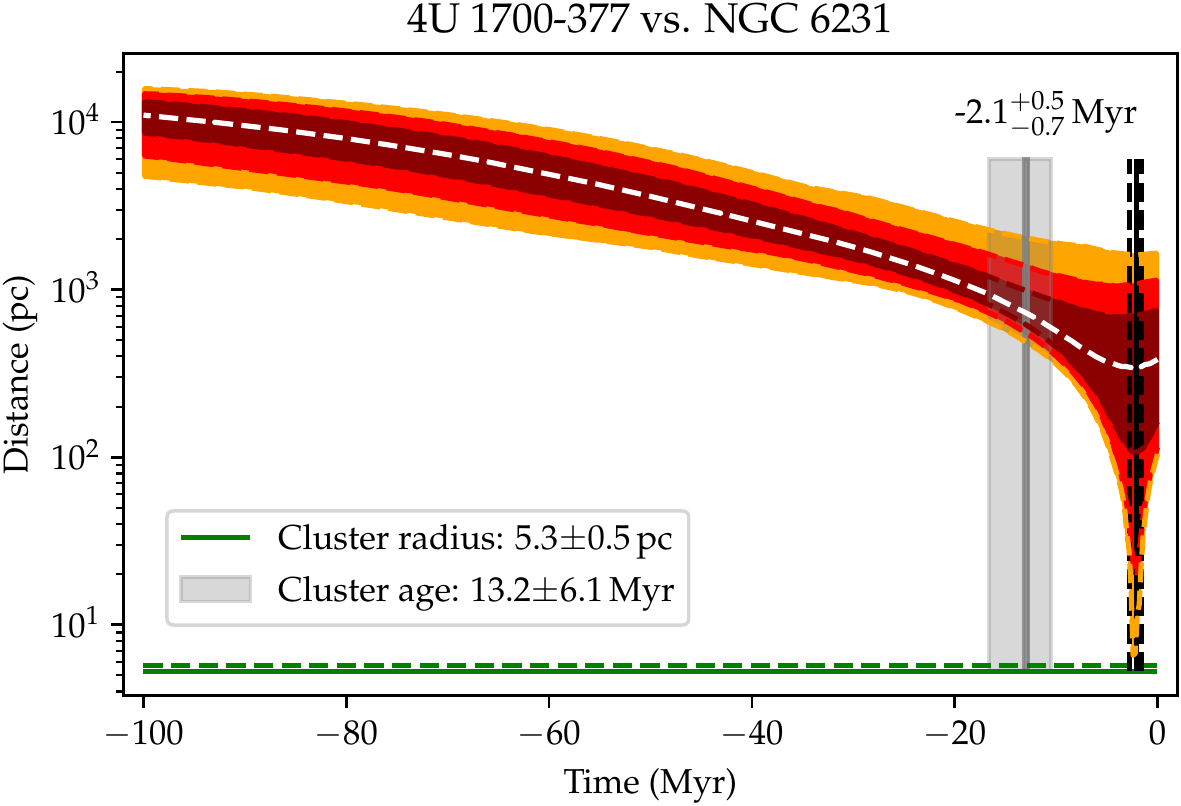}
    
    \includegraphics[width=0.4\columnwidth]{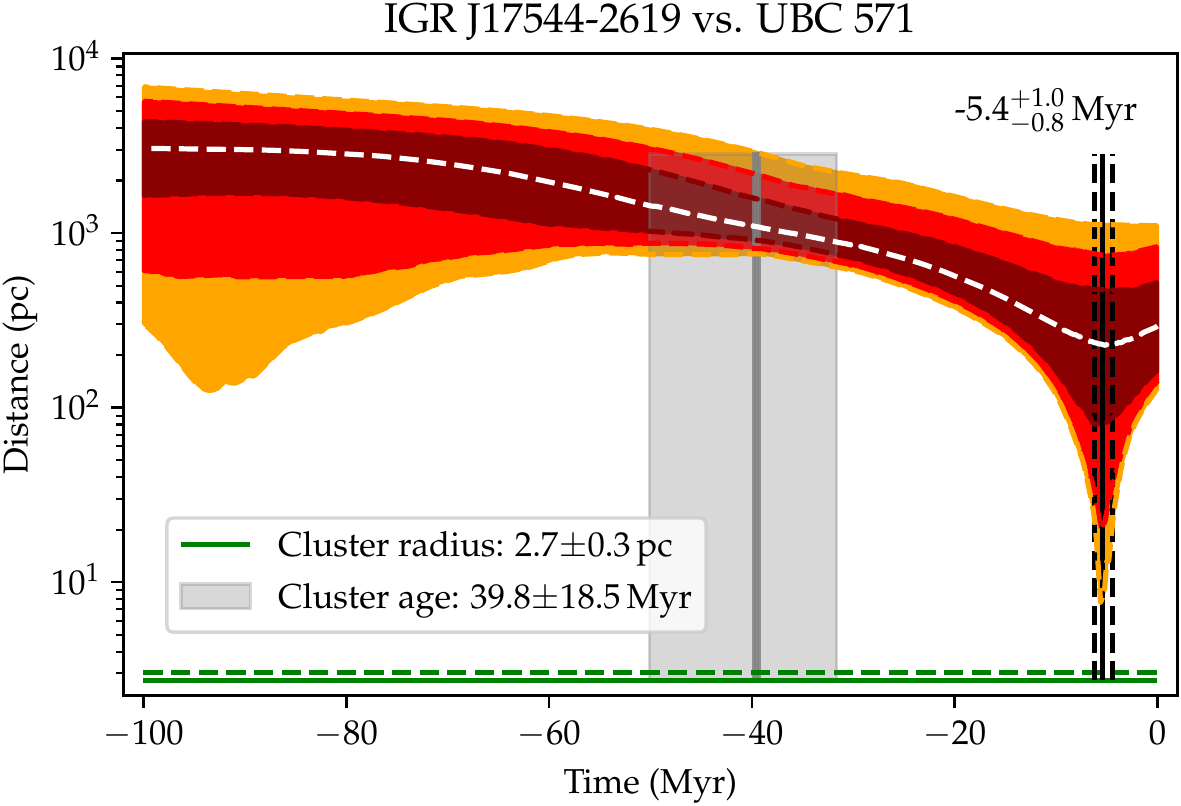}
    \includegraphics[width=0.4\columnwidth]{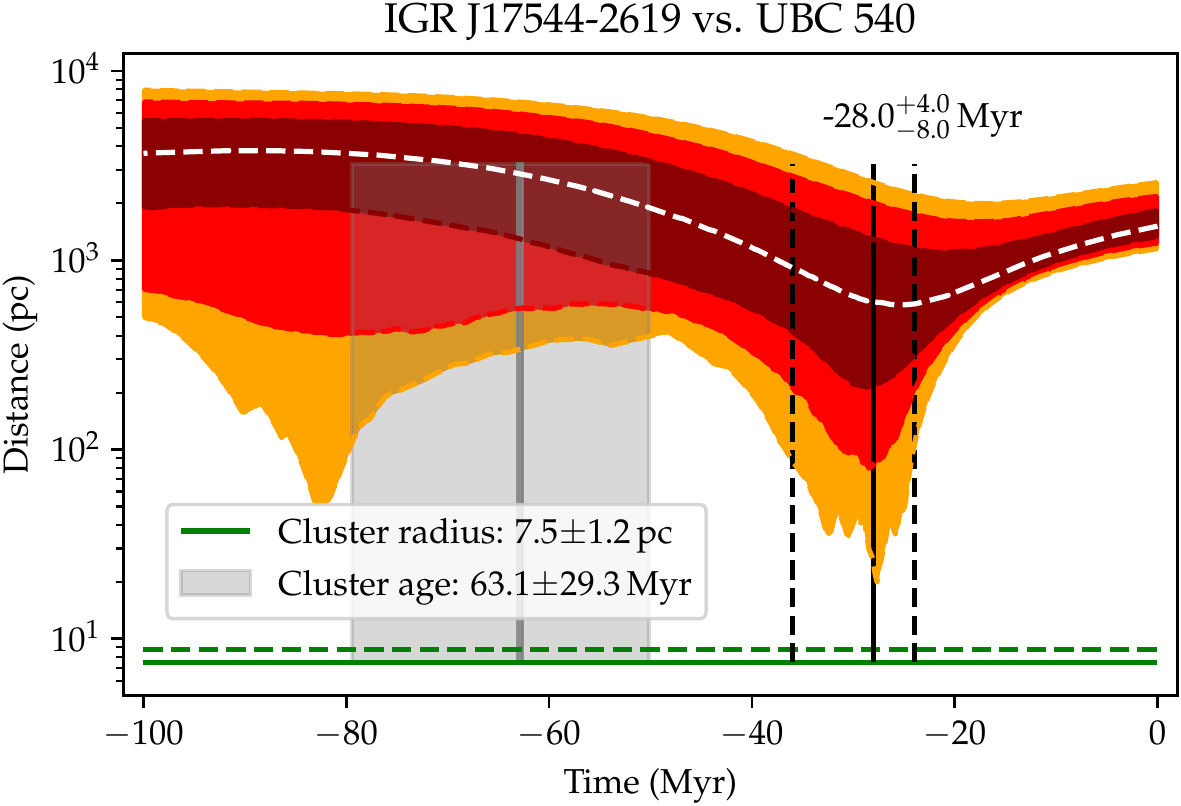}
    
    \includegraphics[width=0.4\columnwidth]{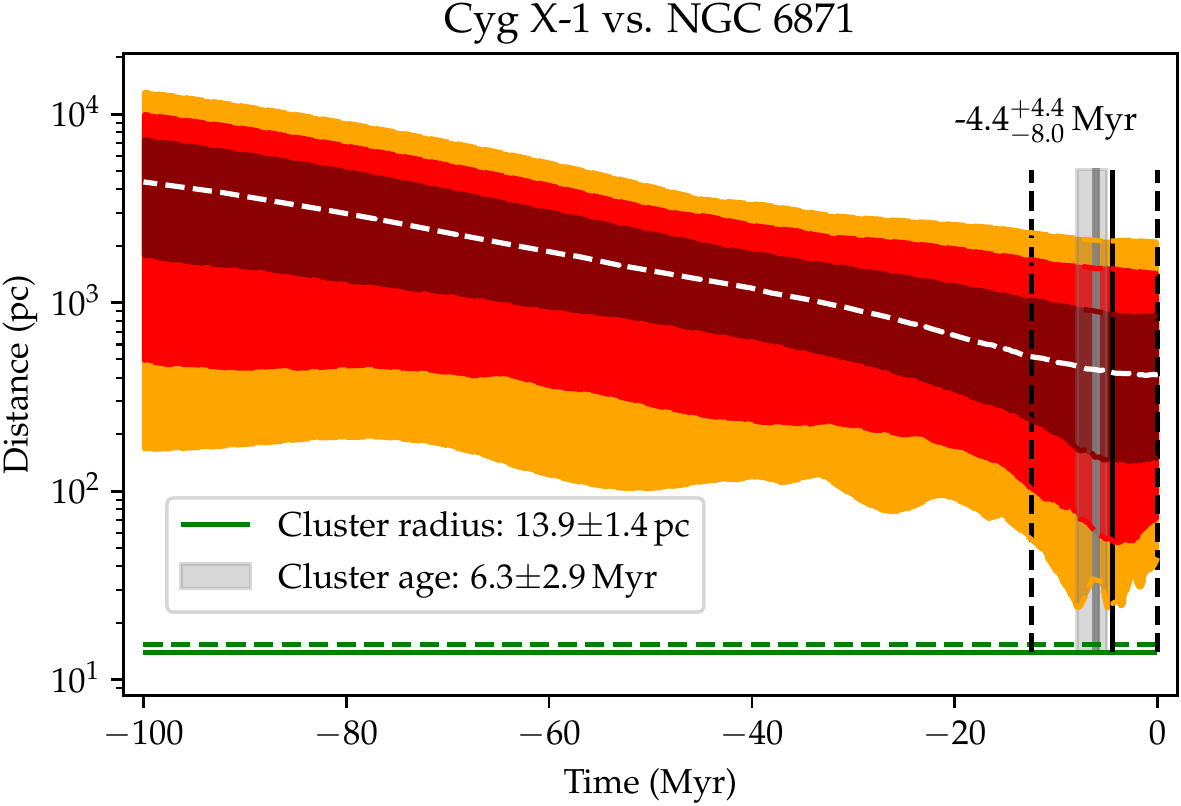}
    \caption{continued.}
    \label{fig:HMclu_2}
\end{figure}

\end{document}